\documentclass[apj]{emulateapj}

\usepackage{graphicx}
\usepackage{subfigure}
\usepackage{color}


\begin{document} 
\begin{abstract}
Using archival SDSS multi-epoch imaging data (Stripe 82), we have searched for flares due to the tidal disruption of stars by super-massive black holes in non-active galaxies. Two candidate tidal disruption events (TDEs) are identified, using a pipeline with high rejection efficiency for supernovae (SNe) and Active Galactic Nucleus (AGN) flares, and minimal selection bias.  The flares have optical black-body temperatures $ 2 \times 10^4 \, K$ and their cooling rates are very low; their observed peak luminosities are $M_g = -18.3$ and $-20.4$ ($\nu L_\nu = 5 \times10^{42},\, 4 \times10^{43} \,{\rm erg}\,{\rm s}^{-1}$, in the rest-frame), qualitatively consistent with expectations for tidal disruption flares.   The properties of the TDE candidates are examined using 
{\em i}) SDSS imaging to compare them to other flares observed in the search, {\em ii}) UV emission measured by \textsl{GALEX} and {\em iii}) spectra of the hosts and of one of the flares.  Our pipeline excludes optically identifiable AGN hosts, and our variability monitoring over 9 years provides strong evidence that these are not flares in hidden AGNs: 1) the luminosity increases in the candidate TDE flares are estimated to be greater by factors 4 and 15 than in any of the variable AGNs monitored, and 2) the TDE candidates' hosts are much quieter in the seasons not including the primary flare than are hosts of AGN flares (less activity on average in each of the non-primary-flare seasons than in 95\% of AGNs, with a cumulative probability estimated to be $\lesssim10^{-5}$ for a flaring AGN to have all other seasons as quiet as for the TDE candidates). 
The spectra and color evolution of the flares are unlike any SN observed to date, and their strong late-time UV emission is particularly distinctive.  These features, along with the high resolution with which they are placed at the nucleus, argue against these being first cases of a previously-unobserved class of SNe or more extreme examples of known SN types.  
Taken together, the observed properties are difficult to reconcile with a SN or AGN-flare explanation, although an entirely new process specific to the inner few-hundred parsecs of non-active galaxies cannot be excluded.  
Based on our observed rate, we infer that hundreds or thousands of TDEs will be present in current and next-generation optical synoptic surveys.   Using the approach outlined here, a TDE candidate sample with {\emph O(1)} purity can be selected using geometric resolution and host and flare color alone, demonstrating that a campaign to create a large sample of tidal disruption events, with immediate and detailed multi-wavelength follow-up, is feasible.  
A by-product of this work is quantification of the power-spectrum of extreme flares in AGNs. 
\end{abstract}

\title{Optical discovery of probable stellar tidal disruption flares}
\shorttitle{Optical discovery of tidal disruption flares}
\author{Sjoert van Velzen\altaffilmark{1,2,3}, 
Glennys R. Farrar\altaffilmark{1,4}, 
Suvi Gezari\altaffilmark{5}, 
Nidia Morrell\altaffilmark{6}, 
Dennis Zaritsky\altaffilmark{7}, \\
Linda \"Ostman\altaffilmark{8}, 
Mathew Smith\altaffilmark{9},  
Joseph Gelfand\altaffilmark{10} and
Andrew J. Drake\altaffilmark{11} } 
\email{s.vanvelzen@astro.ru.nl}
\altaffiltext{1}{Center for Cosmology and Particle Physics, New York Univerisity, NY 10003, USA}
\altaffiltext{2}{Astronomical Institute ``Anton Pannekoek'', University of Amsterdam, Postbus 94249, 1090 GE Amsterdam, The Netherlands}
\altaffiltext{3}{Department of Astrophysics/IMAPP, Radboud University, P.O. Box 9010, 6500 GL Nijmegen, The Netherlands}
\altaffiltext{4}{Department of Physics, New York University, NY 10003, USA}
\altaffiltext{5}{Department of Physics and Astronomy, Johns Hopkins University, Baltimore, MD 21218, USA}
\altaffiltext{6}{Carnegie Observatories, Las Campanas Observatory,  Casillas 601, La Serena, Chile}
\altaffiltext{7}{University of Arizona, Tucson,  AZ 85721, USA}
\altaffiltext{8}{Institut de F\'isica d'Altes Energies, Universitat Aut\`onoma de Barcelona, E-08193 Bellaterra (Barcelona), Spain}
\altaffiltext{9}{Department of Mathematics and Applied Mathematics, University of Cape Town, Rondebosch, 7701, SA}
\altaffiltext{10}{New York University - Abu Dhabi, PO Box 129188, Abu Dhabi, United Arab Emirates}
\altaffiltext{11}{California Institute of Technology, 1200 E. California Blvd, CA 91225, USA}

\maketitle
\section{Introduction}\label{sec:intro}
When a star passes too close to a super-massive black hole, the tidal shear overcomes the star's self-gravity and the star is disrupted.  Much of the stellar debris is ejected from the system, but some fraction remains bound to the black hole and is accreted, resulting in a week- to year-long electromagnetic flare \citep{Rees88}. The fall-back rate is expected to follow a power-law of index \mbox{$-5/3$} \citep{Rees88,Phinney89}, but see \citet*{Lodato09} for predicted deviations from this canonical scaling; the light-curve does not in general show the same behavior, \citep[e.g.,][]{strubbe_quataert09,Lodato_Rossi10}.  For $M_{\rm BH}\lesssim 10^{7} M_{\odot}$ the initial fall-back rate is super-Eddington and the emission is usually assumed to have a black body spectrum.  The predicted temperatures range from {$\sim 10^4$ K} \citep{loebUlmer97} to, more commonly, {$\sim 10^5$ K} \citep[e.g.,][]{Ulmer99}.  A radiatively driven wind, existing as a consequence of the super-Eddington fallback, may dominate the emission of the tidal disruption event (TDE) at early times \citep{strubbe_quataert09}.  Model predictions generally depend on parameters which are quite uncertain, so observational input is needed to constrain the modeling.  

Detections of tidal disruption events are of interest for a number of reasons, including:  1) The light emitted after the disruption depends sensitively on the black hole mass and spin, hence a large sample of TDEs will allow properties of back holes to be studied without relying on scaling relations with global parameters of galaxies \citep{Gebhardt00,Ferrarese_Merritt00,Graham01,Marconi_Hunt03}.  2) TDEs may be our only probe to obtain a large sample of dormant super-massive black holes \citep{Frank_Rees76, Lidskii_Ozernoi79}.  3) A particularly intriguing application is testing the existence of intermediate mass black holes in globular clusters and dwarf galaxies \citep{Ramirez-Ruiz09}.  4) For black holes with mass $M_{\rm BH}\gtrsim 10^{8} M_{\odot}$, the tidal disruption radius lies inside the Schwarzschild radius \citep{Hills75}, hence a TDE survey that covers a sample of galaxies with a wide enough central black hole mass range is in principle sensitive to whether super-massive black holes do have an event horizon.  5) Detailed observations of the emission from a large sample of tidal disruption events will provide a new arena for testing our understanding of accretion physics and may constrain properties of the disrupted stars. 

A number of candidate TDEs have been identified in X-ray surveys \citep{Bade96, KomossaBade99,donley02, Esquej08, Cappelluti09,Maksym10} ---for a review see \citet{Komossa02}--- and in the UV \textsl{GALEX} Deep Imaging Survey \citep{Gezari06,Gezari08,Gezari09}. However establishing that a candidate tidal flare found in UV and X-ray surveys is not an exceptionally variable AGN is hampered because, although the amplitude of AGN variability at these wavelengths is much greater than in the optical \citep{Maoz05, Saxton11}, the range of variation has not yet been well-characterized. With the advent of optical transient surveys, such as the Palomar Transient Factory \citep[PTF;][]{Law09}, Pan-STARRS \citep{Chambers07}, Catalina Real-Time Transient Survey (CRTS) and later the Large Synoptic Survey Telescope \citep[LSST;][]{Ivezic08}, there should be many candidate TDEs.  The challenge is to eliminate the far more common flares from variable AGNs and SNe to a) convincingly exclude these backgrounds as explanations for individual events and b) efficiently produce a high-purity sample of TDE candidates such that expending resources following up uninteresting events can be reduced to an acceptable level.  

Until now, the feasibility of identifying probable tidal flares with an optical transient survey alone has not been demonstrated.  Here we present two candidate stellar tidal disruption events detected in archival multi-epoch imaging data of the Sloan Digital Sky Survey (SDSS), showing that the difficulties of identifying such events in an optical survey can be resolved.  There are several important advantages of using the SDSS data.  First is the very large sample of galaxies, with many having spectra;  this sample enables one to classify flares into well-defined categories, such that the validity of our methodology can be demonstrated.  Second, the spatial resolution of SDSS is adequate to exclude a very high fraction of SNe from the nuclear-flare sample, allowing SNe to be rejected without imposing cuts based on flare properties.  Third, SDSS observed the galaxies in Stripe 82 over typically 7 seasons with a mean of 70 observations in all; this allows hidden AGNs to be rejected based on variability in the non-peak seasons.  Having observations in three or more filters is also useful, providing color information that is valuable in confirming that the flares are not due to AGNs or known types of SNe. 

The selection pipeline we employ reduces the background from SNe and variable AGNs by two orders of magnitude or more, but positive determination that the two flares which pass the pipeline are in fact TDEs requires detailed comparison between the properties of the flares and those of AGNs and SNe.  This we do with multiple tools.  We have obtained spectra of both hosts and of one of the flares, found archival \textsl{GALEX} post-flare observations of both host galaxies and a pre-flare observation of one of them, and archival Catalina Real-Time Transient Survey data to extend the light curve of one of the flares to earlier and later times than observed by SDSS.  Analysis of these observations and comparison between the properties of the TDE candidates and SNe and AGNs is reported.

In addition to demonstrating the feasibility of optically-initiated TDE surveys, discovery of these two flares gives needed observational insight into the tidal disruption phenomenon.   An important consequence of our method of detection -- which does not rely on flare properties beyond requiring them to be nuclear to reject SNe and uses host properties alone to reject AGNs -- is that selection bias is minimized.  With just two events we have only begun to scratch the surface, but the properties of the flares can already test and inform theory.   Our candidate TDEs are both much more luminous in the optical than predicted \citep{strubbe_quataert09}, although with adjusted parameter choices the fit can be improved and some ingredients may still be missing in these early, simple models.    

The outline of the remainder of this paper is as follows. In Section \ref{sec:pipeline} we describe our pipeline for TDE selection. In Section \ref{sec:properties} we present the observed properties of the final products of the pipeline: two candidate tidal disruption flares. Follow-up observations with other instruments are reported.  A detailed comparison to flares of active galaxies and supernovae is given in Section \ref{sec:compare}, leading to the conclusion that the two TDE candidates are indeed likely to be tidal disruption events. 
Section \ref{sec:discussion} compares our TDEs to candidates reported in other wavelengths and compares their observed properties to theory.  The implications of this work for detecting TDEs and obtaining a relatively pure sample of them in future optical surveys is discussed in Section \ref{sec:future}.  We close with a summary (Section \ref{sec:summary}).

All magnitudes are quoted in the AB system \citep{oke74} and are corrected for Galactic extinction \citep{Schlegel98}. We adopt a standard cosmology with $H_0 = 72\, {\rm km}\,{\rm s}^{-1}{\rm Mpc}^{-1}$, $\Omega_m=0.3$ and $\Omega_\Lambda=0.7$.
\bigskip

\section{Identification of tidal flares}\label{sec:pipeline}

\begin{figure}[t]
 \includegraphics[ trim =0mm 80mm 0mm 0mm, clip, width=.50 \textwidth]{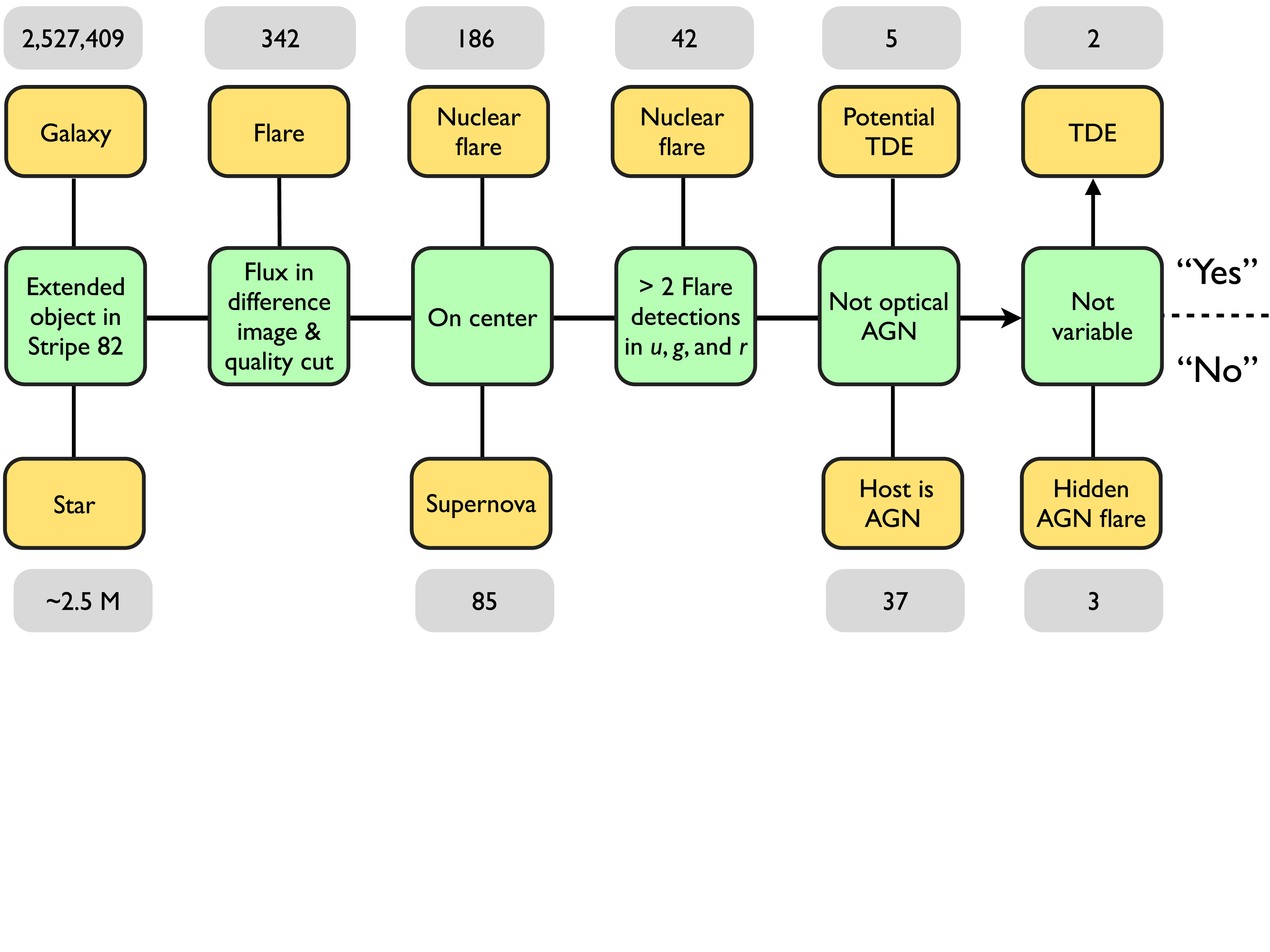} 
\caption{Flow chart summarizing the flare and TDE identification pipeline, discussed in Section \ref{sec:pipeline}. The numbers in the grey boxes indicate the number of objects in each class. The first three steps are discussed in Sections \ref{sec:catalog_cuts}--\ref{sec:h_d};  the last three steps summarize the selection for TDE candidates discussed in Sections \ref{sec:pot_tde}--\ref{sec:rel_flux}.}\label{fig:flow}
\end{figure}

The flowchart in Fig. \ref{fig:flow} summarizes our search graphically. In the following sections we discuss the steps in this chart in more detail; for additional details see S. van Velzen, G. R. Farrar et al. (2011).
To achieve the goal of obtaining a large and uniformly selected sample of tidal disruption events (TDEs), it is necessary to accurately identify their flares in large volumes of optical data.  Here we use SDSS observations of $ 2.5 \times10^{6}$ galaxies in ``Stripe 82'', that have been observed typically 70 times over a nine year baseline. In Section \ref{sec:sdss} we explain how a sample of 342 flares was extracted from this data set. Supernovae or other stellar flares are not of interest in a search for TDEs, and in Section \ref{sec:select_nuc} we explain how they are removed by discarding flares offset from the center of the host.  The final sample for our study consists of 42 nuclear flares having more than two observations in the flaring state.  

The rejection of active galactic nuclei (AGN) is discussed in Section \ref{sec:AGN_rejection}. Eliminating host galaxies with spectroscopic and photometrically identified AGNs removes all but 5 nuclear flares.  The critical challenge is to identify and exclude host galaxies with active nuclei which are too weak or heavily obscured to be identified by usual spectroscopic and photometric criteria.  We can identify variable unrecognized AGNs directly by their variability, thanks to SDSS's multi-year monitoring of the hosts.  Three of the five candidates do display variability in other seasons, consistent with the variability we measure in our identified AGNs, while two others do not, at the $< 10^{-5}$ CL level (see Section \ref{sec:agn_compare}). On geometric grounds alone, the probability these two flares are SNe is $< 0.5\%$ (see Section \ref{sec:sn_compare}), assuming the distribution of SNe follows the stellar light. Thus these two flares are strong candidates to be stellar tidal disruption events and we refer to them as TDE1 and TDE2 below. 

\subsection{Selecting flares in SDSS observations}\label{sec:sdss}
The $\sim300\, \mathrm{deg}^2$ Sloan Digital Sky Survey \citep[SDSS;][]{york02} multi-epoch imaging data Stripe 82 \citep{sesar07,bramich08,frieman08, Abazajian09} was the starting point of our search. Using the SDSS morphological star-galaxy separation \citep{lupton_gunn01, stoughton02} and the standard checks on quality flags \citep{stoughton02}, we extracted $\sim 2.5\:10^{6}$ galaxies from this data set. The typical number of observations per galaxy is 70.

\subsubsection{Catalog cuts}\label{sec:catalog_cuts}
\begin{deluxetable}{ c c }
  \tablewidth{0pt}    
  \tablecolumns{2}
  \tablecaption{Summary of catalog cuts.\label{tab:catcuts}}
  \tablehead{\colhead{Cut} & \colhead{SDSS band} }
  \startdata
  $m_{\mathrm{mean}} < 22.5$ &  $\geq 3$ \\ 
  identified as extended in co-add & all \\
  $\chi^2/{\rm DOF} > 5$ &  \emph{g},\emph{r}, or \emph{i} \\
  $ F_{\mathrm{peak}} / F_{\mathrm{mean}} >1.1$ & $\geq 2 $  \\ 
  $ (F_{\mathrm{peak}}-F_{\mathrm{baseline}}) / \sigma_{\mathrm{peak}} > 7 $ &  $\geq 2$\\ 
  $ (F_{\mathrm{peak}}-F_{\mathrm{baseline}}) / \sigma_{\mathrm{\rm rms}} > 3 $ &  $\geq 2$\\ 

  \enddata
  \tablecomments{The first two cuts are designed to obtain a clean, flux limited sample of galaxies. 
The co-add runs are obtained by addition of nearly all Stripe 82 imaging data \citep{Abazajian09},  J. Annis et al. (2009), in preparation. 
The goal of the remaining cuts is to select flares. For all cuts we use the Petrosian flux \citep{stoughton02}. The subscript `mean' is the inverse-variance-weighted mean using all observations, while the subscript `baseline' refers to the non-flare observations only.  $\chi^2$ is calculated using mean flux as a model for the galaxy light curve. These cuts select $\sim2 \times 10^4$ candidate flares from the $\sim 2\times10^6$ galaxies with $m<22.5$ in Stripe 82.}
\end{deluxetable} 

In order to focus our analysis resources and time most efficiently, we first selected candidate flares from the cataloged parameters, which reduced the number of galaxies by two orders of magnitude.  Galaxies were required to have $m<22.5$ in at least 3 of the 5 SDSS bands \citep[\emph{u,g,r,i,z};][]{fukugita96,smith02}.  
A minimum flux increase of 10\%, measured at the 7-$\sigma$ level, is the most important requirement that was imposed to find flares. The cuts on the cataloged parameters were chosen to be ``soft'' (i.e., high efficiency, low purity) and yielded 21,383 potential flares. See  Table \ref{tab:catcuts} for a summary of the catalog cuts.

\subsubsection{Difference imaging}
Next, with this data set that is a factor of 100 smaller than the starting one, we applied a more rigorous analysis method: image subtraction. Given that there are typically 70 observations of a galaxy, we have many images of the host of the flare; from these we selected the observations with the best seeing and lowest sky level to be the reference images for subtraction. First we implemented a simple and relatively fast ``direct subtraction'' method, which allows quick determination that 8834 flare candidates are spurious because they show no flux in the difference image. 
For each of the remaining galaxies hosting a flare, nine reference images (of size $1'\times 1^{\prime}$) were cross convolved using a modified version of the software by \citet{yuan_akerlof08} and subtracted to obtain nine difference images, for each filter in each night. The convolution kernels were determined using nine reference stars that were selected to lie close to the host galaxy. 
The mean flux of the nine difference images and its standard deviation were computed for each pixel, after applying a clipping algorithm to reject pixels that lie more than 3$\sigma$ from the median.  The flux, $F$,  in the mean difference image was computed using an aperture of 2 times the full-width-half-maximum (FWHM) of the point spread function (PSF) measured in the mean image of the convolved reference stars. 
The minimum flux in the difference image was required to be $m<22$. In addition, we require for each band that the flare is detected at the 7-$\sigma$ level.
We further require that the shape of the source in each difference image is well-fit by the PSF extracted from the mean convolved reference stars for that image. 

Of the flares meeting the above conditions, 583 satisfy the requirement of detections in at least two bands. Good agreement on the flux in the difference image is obtained for the light curves produced by \citet{Holtzman08} of supernovae from the SDSS SN Survey \citep{frieman08,sako08} which were detected independently by our pipeline. 

\subsubsection{Rejecting moving objects}
Solar system objects, when seen in front of a galaxy, can fake a flare. SDSS checks for moving objects by measuring the position of the centroid across different filters \citep{ivezic01}. We eliminate moving objects flagged by SDSS \citep{stoughton02}, and identify 32 additional solar system objects using the position of the centroid in the difference image. By requiring detections in the difference image in at least two observing nights, false flares due to solar system objects are eliminated and we are left with 419 flares. 

\subsubsection{Manual rejection}
\begin{figure}
\subfigure[Rejected: bad subtraction, one noisy filter]{
{\includegraphics[width=80 pt]{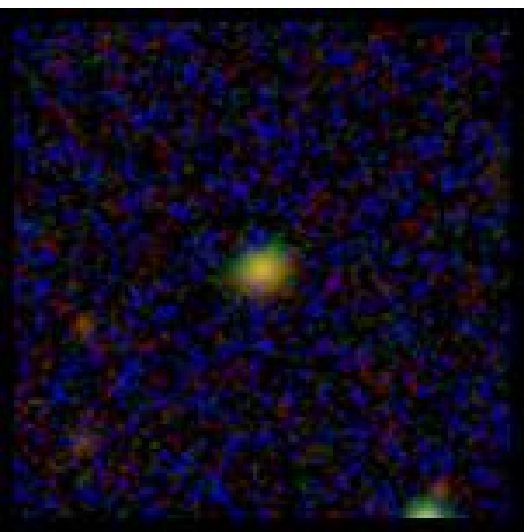}}
{\includegraphics[width=80 pt]{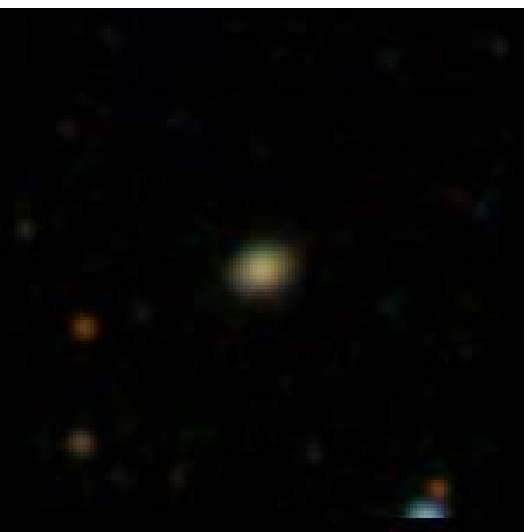}}
{\includegraphics[width=80 pt]{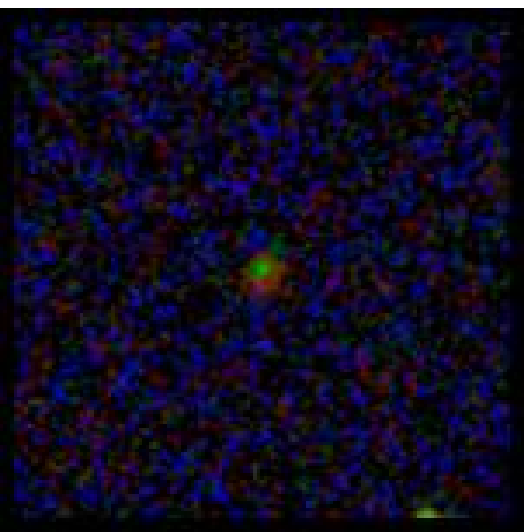}} } \\ 
\subfigure[Rejected: bad subtraction, peak displaced between filters]{
{\includegraphics[width=80 pt]{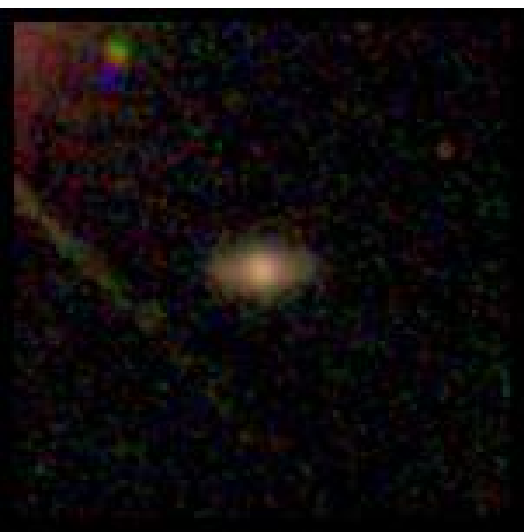}}
{\includegraphics[width=80 pt]{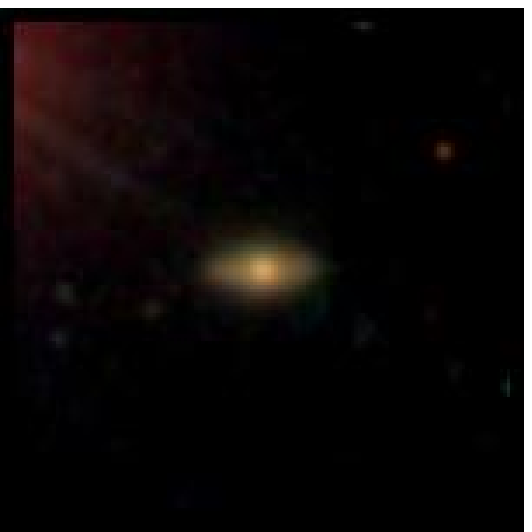}}
{\includegraphics[width=80 pt]{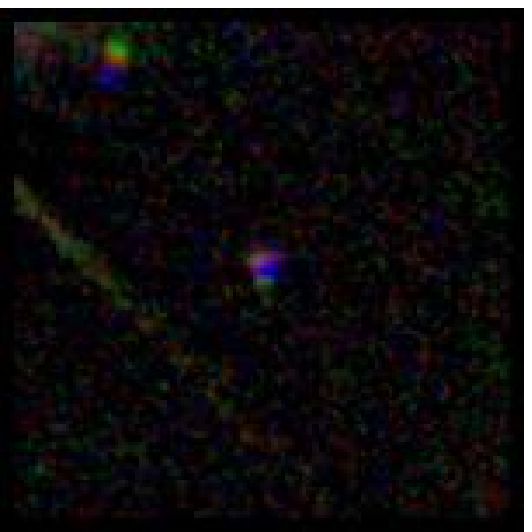}} }
\subfigure[Good subtraction: near the flux limit]{
{\includegraphics[width=80 pt]{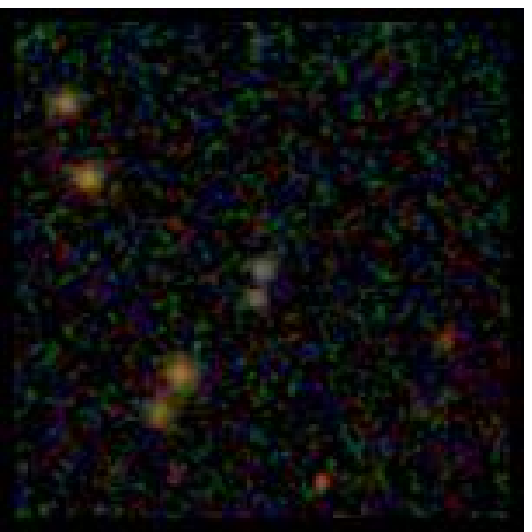}}
{\includegraphics[width=80 pt]{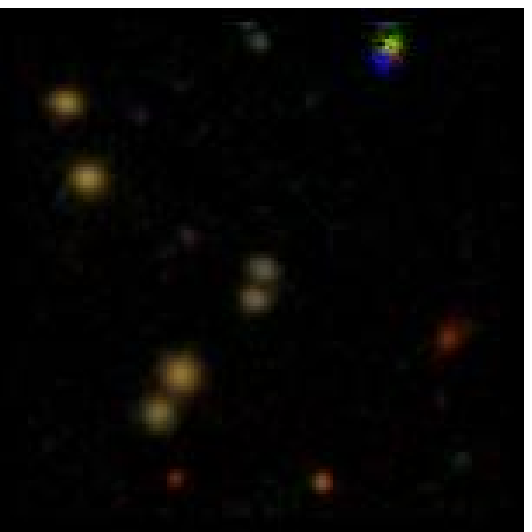}}
{\includegraphics[width=80 pt]{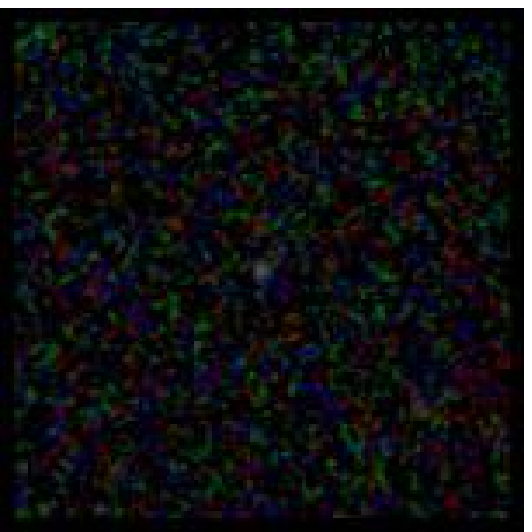}} } \\ 
\subfigure[Good subtraction: off-center flare (SN) ]{
{\includegraphics[width=80 pt]{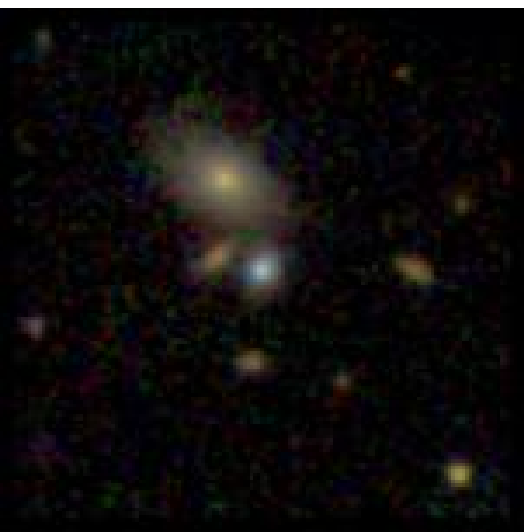}}
{\includegraphics[width=80 pt]{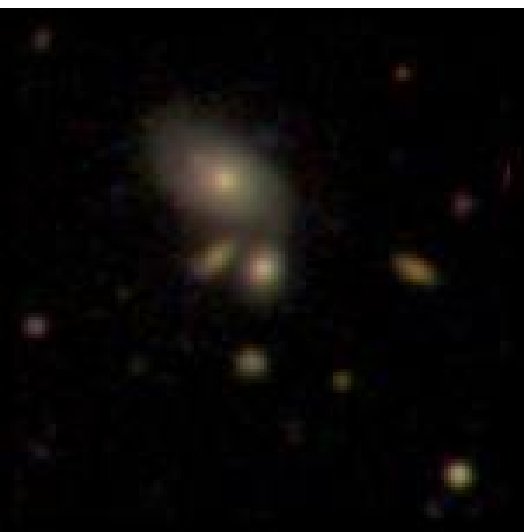}}
{\includegraphics[width=80 pt]{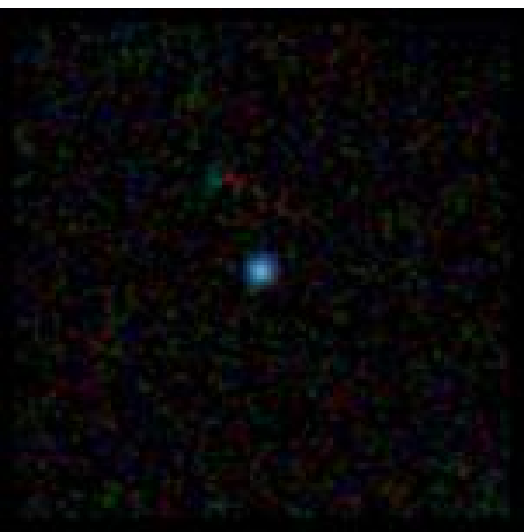}} } \\ 
\caption{Examples of bad and good image subtractions. Image size is $1' \times 1'$. Left to right: flare image, mean reference image and difference image.  The bad subtractions, images (a) and (b),  were rejected by manual inspection; image (c) shows the quality of a subtraction that is relatively faint, $m_r = 21.5$ in the difference image. Image (d) shows an example of an off-center flare detected at $r=0\farcs4\pm 0\farcs02$.}\label{fig:example_ims}
\end{figure}

We remove galaxies with a spectroscopic redshift $> 1.2$, because they are mis-classified as extended. This requirement removes 7 quasars from the sample. Finally, we inspect the difference images of the remaining 412 flares to search for bad subtractions or other anomalies. The number of flares rejected after this inspection is four, which is quite low considering the large number of galaxies processed by the image subtraction pipeline ($\sim2 \times 10^4$). In Fig. \ref{fig:example_ims} we show examples of rejected and good subtractions.  Additional quality cuts on the distance between the flare and the center of its host are discussed below. 

\subsection{Selection of nuclear flares}\label{sec:select_nuc}
An accurate measurement of the distance between the center of the host galaxy and the flare is crucial for obtaining a clean nuclear flare sample (i.e., removing SNe). In Section \ref{sec:d} we discuss the details for accurately measuring this distance and in Section \ref{sec:h_d} we explain how the host-flare distance is used to define the nuclear flare sample.

\subsubsection{Precise determination of host-flare distance}\label{sec:d}
\begin{figure}
\begin{centering}
 \includegraphics[ trim =0mm 0mm 0mm 75mm,  width=.5 \textwidth]{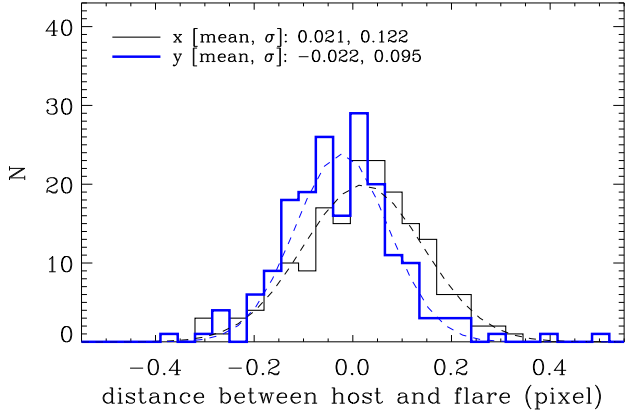} 
\caption{Histogram of the distance between the center of the host and the flare for all 186 nuclear flares (eq. \ref{eq:cut}). The dashed lines display the best-fit Gaussian distribution. As expected for nuclear flares, the distribution peaks at zero. The mean accuracy on the distance between the host and the flare for the nuclear sample is about $0.15$ pixel or $0\farcs06$. This is similar to the SDSS single epoch astrometric accuracy for point sources brighter than $m_r\sim 20$ \citep{pier03}.}\label{fig:h_d_oncen}
\end{centering}
\end{figure}
 In this section we consider first the accuracy with which we determine the position of the flare with respect to the Sersic center of the host, and then the extent to which the Sersic center of the host is an accurate determination of the true galactic center.

The location of the flare relative to the host is found by fitting a point source corrected by the PSF to the difference image. We repeat these measurements for all reference images in all nights with detections in the $g$, $r$ or $i$ bands; this yields at least $9\times2\times2=36$ measurements of the distance $d$ between the flare and the center of the host. The uncertainty on this distance in pixel coordinates, $\sigma_x$ and $\sigma_y$, is obtained from the standard deviation of these measurements. As seen in Fig. \ref{fig:h_d_oncen}, the typical measurement uncertainty on the distance between the host and the flare is $0.1$ pixel. We use this as a lower limit for $\sigma_{x,y}$. To divide flares into categories of clearly nuclear, clearly non-nuclear or ambiguous, we define $\sigma_d \equiv \sqrt{(x\sigma_x)^2+(y\sigma_y)^2}/d$, which is a measure of the uncertainty on the host-flare distance.

\begin{figure}
\begin{center}
 \includegraphics[ trim =0mm 0mm 0mm 75mm, width=.5 \textwidth]{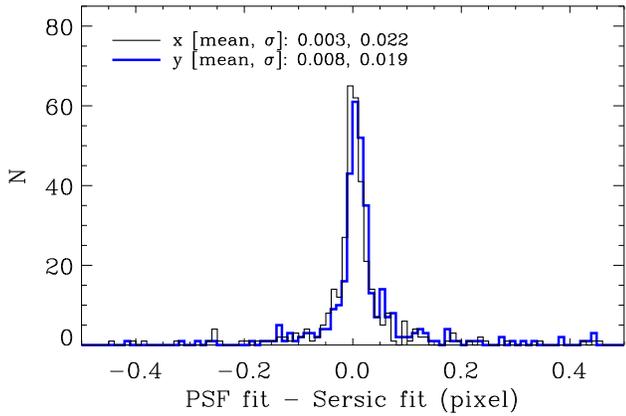} 
\caption{The distance between the center of the galaxy obtained by fitting a Sersic profile and the center obtained by fitting a PSF. Irregularities in the galaxy may push the center obtained by the Sersic fit away from the true geometrical center. To enhance this potential systematic uncertainty, we subtract 50\% of the best-fit Sersic model galaxy from the original image before fitting the PSF to find the center. We find that for almost all galaxies the two methods of locating the center agree very well: the Sersic profile is not significantly affected by irregularities. (1 pixel corresponds to $0\farcs4$) }\label{fig:sersic_sys}
\end{center}
\end{figure}

In the above, $d$ is defined with respect to the coordinates of the optical center of the host obtained by fitting a Sersic profile to the convolved reference image.  However a deviation from azimuthal symmetry in the galaxy (e.g., an H II region) may cause the center found by fitting a Sersic profile to be offset from the true center, thus causing an error in determining the separation of the flare from the galaxy center. We have investigated the impact of this on our accuracy, galaxy-by-galaxy, by comparing the center obtained from a Sersic fit to the center obtain by fitting the PSF to the galaxy. To enhance any deviation, we subtract 50\% of the best-fit azimuthally symmetric model galaxy from the original image before fitting the PSF. As shown in Fig. \ref{fig:sersic_sys}, the two methods on average yield the same answer. Visual inspection of the few galaxies with large ($>0.1$ pixel, i.e., $0\farcs04$) deviations between the two ways of measuring the center shows that they indeed have more complicated shapes or consist of close pairs of galaxies. In those rare cases that the magnitude of the difference between the PSF and Sersic centers in either coordinate is larger than the uncertainty on the separation between the flare and Sersic center, we take the flare positional uncertainty $\sigma_{x,y}$, to be the former.  

Now, with the positional uncertainties understood, we proceed to define our flare sample.  First, we set the maximum distance between the host and the flare to $d<1''$ and remove all 
transients detected beyond this radius from our flare sample. In addition, we demand that the distance between the host and the flare is measured with a minimum accuracy of $\sigma_d< 0\farcs1$. After applying these cuts, we are left with flares in 342 galaxies. 

\subsubsection{Separating nuclear flares from non-nuclear flares}\label{sec:h_d}
\begin{figure}
\begin{center}
\includegraphics[ trim =0mm 15mm 20mm 95mm, clip, width=.50 \textwidth]{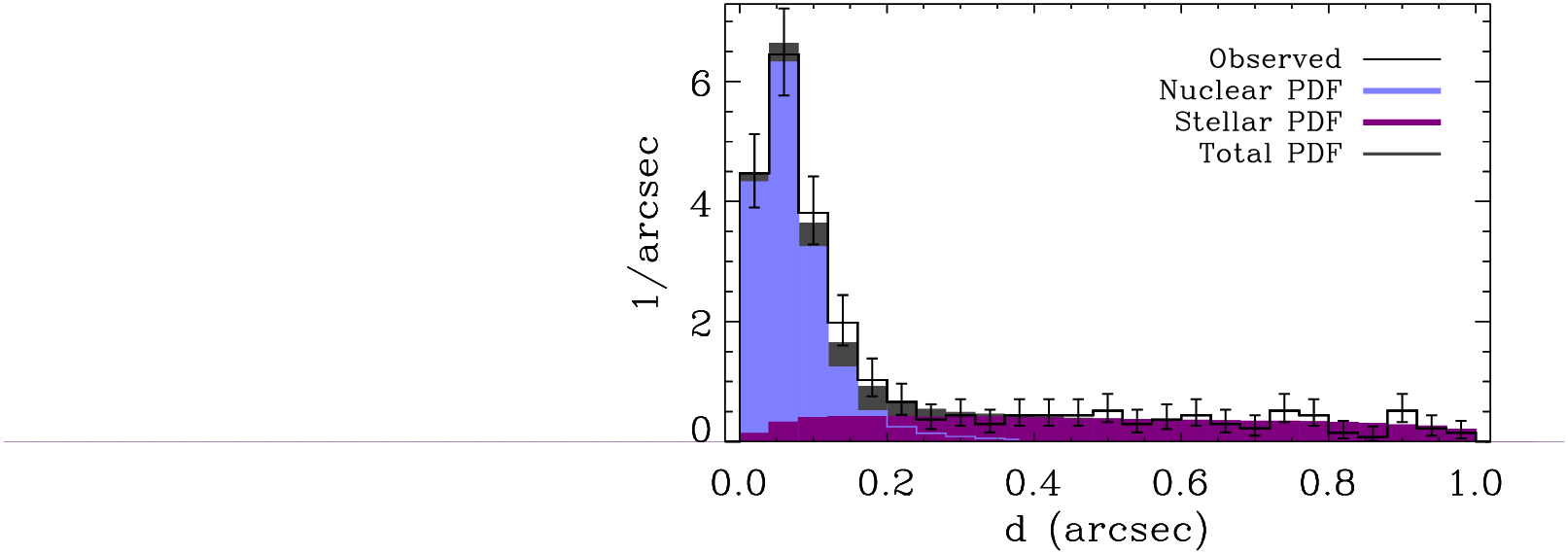} 
 \includegraphics[ trim =0mm 0mm 20mm 95mm, clip, width=.50 \textwidth]{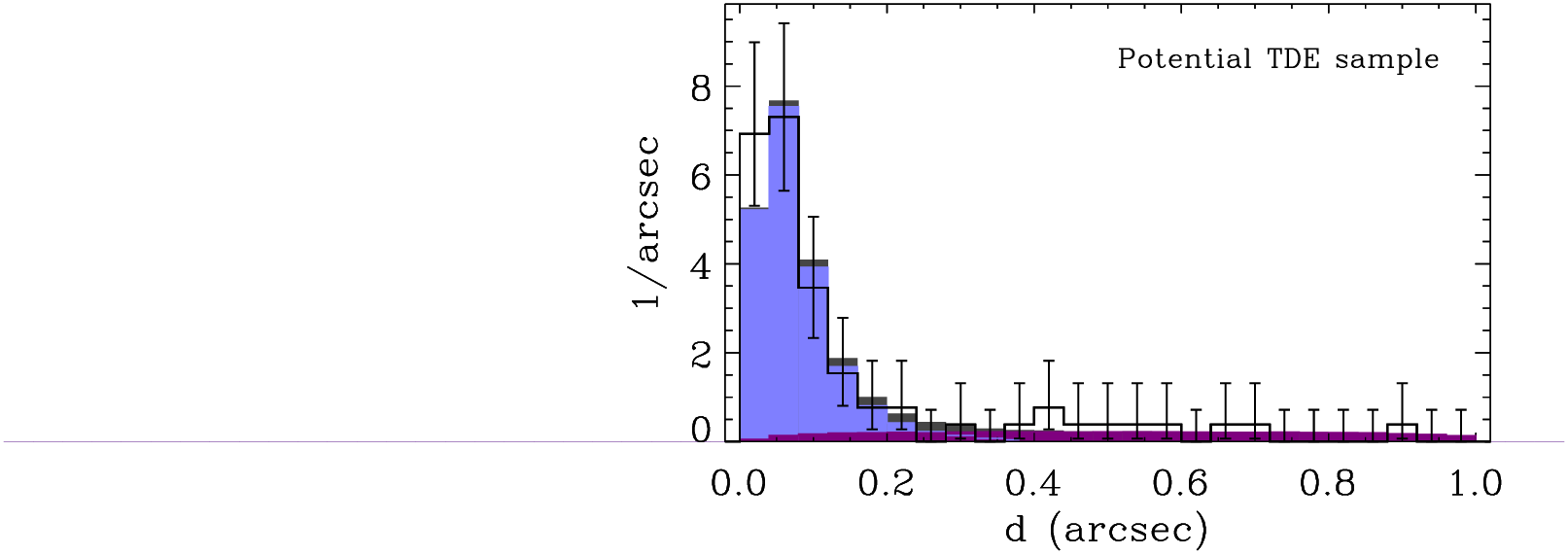} 
\caption{The observed distribution of host-flare distances (thin black histogram, with error bars given by Poisson statistics),  the fit (filled black histogram) with the  decomposition into nuclear flares (blue histogram) and SNe (purple histogram), all normalized to unit area. The only free parameter in the fit is the fraction of stellar-distributed flares, $P($SN). The top panel shows the $d$ distribution for the full sample of 342 flares; the bottom panel shows the flare sample that remains after requiring at least 2 observations after the peak of the flare, imposed for TDE analysis.}\label{fig:sn_mc-0}
\end{center}
\end{figure}
In this section we use the distance $d$ between the center of the host galaxy and the flare to divide the sample of 342 flares into two well-separated samples of 186 nuclear flares and 85 off-center flares, plus 71 flares that are not used because their classification is ambiguous.  

We begin by determining the overall fraction of SNe in the full sample of 342 flares, $P({\rm SN})$.  To do this we model the distribution in $d$ as a sum of nuclear flares and stellar-distributed flares (i.e., SNe). The unsmeared probability density function (PDF) for nuclear events is a delta function at $d=0$. 
We assume that stellar flares trace the stellar light, as is justified by the discussion in Section \ref{sec:sn_compare}.  The flare detection efficiency as a function of $d$ has been modeled and tested against observations, as will be reported in detail elsewhere; for the purposes of this paper it is sufficient to note that the detection efficiency should be at most weakly dependent on $d$, except for a possible reduction in efficiency near the nucleus.  We confirm this \emph{ a posteriori} below.

Under these assumptions, the unsmeared PDF for stellar-distributed flares at distance $d+\Delta d$ from the host center is given by the sum total galaxy flux in this interval.  For each host galaxy, the surface brightness profile, $F(d)$, is taken to be that of the PSF-corrected model galaxy corresponding to the best-fit Sersic parameters in the $r$-band.  We smear the nuclear and stellar PDFs of each galaxy according to the measured uncertainty in the flare separation for that galaxy, $\sigma_x$ and $\sigma_y$. 

We make an unbinned maximum likelihood fit to the sum of the resulting stellar-distributed PDF, times $P(\rm SN)$, plus the nuclear PDF, times $1-P(\rm SN)$, to the observed $d$ distribution, with the result $P({\rm SN}) = 0.34 \pm 0.04$ (90\% CL).  The observed and predicted distributions are shown in  Fig. \ref{fig:sn_mc-0} (top panel). The quality of this fit, in which the SN fraction $P(\rm SN)$ is the only free parameter, validates the assumption that the detection efficiency is independent of $d$ for $d > 0\farcs2$ and shows that we understand the
distribution of host-flare separations.  

Based on Fig. \ref{fig:sn_mc-0}, we formulate the cut for nuclear flares:
\begin{eqnarray}
\mathrm{cut} \equiv (d/\sigma_d<2)\: {\rm and} \: (d<0\farcs2) \quad .\label{eq:cut} 
\end{eqnarray}
This gives 186 on-center flares, with a loss of $15\%$ of the true nuclear flares. 
The probability of a background event (i.e., a stellar-distributed flare within this cut), can be computed directly from the $d$-histogram and is $P({\rm SN}| \mathrm{cut}) =  0.052$, giving an expected number of background events in the total nuclear flare sample of 10. This result is independent of whether the $g$, $r$ or $i$-filter is used to obtain the surface brightness profile of the galaxies.  This procedure gives an upper bound on the contamination of the SNe in the nuclear sample, since a possible reduction in detection efficiency for $d<0\farcs2$ would result in a lower predicted number of SNe in the nuclear sample (and larger rate of nuclear flares relative to SNe, generating the observed flare sample).

A high-purity sample of flares which are clearly off-center from their host is obtained by requiring 
\begin{eqnarray}\label{eq:sn_cut}
(d/\sigma_d>3)\: {\rm and} \: (d>0\farcs2) \quad.
\end{eqnarray}
This selects 85 flares, of which 27\% have been spectroscopically identified as SNe by the  Sloan Digital Sky Survey-II Supernova Survey \citep{frieman08}.  We assume that the remaining off-center flares are SNe as well and take these 85 flares as our SN sample below.  It is important to note that -- because this sample is obtained by a cut on the host-flare distance only -- the properties of the SNe in this sample are not subject to the selection biases that exist for spectroscopic SNe surveys, and are representative of the properties of SNe that appear in the nuclear sample (to the extent that SNe properties are independent of their location in the host galaxy).  

\begin{figure}
\begin{center}
 \includegraphics[trim =0mm 0mm 20mm 80mm,  width=.44 \textwidth]{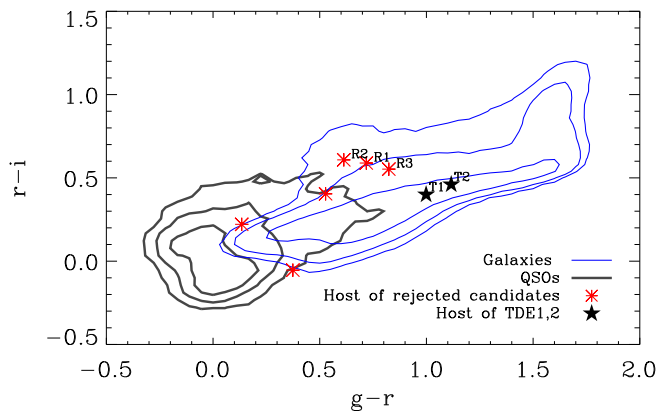} 
\caption{Color-color diagram of the PSF flux \citep{stoughton02} for the $2.5\times10^6$ extended objects in Stripe 82 that were used in this work (thin blue line) and the SDSS QSO sample \citep{schneider07} with $z<1$. For both samples, contours encompass 50, 80 and 90\%. The  TDE candidates that fall outside the QSO locus but are rejected based on additional variability are labelled R1--R3; TDE1 and TDE2 are labelled T1 and T2, respectively. }\label{fig:gal_qso_locus}
\end{center}
\end{figure}

\begin{figure*}[t!]
\begin{center}
 \includegraphics[trim =0mm 0mm 5mm 42mm,  width=.47 \textwidth]{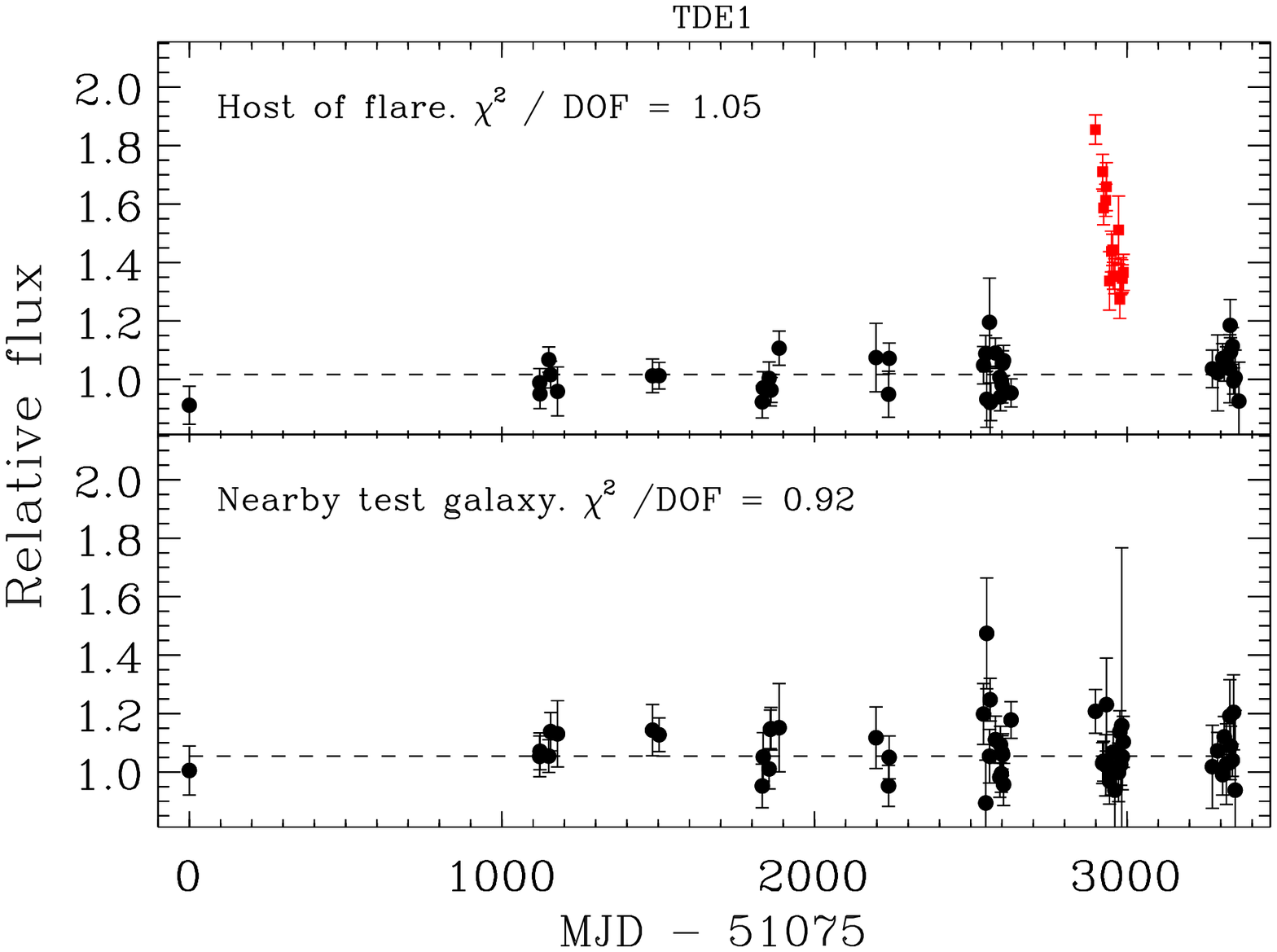}
 \includegraphics[trim =3mm 0mm 2mm 42mm,  clip, width=.47\textwidth]{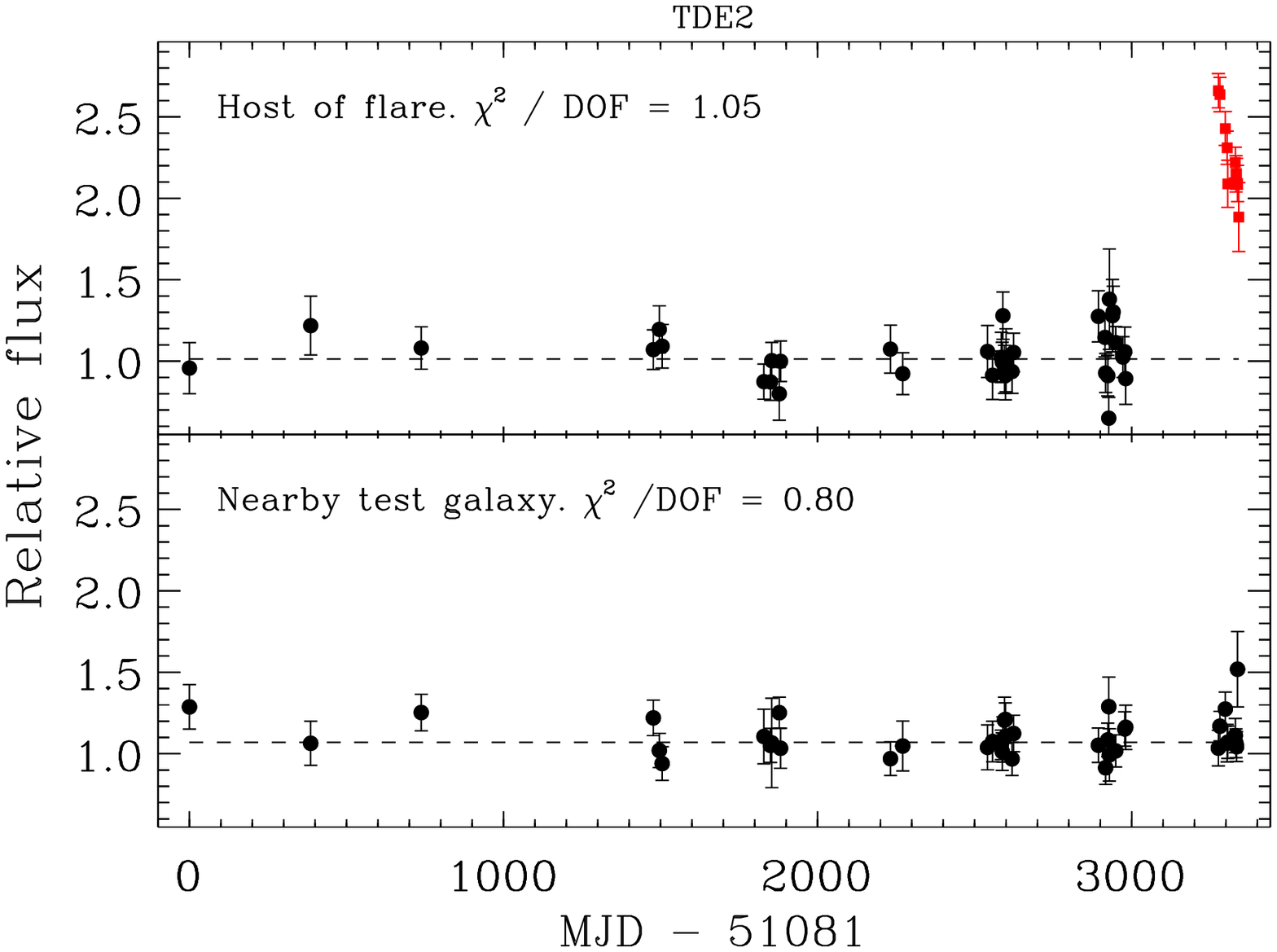}  

 \includegraphics[trim =0mm 0mm 5mm 42mm,  width=.47 \textwidth]{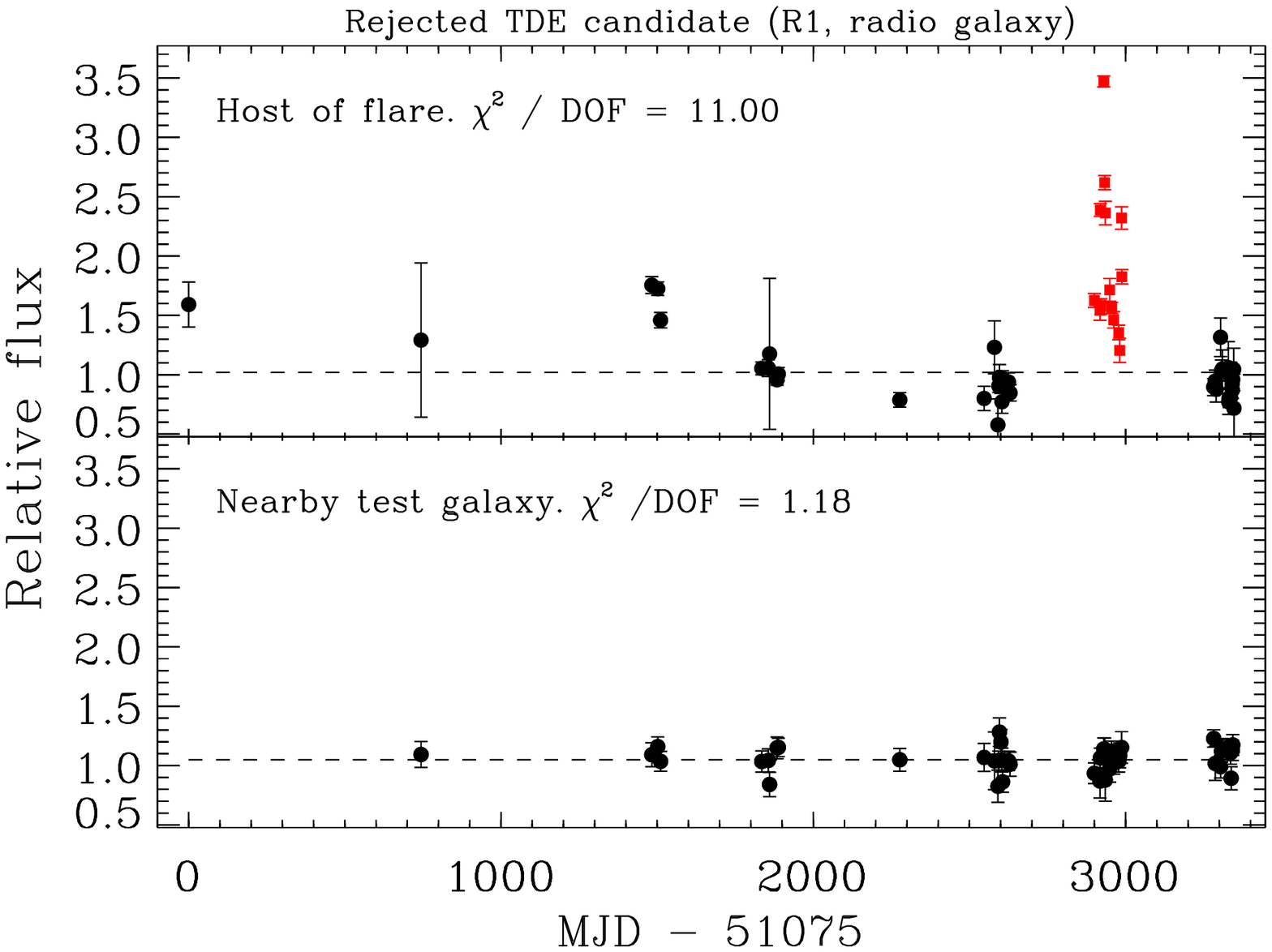}
 \includegraphics[trim =3mm 0mm 2mm 42mm,  clip, width=.47 \textwidth]{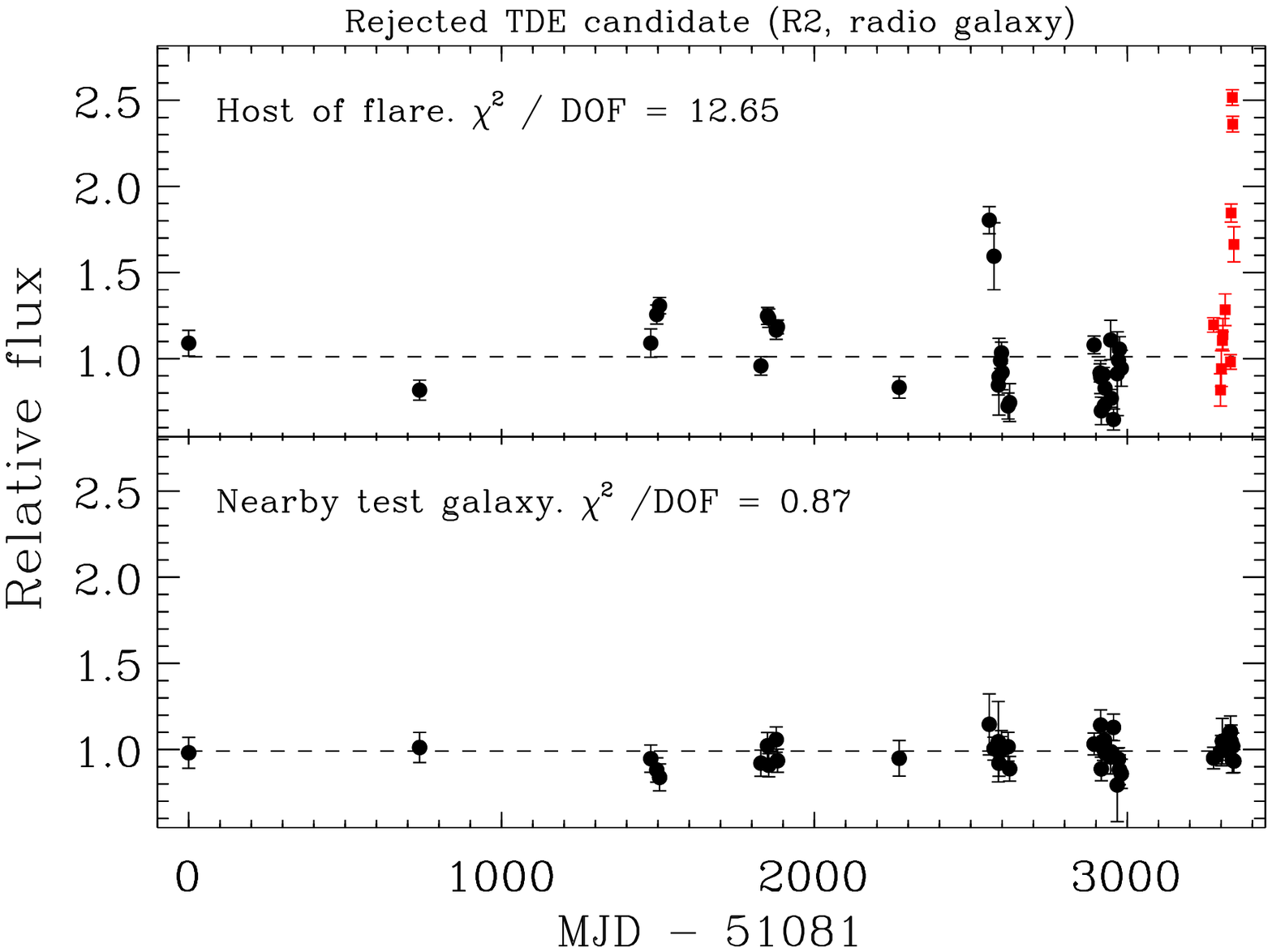}  

 \includegraphics[trim =0mm 0mm 5mm 42mm,  width=.47 \textwidth]{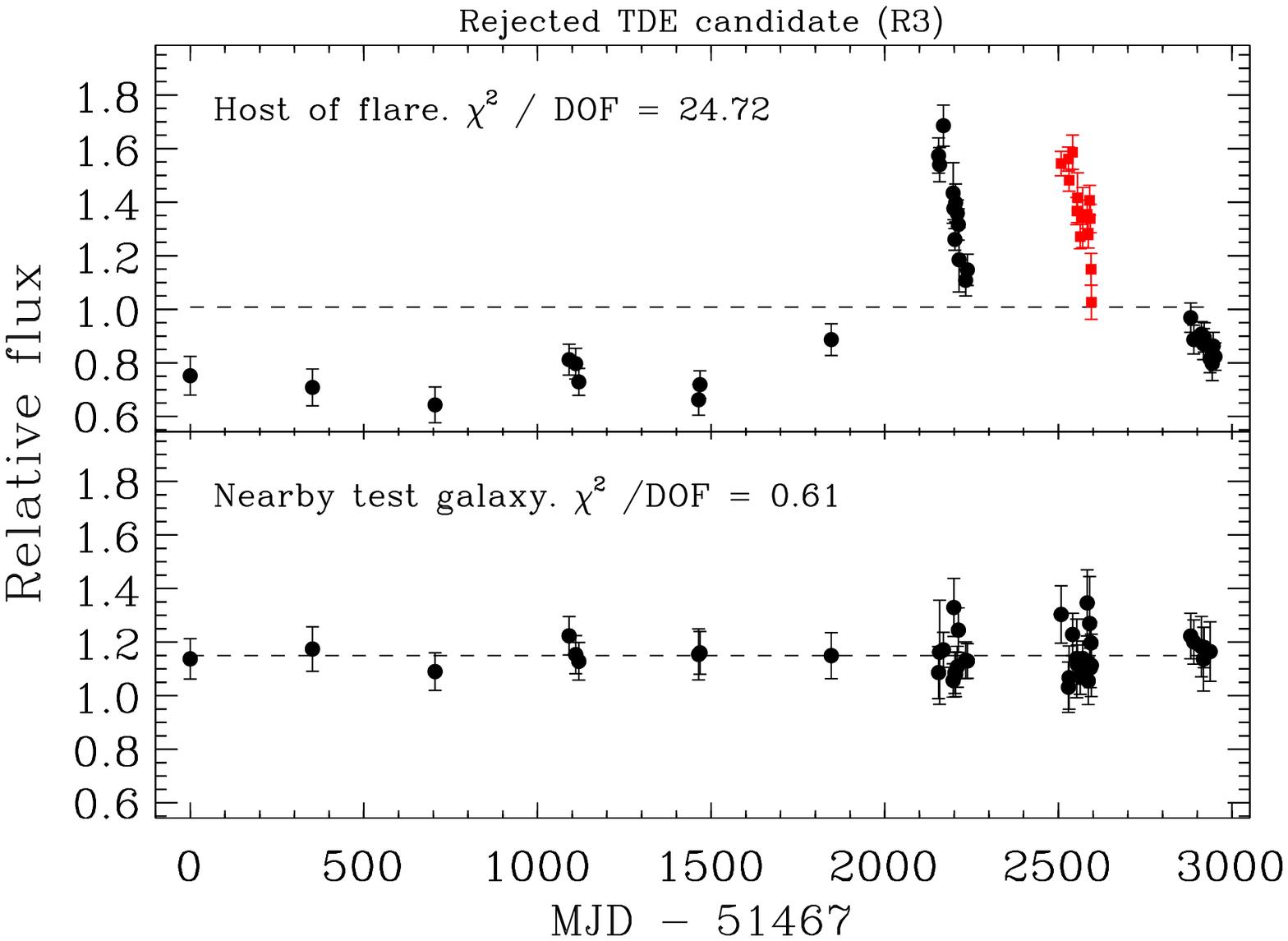}  
 \includegraphics[trim =3mm 0mm 2mm 42mm,  clip, width=.47 \textwidth]{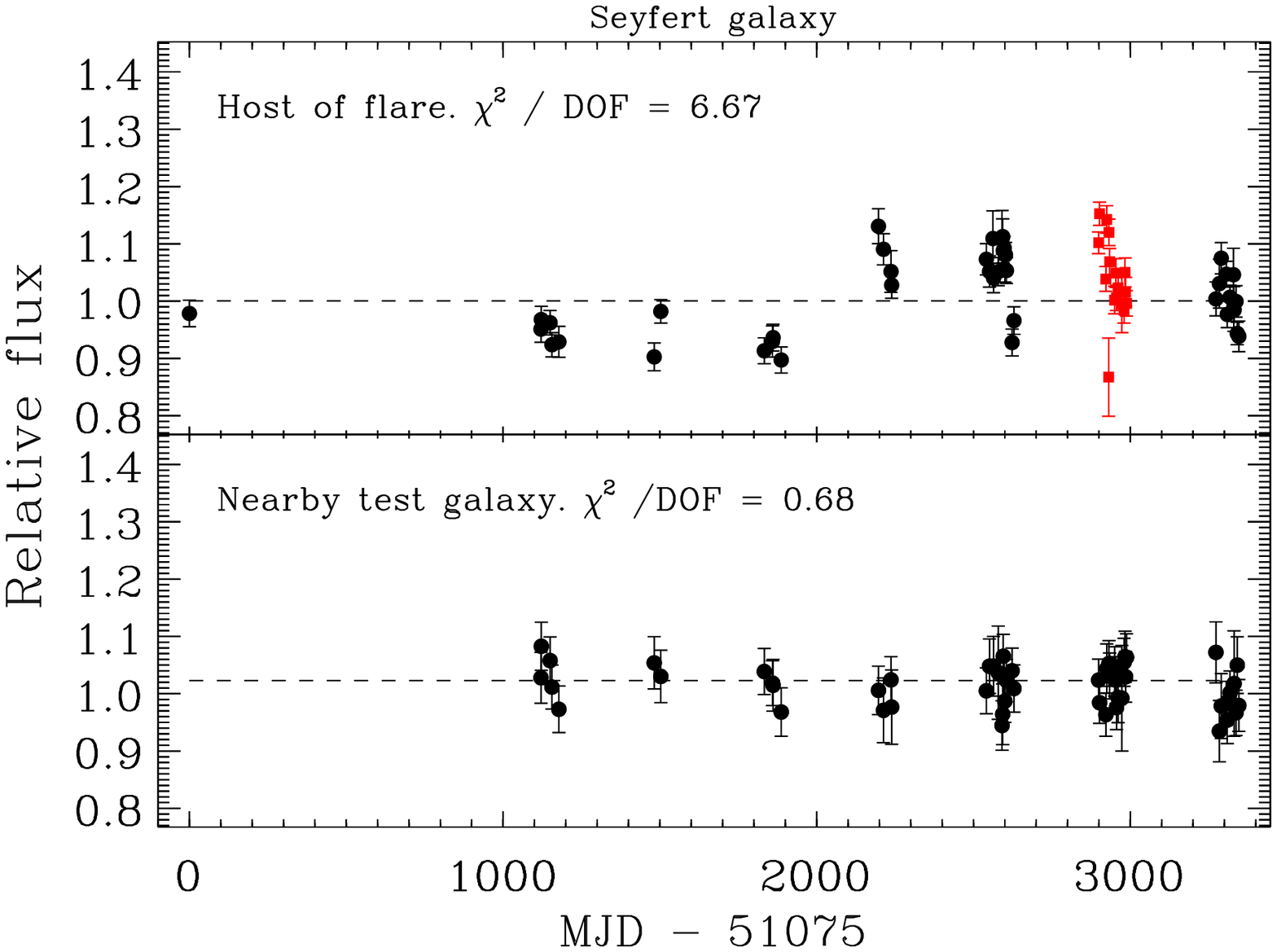}  
\end{center}
\caption{Relative photometry light curves, using the Petrosian $g$-band flux, of the hosts of potential TDEs and one Seyfert (see Section \ref{sec:agn_compare}). Each observation is scaled using the flux of nearby (distance $< 20'$) galaxies that are of similar magnitude ($\Delta m < 1$). The lower portion of each panel displays the light curve of one of the ten test galaxies (selected to have $\Delta m <0.3$). The observations in the season that contained the flare (indicated with red boxes) are excluded in the calculation of $\chi^2$.Variability in the non-flare season is evident for all galaxies expect the host of TDE1 and TDE2. }\label{fig:rel_lc}
\end{figure*}

\subsubsection{Potential TDE sample}\label{sec:pot_tde}
To be able to obtain a well-measured decay rate -- a valuable diagnostic in separating SNe, AGN flares, and TDE flares -- we require at least two detections after the peak of the flare in the $u$, $g$ and $r$ bands.  (At this step, the nuclear -- potential TDE -- sample is reduced to 42 flares  and SNe sample to 12 flares.)  Repeating the $d$-distribution analysis of Section \ref{sec:h_d}, the expected number of residual SNe in the nuclear flare sample is 0.9.
 The corresponding histogram is shown in Fig. \ref{fig:sn_mc-0} (bottom panel). 

\subsection{AGN rejection}\label{sec:AGN_rejection}
AGNs are well-known to be variable and we expect that the majority of the 42 nuclear flares that remain at this stage of the analysis originate from active black holes. Fortunately, AGN hosts are readily identified using SDSS spectra of the host galaxy and, with a small loss in efficiency, can be identified photometrically too. After applying these methods to identify AGNs in sections \ref{sec:AGNspec} and \ref{sec:AGNphot}, five potential TDE remain.  Motivated by previous work on Stripe 82 showing that at least 90\% of unresolved, spectroscopically confirmed quasars are variable at the 0.03 mag level (rms) \citep{sesar07}, in section \ref{sec:rel_flux} we use variability beyond the flare season to reject three more AGNs, leaving only two flares, we label these TDE1 and TDE2.

\subsubsection{Spectroscopic AGN Identification}\label{sec:AGNspec} 
SDSS spectra with a median S/N greater than 3 are available for about 2/3 of the hosts of the 186 nuclear flares. The Princeton Reductions\footnote{http://spectro.princeton.edu} classify galaxies with spectra into three classes using a Principal Component Analysis: STAR, GALAXY or QSO.  Here, QSO does not refer to the classical $M_B<-21$ luminosity cut, but implies that the spectral energy distribution (SED) is dominated by an AGN-like spectrum inside the fiber ($3''$) that was used to obtained the spectrum. Of the 42 nuclear flares suitable for our TDE analysis, 32 are in hosts that are classified as QSOs; we eliminate these from our candidate sample.  We also reject hosts in the class GALAXY which we can identify as Seyfert galaxies based on either of the following criteria: 
\begin{itemize}
\item Galaxies that show broad ($>200$ km/s line-width, at the 7-$\sigma$ level) H$\alpha$, H$\beta$, $\mathrm{[OIII]}\lambda5007$, $[\mathrm{OII}]\lambda3727$, or $\mathrm{MgII}\lambda2799$ emission lines. We only consider lines detected at the 3-$\sigma$ level with rest-frame equivalent width $>5 \mbox{\AA}$. 
\item Galaxies that can be classified as Seyferts using the \citet*{baldwin81} diagram:
\begin{equation}\label{eq:BPT}
 \log\left(\frac{\mathrm{[OIII]}\lambda 5007}{\mathrm{H}\beta}\right) > 
  \frac{0.61}{\log( \mathrm{NII}\lambda6583/\mathrm{H}\alpha ) - 0.05} + 1.3
\end{equation} 
\citep{kauffmann03}. We apply this formula only if all four emission lines are measured at the 3-$\sigma$ level. 
\end{itemize}
This eliminates two more objects from the TDE analysis, reducing the sample of potential TDEs to 8 flares.

\subsubsection{Photometric AGN Identification}\label{sec:AGNphot}
We use the photometric properties of the hosts of the 8 flares that remain at this point to test if they are QSOs. We define a locus that contains 90\% of the 20,710 spectroscopic QSOs identified by SDSS \citep{schneider07} with $z<1$ and reject all hosts that fall inside;  23\% of the $\sim 2.5 \times 10^6$ galaxies that were in the original sample and processed in the flare search fall inside this locus. This cut on QSO locus removes 3 flares from the potential TDE sample, see Fig. \ref{fig:gal_qso_locus}. 

\subsubsection{Additional AGN variability}\label{sec:rel_flux} 
Figure \ref{fig:gal_qso_locus} shows that the hosts of five flares in the potential TDE sample fall outside the QSO locus. We investigated the light curves of these hosts to look for additional variability beyond the season that contains the flare.  Having an efficient but accurate means to quantify the flux variability is applicable to a number of studies, so in this subsection we give details of the method we use to produce relative flux light curves.

As explained in section \ref{sec:catalog_cuts}, we use the SDSS Petrosian flux to search for flares in a galaxy light curve. The Petrosian flux is designed to yield  a robust estimate of the total flux of a galaxy \citep{blanton01, stoughton02}, yet night to night differences in seeing or other observational conditions introduce a small amount of jitter to our light curves. In most cases, nearby and similar galaxies suffer the same attenuation (e.g., worse seeing causes a lower Petrosian flux for all faint galaxies in the field), hence this jitter can be reduced by normalizing the observations of each night using a set of reference galaxies.  For each host of a flare, we select this set by looking for nearby (within $ 20'$) galaxies, whose apparent magnitude difference (in a given filter) with respect to the target host is less than one, and for which the rms variance in the magnitude in all observations is smaller than 30\%; this typically yields 70--100 galaxies for each host.  For each point in the light curve of the host, we determine an overall flux correction by averaging the rescalings required to bring the flux of each reference galaxy to equality with its mean flux. The above procedure can be applied to any filter; for our variability studies we use the Petrosian $g$-band flux. 

To test the quality of the resulting light curves, we select ten test galaxies from among the reference galaxies, chosen to be even more similar to the host of the flare by requiring $|\Delta m| <0.3$. Sometimes, fewer than ten test galaxies are found; if none are found, no relative flux light curve is produced for this host.  SDSS provides errors -- let us call them $\sigma_{{\rm SDSS} \, i,t}$ -- for each observation $i$ of each test galaxy $t$.  We test empirically whether these error bars are a good representation of the flux uncertainty after rescaling, by computing  $\chi^2_t$, the reduced chi-squared assuming a constant flux of the light curve of the each test galaxy. If all of the uncertainty in the flux measurement were captured in the $\sigma_{{\rm SDSS} \, i,t}$, the $\chi_t^2$ values would obey a $\chi^2$ distribution but they do not.  We find that the errors reported by SDSS sometimes under-predict the true uncertainty of the flux.  We therefore use the fact that the expectation value of $\chi^2$ per degree of freedom should equal 1 for the test galaxies, to obtain a better estimate of the measurement uncertainty on the relative flux the host or any of its test galaxies, e.g.,  $\sigma_{i} = \sigma_{{\rm SDSS} \, i} \sqrt{<\chi_t^2>}$.

\begin{figure}
\begin{centering}
     \includegraphics[trim = 0mm 0mm 5mm 57mm, width=.5 \textwidth]{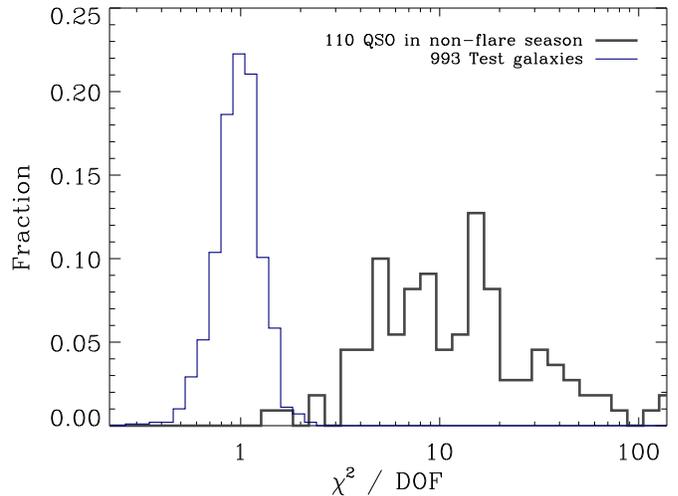}
     \caption{Histogram of $\chi^2$/ DOF obtained for the relative photometry light curves fit to a constant flux, showing the difference between QSOs and ordinary galaxies. We use the non-flaring seasons for the QSO, and all seasons for the nearby test galaxies (see Section \ref{sec:rel_flux}). For three spectroscopically identified Seyfert galaxies with flares we find: $\chi^2/{\rm DOF} = \{5.14,\, 2.51,\,  1.54\}$. The lowest value in this list is larger than 99\% of the test galaxies. We may thus conclude that flares from Seyfert galaxies can also be identified based on additional variability of their host.}\label{fig:non-flare-chi2} 
\end{centering}
 \end{figure}

The relative flux light curves produced by this method using the Petrosian flux in the $g$-band for the five potential TDEs are displayed in Fig. \ref{fig:rel_lc}.  Three of the hosts clearly show additional flaring activity:  $\chi^2$ per degree of freedom of the flux calculated from their relative flux light curves excluding the season with the primary flare is $>10$; for reference, we label these R1--R3.   However two hosts show no additional variability, with each having $\chi^2/{\rm DOF}=1.05$. We show in Fig. \ref{fig:non-flare-chi2} that this is significantly smaller than measured for any QSO or Seyfert galaxy in our sample.  These are the hosts of the flares we designate as TDE1 and TDE2.  In Section \ref{sec:agn_compare} we use the observations of QSOs and Seyferts to quantify the probability that TDE1 or TDE2 are flares from AGN which had quiet years in the other observing seasons.

\bigskip

\section{Properties of the identified tidal flares}\label{sec:properties}

\begin{deluxetable*}{c c c c c cc c c c}
  \tablewidth{0pt}    
  \tablecolumns{8}
  \tablecaption{Properties of the TDEs.\label{tab:tde_prop}  }
  \tablehead{\colhead{Name}  & \colhead{$z$} & \colhead{$M_{g}$}& \colhead{$L_g $ } & \colhead{$d\ln{L_g}/dt $} & \colhead{$T$} & \colhead{$d\ln{T}/dt\,$} &\colhead{$d$} &\colhead{$d_{68}$}\\
    {} &  {} & {} & {($\times 10^{43}$ erg/s)} & {($\times 10^{-2} \,{\rm day}^{-1} $)} & {($\times 10^4 \, K$)} & {($\times 10^{-3}\,{\rm day} ^{-1}\,$)} & {(arcsec)} & {(arcsec)}}
  \startdata
  TDE1&$0.136\pm 0.001$ & $-18.3\pm0.04$ & $ 0.54 \pm 0.02 $ & $-1.7 \pm 0.1$ & $2.4_{-0.2}^{+0.3}$ & $-2 \pm 4$&$0.058$ & $0.124$ \\
  TDE2&$0.251\pm0.002$ &  $-20.4\pm0.05$ & $ 4.1 \pm 0.2 $ & $-0.8 \pm 0.1$ & $1.82_{-0.06}^{+0.07}$ & $-3 \pm 2$&$0.068$ & $0.075$\\

  \enddata
  \tablecomments{The rest-frame $g$-band peak observed absolute magnitudes, luminosities and black body temperatures of the flares, measured in their SDSS difference images.  We exploited the absence of significant color evolution to improve the accuracy of the black-body temperature determination by using the mean flux in each band over the SDSS light curve. In the eighth column, $d$ denotes the distance between the center of the host and the flare (see Section \ref{sec:pipeline}) and $d_{68}$ the 68\% confidence radius. The host properties (including coordinates) are given in Table \ref{tab:host_prop}.}
\end{deluxetable*}

\begin{figure}
\subfigure[TDE1]{
{\includegraphics[width=80 pt]{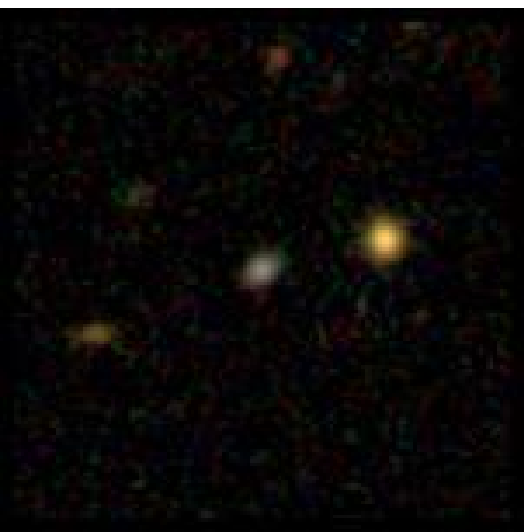}}
{\includegraphics[width=80 pt]{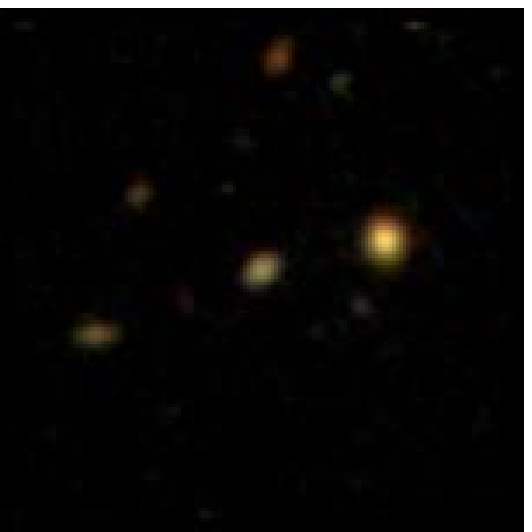}}
{\includegraphics[width=80 pt]{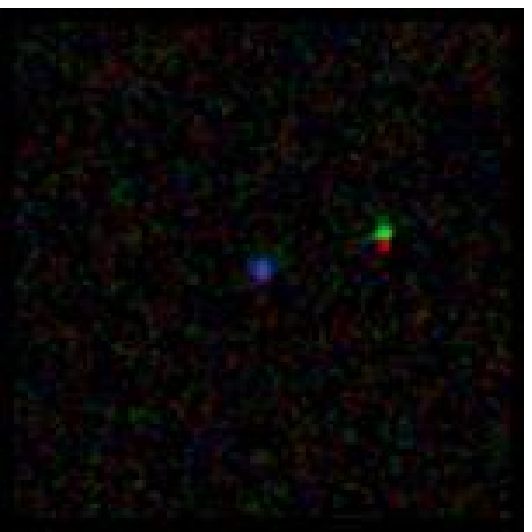}} } \\ 
\subfigure[TDE2]{
{\includegraphics[width=80 pt]{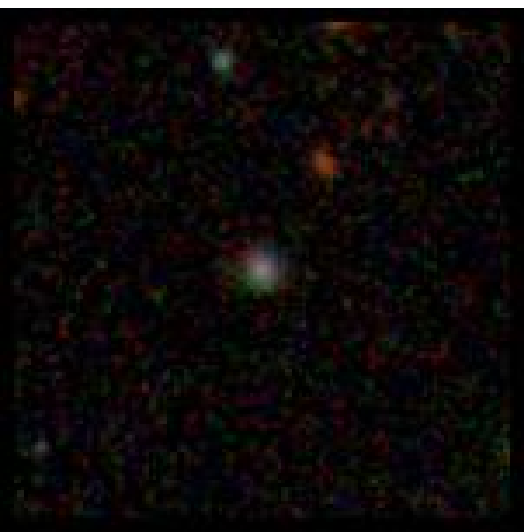}}
{\includegraphics[width=80 pt]{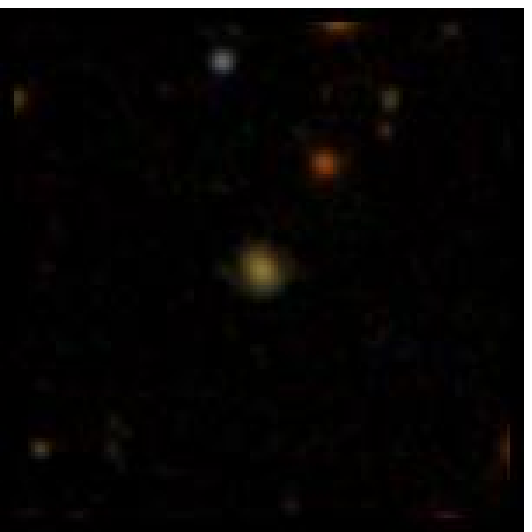}}
{\includegraphics[width=80 pt]{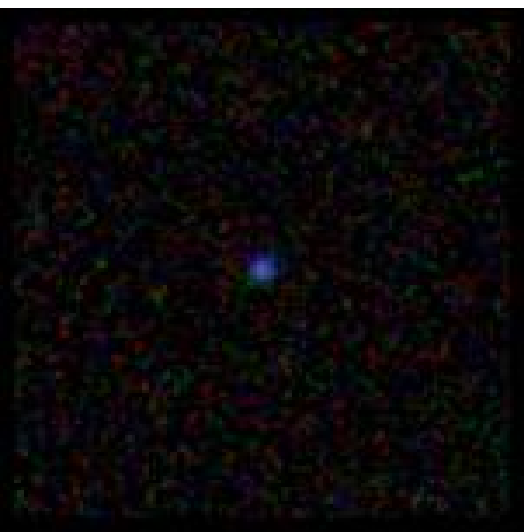}}} \\ 
\caption{SDSS images ($1'\times1'$) of the TDE flares and their host galaxies.  Left to right: flare image, mean reference image and difference image for TDE1 and TDE2. We see that the hosts can be classified as E/S0. The difference image of TDE1 shows a subtraction artifact at the location of the bright point source, which is not subtracted perfectly because our difference imaging method is optimized for the center of the field.}\label{TDEimages}
\end{figure}

In this section we report on the properties of the two stellar tidal disruption flare candidates, based on Stripe 82 SDSS imaging data and observations using other instruments.   Although we did not rely on properties of the flaring state to {\em select} TDEs, to avoid unnecessarily biasing the selection and because theoretical predictions are uncertain, the TDEs identified by our pipeline prove to be quite distinct from SNe and flaring AGNs.  This increases confidence that TDE1,2 are not examples of familiar phenomena occurring in improbable circumstances, such as SNe accidentally close to the nucleus or AGNs which flare dramatically in the midst of a multi-year quiet phase.  

We begin with an overview of the two events by giving their light curves, cooling rate and flare colors based mainly on the SDSS Stripe 82 observations.  The SDSS images of the TDEs and their host galaxies are given in Fig. \ref{TDEimages}.  Tables \ref{tab:tde_prop} \& \ref{tab:host_prop} summarize the properties of the TDEs and their hosts, respectively.  Then we describe the observations of TDE1 and TDE2 obtained with other telescopes.  Finally, we combine the observations to quantify attributes of the host and flare relevant to the possibility they may be produced by a supernova or variable AGN.  In Section \ref{sec:compare}, we attempt to account for the totality of these observations with supernovae and variable AGN hypotheses, but find that no known phenomenon other than tidal disruption is compatible with all of the observations. 

\begin{figure}
     {\includegraphics[trim = 0mm 40mm 0mm 40mm, width=.5 \textwidth]{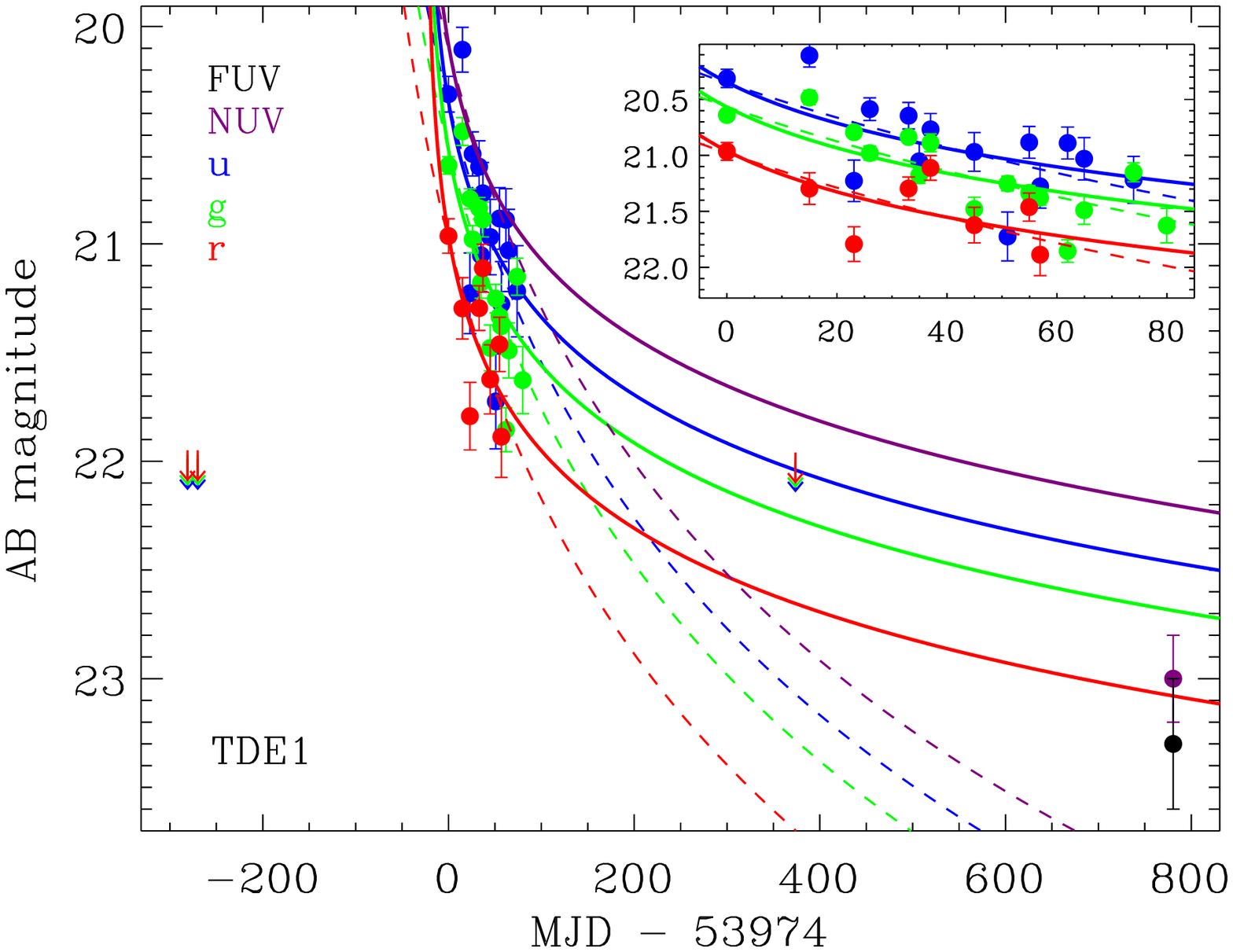}}
     {\includegraphics[trim = 0mm 0mm 0mm 0mm,  width=.5 \textwidth]{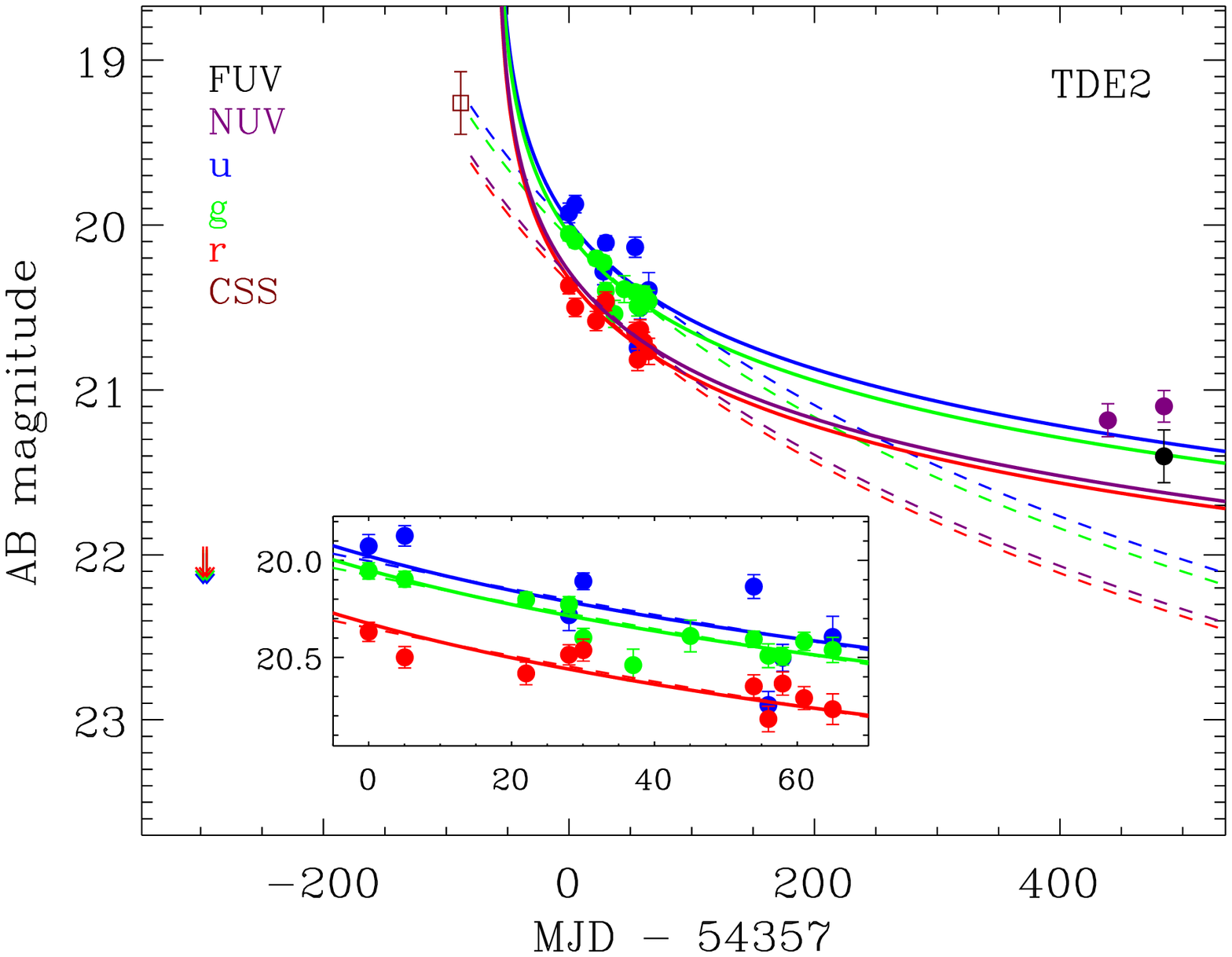}}
     \caption{ UV and optical light curves for TDE1 and TDE2 as a function of days since observed peak.  The SDSS difference image flux (i.e., host subtracted) in the r,g, and u bands is shown with red, green, and blue solid circles.    The orange open square indicates the mean of the 3 CSS observations 3 months before the first SDSS flare observation (see Section \ref{sssec:CRTS}).  The dashed (solid) lines display the result of fitting a $(t-t_D)^{p}$ power law decay with $p = {-5/3} \, {(-5/9)}$ to the SDSS observations only.  The corresponding NUV curves (purple) are obtained from the black body fit to the mean optical colors of the flare, assuming no cooling.  Because the UV baseline of the host of TDE1 is unknown, we show the \textsl{GALEX} aperture flux for TDE1, while for TDE2 we show the \textsl{GALEX} difference flux.
      }\label{fig:lc}
 \end{figure}
 
 \begin{figure}
{\includegraphics[trim =  0mm 17mm 0mm 15mm, clip, width=.5 \textwidth]{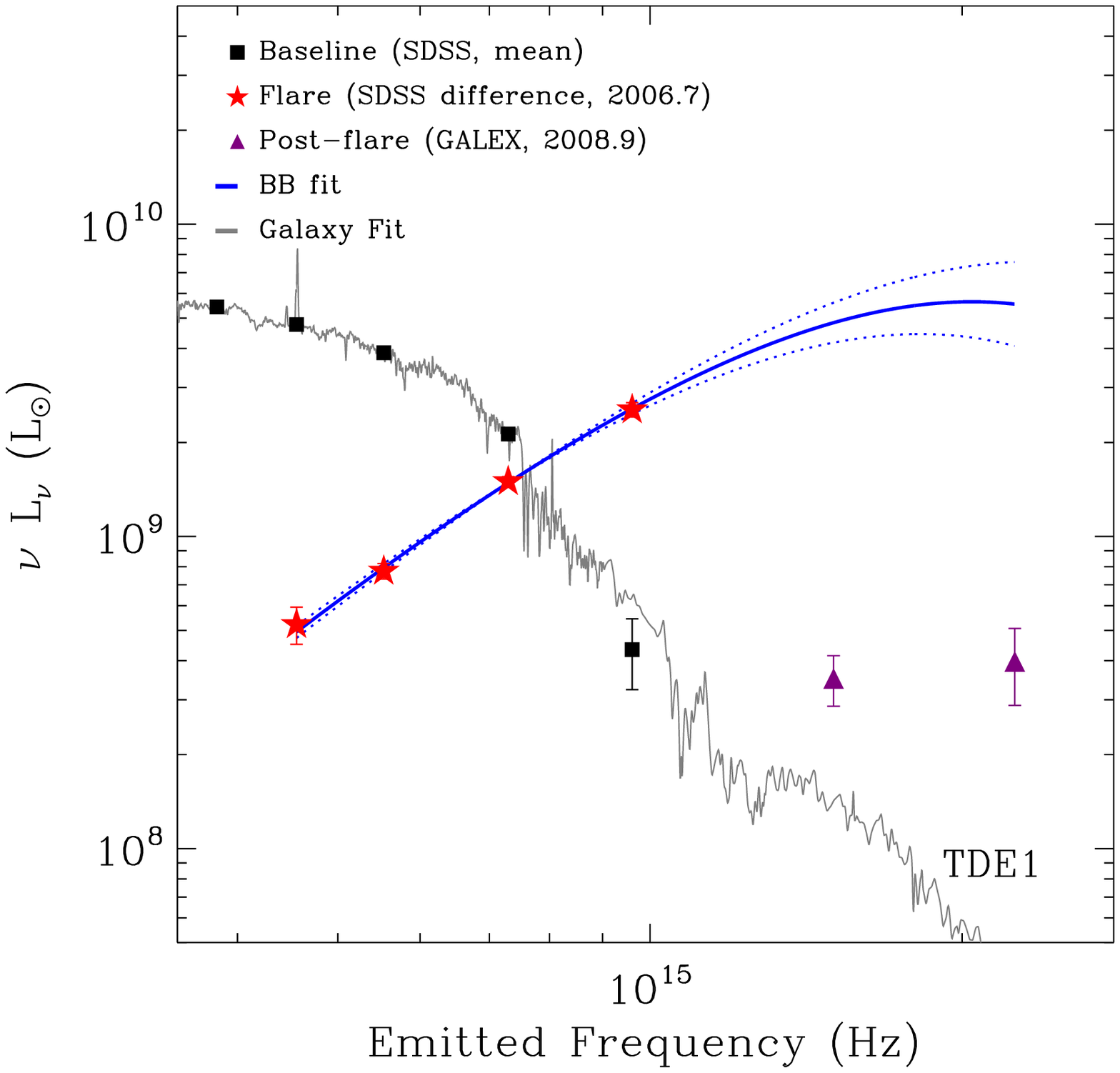}}
{\includegraphics[ trim =  5mm 0mm 00mm 15mm, width=.5 \textwidth]{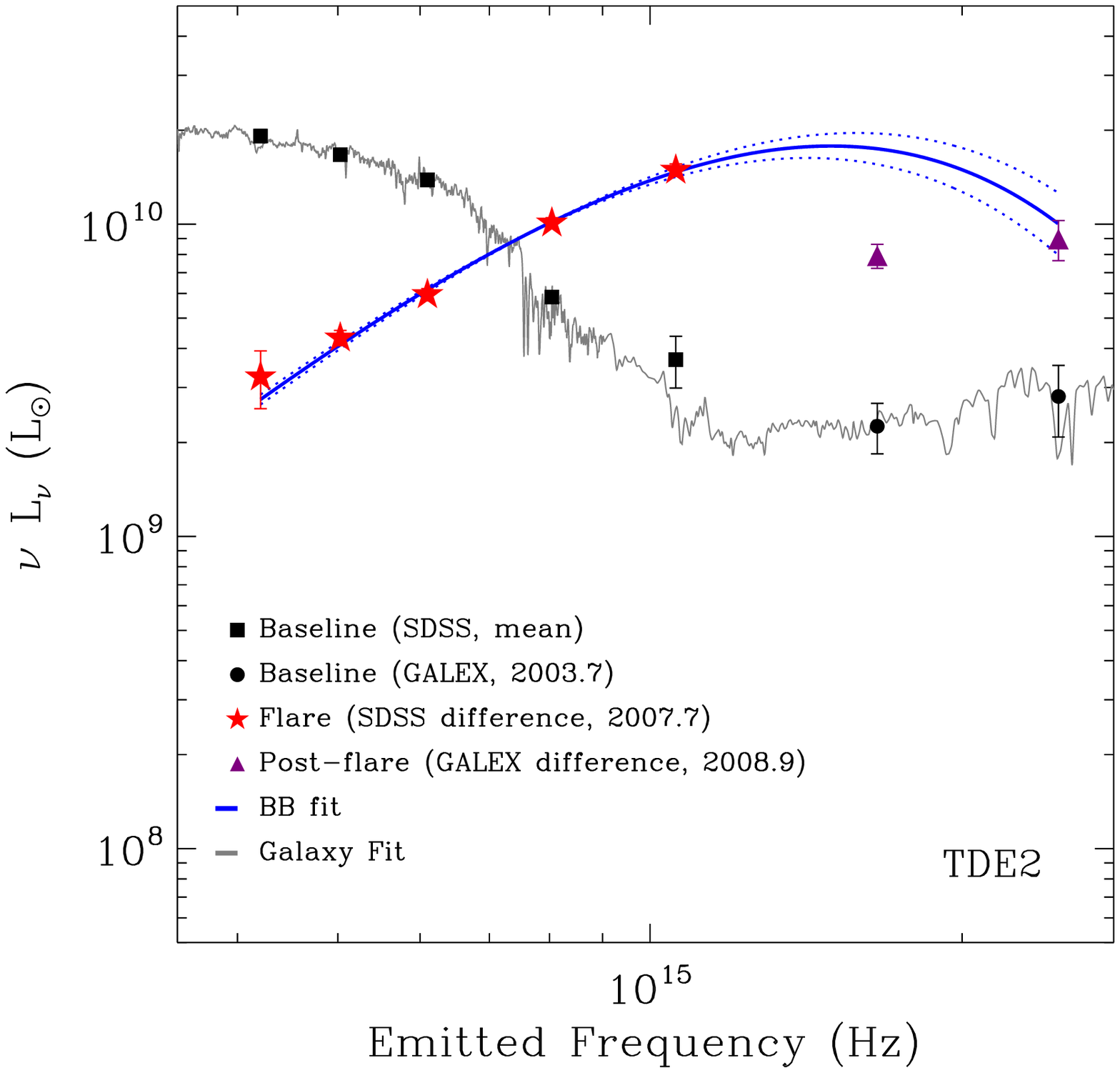}}
     \caption{The SED for TDE1 (top) and TDE2 (bottom).  The optical baseline flux of the host is shown (black squares) with the best-fit combination of eigen spectra \citep{blanton07} (grey line). For TDE2, the  \textsl{GALEX} observations before the flare (black circles) are also used in this fit. 
The flux of the flare in the optical difference images is shown with red stars.  It is well-fit by a black body shown as a blue line, with dotted lines indicating the 1-$\sigma$ uncertainty. The post-flare UV flux is shown with purple triangles; for TDE1 it is the total flux (no baseline being available) while for TDE2 the difference flux is shown.  The detailed interpretation of the post-flare UV detections are unclear; the dynamics of the tidal debris and associated accretion are complex -- a single black body need not be valid and the cooling behavior is uncertain. 
}\label{fig:sed}
 \end{figure}

\subsection{SDSS observations}\label{sec:prop_sdss}
Figure \ref{fig:lc} shows the $u$, $g$, and $r$-band light curves for the flares (i.e., the difference images).  Also plotted are the FUV and NUV fluxes from \textsl{GALEX} (see Sections \ref{sec:tde1} \& \ref{sec:tde2} for details on these \textsl{GALEX} observations), and the flux estimated from the CRTS optical observation of TDE2 3 months prior to the first SDSS observation in the flaring state (see Section \ref{sssec:CRTS}).   Both flares were detected by SDSS in the first observation of a Stripe 82 season, so were most likely past their peaks when first detected; this is confirmed by the CRTS detection in the case of TDE2 (see Section \ref{sssec:CRTS}).  For comparison, we show fits of the SDSS data to $f_{\nu}(t) \propto (t-t_{D})^{p}$, where $t_{D}$ is the time of disruption. The SDSS optical observations do not cover a long enough period to break the degeneracy between $t_D$ and $p$.  For $p = {-5/3}$ (the fallback rate of the debris \citep{Rees88}), the inferred time delay between when the disruption occurred and when it was first observed by SDSS is 107 and 220 days for TDE1,2 respectively, while for $p = {-5/9}$, as predicted by \citet{strubbe_quataert09} for the initial super-Eddington outflow phase, these become 24 and 55 days.  In Fig. \ref{fig:lc} we show the $p = {-5/3}$ and ${-5/9}$ extrapolations assuming constant temperature to obtain the NUV magnitude, for orientation. 

\begin{figure*}
\begin{center}
\includegraphics[ trim = 0mm 2mm 0mm 50mm, width=.99 \textwidth]{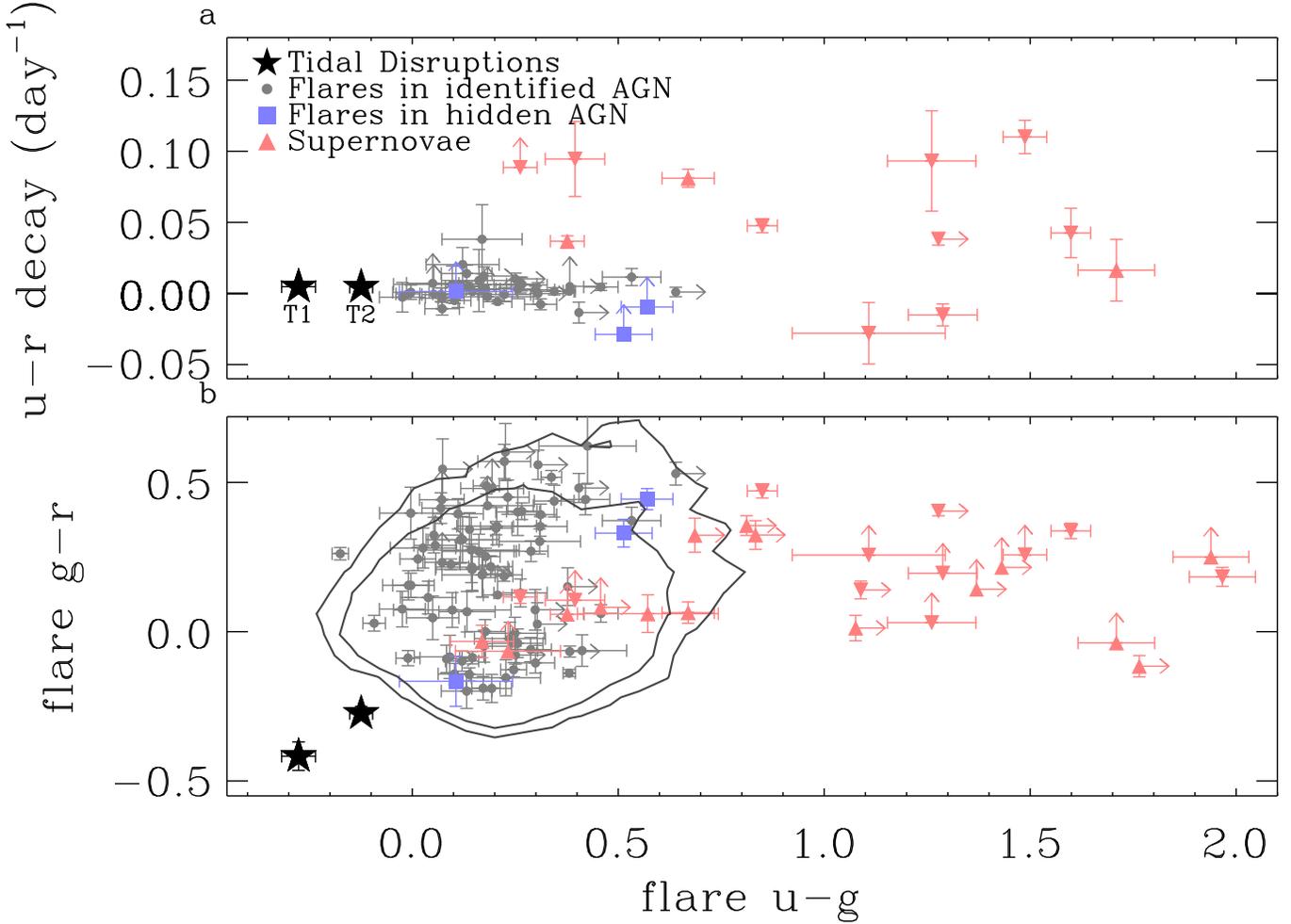}
\caption{ a)  The cooling time measured by fitting $u-r$ as a function of time. The two TDE candidates are incompatible with being ordinary SNe because they show no signs of cooling and have very high $u$ to $g$ flux ratios. b) Color-color diagram using the mean colors of the decaying part of the light curve. Contours containing 90 and 95\% of 14,776 nearby ($z<0.8$) spectroscopic QSOs \citep{Richards04} are also shown.  Fewer objects appear in the upper panel because two simultaneous detections in both bands are required to measure the cooling time.  For SNe and potential TDEs, the flux shown is that of the difference image; AGNs are shown in their high state.  Colors are obtained from the error-weighted mean of all observations of the flare.  Blue boxes mark flares from hosts that are not identified as AGNs based on their spectra or color, but whose variability in other seasons shows they are, in fact, AGNs. The supernovae in this work are selected purely geometrically by being off-center (Eq. \ref{eq:sn_cut}) and thus their properties are unbiased. The SNe that survive the TDE quality cuts (section \ref{sec:pot_tde}) are indicated with a downward pointing triangles. }\label{fig:cc}
\end{center}
\end{figure*}

The photometric observations are well-fit by a rest-frame black-body spectrum with temperatures given in Table \ref{tab:tde_prop}, as can be seen in Fig. \ref{fig:sed}.  An estimate of the cooling rate was obtained by least-squares fitting for the slope of color as a function of time, using only SDSS observations starting with the peak of the flare. We also computed the mean colors by averaging all observations of the flare.  Comparing the mean color to the cooling (Fig. \ref{fig:cc}a), gives strong evidence that the TDE flares are not (ordinary) supernovae:  they are very much bluer than any observed off-center flare in our sample and show negligible cooling, whereas SNe are either blue and rapidly cooling or red with little cooling because they start hot and cool very rapidly.  Furthermore, Fig. \ref{fig:cc}b shows that the colors of the TDE1,2 flares are strikingly different from those of QSOs;  if these were AGN flares, the spectrum of the flare itself (with the galaxy subtracted) should more closely resemble the spectrum of a QSO flare. 

\begin{deluxetable*}{l l c c c c  c c} 
  \tablewidth{0pt}
  \tablecolumns{8}
  \tablecaption{ Properties of  the hosts of the TDEs.\label{tab:host_prop}}
  \tablehead{\colhead{Flare} & \colhead{SDSS ID} & \colhead{RA} & \colhead{Decl.} & \colhead{$M_r$} &  \colhead{ $u-g$} &  \colhead{$g-r$} & \colhead{$M_{\rm BH}$}\\
    {} & {} & {(J2000)} &{(J2000)} &  {} & {} & {} &{($M_{\odot}$)} }
    \startdata      
    TDE1&J234201.40$+$010629.2& $350.95257$&$-1.1361928$&$-19.85\pm 0.02$&$1.95\pm0.3$&$0.73\pm 0.02$& $(6-20) \times 10^{6\pm0.3}$\\
    TDE2&J232348.61$-$010810.3&$355.50586$&$\phantom{-}1.1081316$&$-21.30\pm0.02$&$0.99\pm0.2$&$0.73 \pm  0.05$ &  $(2-10) \times 10^{7\pm0.3}$\\
    \enddata
    \tablecomments{The host magnitude and colors are obtained from the K-corrected \citep{blanton07}, inverse-variance-weighted mean Petrosian magnitude of the non-flare seasons. The black hole mass is estimated using the correlation between $M_r$ and $M_{\rm BH}$ \citep{haringRix04,Tundo07}, using two different estimates for the bulge magnitude (see Section \ref{sec:prop_sdss}).}
\end{deluxetable*}

To estimate the black hole mass ($M_{\rm BH}$) of the host of TDE1,2 we use the \citet{haringRix04} black hole mass-bulge mass relation, calibrated for the SDSS $r$-band by \citet{Tundo07}. Unfortunately, the  S/N in the co-added images of the host of the flare is too low to measure the bulge magnitude by decomposing the images into a bulge and disk component. We therefore estimate the bulge magnitude using two different assumptions for the bulge-to-total ratio ($B/T$). \emph{(i) }The typical ratio for S0 galaxies, $B/T=0.55$ \citep{Aller_Richstone02}. \emph{(ii)} The ratio that follows from the correlation between $B/T$ and the concentration index (i.e., R90/R50, where R90 and R50 are the radii enclosing 90 and 50 per cent of the galaxy flux) \citep{Gadotti09}, which yields $B/T =0.16, 0.13$ for TDE1,2. Using these estimates to obtain a range for the bulge luminosity of the host we obtain $M_{\rm BH} = (6-20) \times 10^{6\pm0.3}\, M_\odot$ for TDE1 and  $M_{\rm BH} = (2-10) \times 10^{7\pm0.3}\, M_\odot$ for TDE2, where the error in the exponent  reflects the scatter in the $M_{\rm BH}$-bulge luminosity relation and the range in the prefactor reflects the uncertainty in the bulge luminosity.

\subsection{TDE1}\label{sec:tde1}
\subsubsection{Additional observations of TDE1}
For TDE1 we have a single \textsl{GALEX} NUV and FUV detection on 2008 October 25, $\sim 800$ days after the optical flare.
The \textsl{GALEX} UV flux values, corrected for the energy lost due to the $6''$ radius aperture, and using the pipeline-generated sky background value in counts per pixel and assuming Poisson errors aperture \citep{Morrissey07}, are: ${\rm FUV}=23.3\pm0.3$, ${\rm NUV}=23.0\pm0.2$. These UV fluxes are plotted in Figs. \ref{fig:lc} and \ref{fig:sed} without host subtraction since we have no pre-flare observation to establish the baseline.

\begin{figure}
\begin{centering}
     \includegraphics[ trim = 15mm 3mm 0mm 7mm, width=.45 \textwidth]{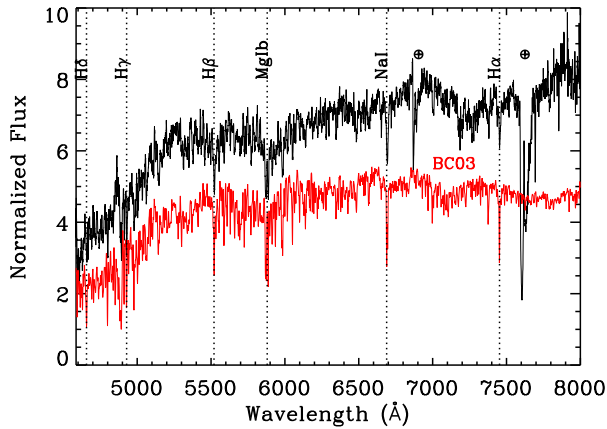}
     \caption{Host spectrum of TDE1. A template \citet{Bruzual_Charlot03} spectrum is shown in red. Note lack of [OIII] emission lines and the H$\alpha$ absorption. Hatch marks indicate the locations of the strong O$_{2}$ telluric $B$- and $A$-band absorption features that were not removed in the data reduction. }\label{fig:obs_spec}
\end{centering}
 \end{figure}

We observed the host galaxy of TDE1 on 2009 November 7 with the MagE  spectrograph on the Magellan II Clay telescope with a spectral resolution of $R=4100$ or \mbox{1.6 \AA}.  The spectrum, shown in  Fig. \ref{fig:obs_spec}, shows Balmer absorption features, as well as Mg Ib and NaI absorption, which yields $z=0.136 \pm 0.001$. No emission lines are detected: $L_{{\rm H}\alpha}< 4.4\times 10^{37}\, {\rm erg}\,{\rm s}^{-1}$ and  $L_{\rm {[OIII]}}<1.3\times10^{38}\, {\rm erg}\,{\rm s}^{-1}$. 

\subsubsection{Origin of UV emission of TDE1}
Star formation contributes little to the observed UV in TDE1.  This follows from the $L_{{\rm H}\alpha}$--$L_{\rm FUV}$ relation \citep{Kennicutt98} which predicts that if the FUV luminosity observed by \textsl{GALEX} $\sim$800 days after the optical flare were due to star formation, the H$\alpha$ luminosity should be $2.4 \times10^{40}$ erg s$^{-1}$ -- two orders of magnitude above the observed upper limit.   The contribution of stellar sources that evade the correlation between FUV and H$\alpha$ (e.g., blue horizontal branch stars) is constrained  by using the color-magnitude relation observed by \citet{Haines08} for galaxies that have been spectroscopically identified as passively evolving.  At the luminosity of the host of TDE1 ($M_r=-19.85$), these galaxies are observed at ${\rm NUV}-r = 5.4$. The scatter in this relation, $\sigma_{{\rm NUV}-r} = 0.37$, can be attributed to different amounts of residual star formation. Thus the UV detection of the host of TDE1, with ${\rm NUV}-r = 3.6$, is a 4.6-$\sigma$ blue outlier to these galaxies.  Moreover, with ${\rm FUV}-r = 3.7 \pm 0.3$, it is distinctly bluer than passive galaxies of similar luminosity, which cluster at ${\rm FUV}-r = 7 $. The UV color, ${\rm FUV} - {\rm NUV} = 0.0 \pm 0.36$, is also bluer than any of the early type galaxies which have been targeted by the SAURON project \citep{jeong09} to study residual star formation. These galaxies are observed to cluster at ${\rm FUV} - {\rm NUV} = 1.5$, with the bluest object at ${\rm FUV} - {\rm NUV} = 0.6$. Recent or residual star formation is thus excluded as a significant contributor to the observed UV emission detected 800 days after the first observation of TDE1. 

Nor can an active nucleus account for the observed UV flux 800 days after the TDE1 flare.  The upper limit on the [OIII] luminosity from the Magellan spectrum places a limit on the optical baseline luminosity of a possible active nucleus. Using the conversion between [OIII] and optical luminosity for type 1 AGNs obtained by \citet{heckman04}, yields an upper limit on the luminosity of an active nucleus in the TDE1 host:  $L_{5000} < 4.2\times 10^{40\pm0.34}\,{\rm erg}\,{s}^{-1}$, where $L_{5000}$ is the monochromatic continuum luminosity at 5000 $\mbox{\AA}$ in the rest frame.  Although the \citet{heckman04} relation is not valid for [OIII] emission from LINERs \citep{heckman80}, we can still use it to obtain an upper limit on the baseline luminosity of a potential AGN in such galaxies, because at similar [OIII] luminosity, the active nuclei of LINER galaxies have an order of magnitude lower bolometric luminosity compared to normal AGNs, as well as an lower optical luminosity for a given bolometric luminosity  \citep{Ho99,Ho04}.
We convert $L_{5000}$ to a $g$-band magnitude (centered at $\lambda =4670\,\mbox{\AA}$), by assuming that the luminosity per unit wavelength at the $g$-band is equal to $L_{5000}$.  This is the conservative approach since $F_{\nu} \propto \nu^{-0.44}$ is typical for an AGN \citep{VandenBerk01}. This yields ${\rm NUV}-g<- 2.9 \pm 0.9$, a NUV to $g$-band ratio greater than the maximal value that can be reached with photons that are in thermal equilibrium; such an extreme color has never been observed for any of the sources detected in both SDSS and \textsl{GALEX} \citep{bianchi07}. Because we obtained our upper limit on the $g$-band luminosity of the accretion disk from the [OIII] line, this limit applies to the baseline luminosity of the AGN, i.e., time-averaged over the light crossing time of the narrow line region, $> 10^2$ yr \citep{Murayama98}. Thus we conclude that the UV flux present 800 days after the optical flare does not originate from the baseline of an AGN.  

\subsection{Additional observations of TDE2}\label{sec:tde2}
\begin{figure}
\begin{centering}
     \includegraphics[ trim = 5mm 50mm 5mm 10mm, clip, width=.5 \textwidth]{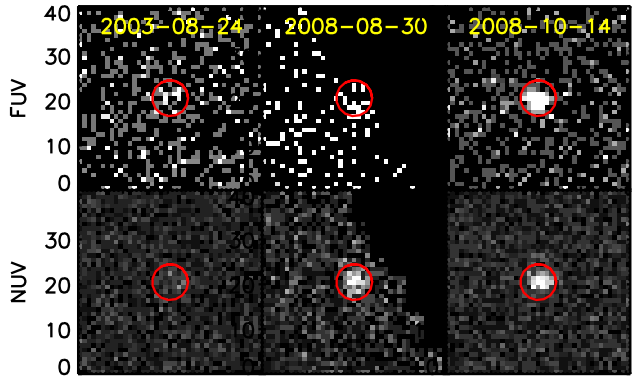}
     \caption{GALEX FUV and NUV images before and during the flare in TDE2.  Red circles show the 6 arcsec radius aperture used to measure the photometry.  The plate-scale of the image is 1.5 arcsec/pixel.  Note that in the second epoch (2008-08-30), the source is at the edge of the field where the photometry is more susceptible to systematic errors due to distortions in the PSF. }\label{fig:GALEX-stamp}
\end{centering}
 \end{figure}

We have \textsl{GALEX} NUV and FUV detections of the TDE2 host in both the pre-flare (2003 24 August) and post-flare (2008 14 October) state.  The TDE2 host was also observed by \textsl{GALEX} on 2008 30 August, but the image was at the edge of the field and the sky was noisy, so we display the NUV value in Fig. \ref{fig:lc}b) but do not use it further.  Figure \ref{fig:GALEX-stamp} shows the \textsl{GALEX} images.  The pre-flare \textsl{GALEX} UV flux values, corrected for the energy lost due to the $6''$ radius \citep{Morrissey07}, are ${\rm FUV}=23.0\pm0.3$, ${\rm NUV}=22.8\pm0.2$. The \textsl{GALEX} post-flare \emph{difference} magnitudes are:  ${\rm FUV}=21.4\pm0.2$, ${\rm NUV}=21.1\pm0.1$. Fig. \ref{fig:sed}b shows the optical-UV SED.

In addition to the UV data, we have two serendipitous and two follow-up optical observations of TDE2 and its host as discussed in greater detail below.  The spectrum of the host galaxy was obtained in a follow-up observation with the William Herschel Telescope (WHT) in November 2010.  Serendipitously, the first points on the SDSS light-curve of TDE2 suggested it might prove to be a supernova and an alert was issued \citep{Pojmanski07}.  As a result, an optical spectrum was obtained with ESO's New Technology Telescope (NTT) on 2007 September 18, 4 days after the first detection of the flare, and radio observations were obtained with the VLA 7 and 92 days after the first SDSS detection;  these observations are discussed in Section \ref{sssec:spectra} below.  Furthermore, a review of Catalina Real-time Transient Survey observations after discovery of TDE2 reveals a detection of the flare 95 days before the beginning of the Stripe 82 observing season. A preliminary report on this data is given in Section \ref{sssec:CRTS}; a more thorough analysis using also a Keck spectrum of the host galaxy obtained in October 2010 will be reported elsewhere.

\subsubsection{TDE2 Host and Flare Spectra}\label{sssec:spectra}
In November 2010, 850 days after the first SDSS detection of the flare, the host galaxy of TDE2 was observed with the WHT for two nights.  On 01 November two ACAM exposures of 900s were taken using a V400 grating and a slit width of $1"$. On 05 November, the host was observed again using the same specifications and exposure times, except that a slit width of $1\farcs5$ was used. The data of both nights (1h total integration time) was reduced separately, using standard IRAF\footnote{IRAF is distributed by NOAO, which is operated by AURA, under cooperative agreement with NSF} routines, to yield four 1D spectra which were then combined. The overall flux normalization was obtained using a standard reference star; the wavelength dependent flux normalization was fixed using the SDSS PSF flux of the host.

The NTT spectrum -- taken 5 days after the first SDSS detection of the flare -- was taken using the ESO Multi-Mode Instrument (EMMI) \citep{SPIE86} in the RILD mode (Red Imaging and Low-Dispersion spectroscopy). The grism that was used, Gr 2, has a wavelength coverage of 3800-9200 {\AA}, 300 grooves/mm, a wavelength dispersion of 1.74 {\AA}/pixel, and a spatial resolution of $0\farcs166$ /pixel (before binning). During the observation a binning of 2x2 was used. The slit width was $1\farcs5$ and the exposure time was 1200 s.  The flux calibration was performed using spectra of spectrophotometric standard stars observed with a $5''$ slit. The standard star spectra were also used to construct a telluric spectrum by isolating the telluric absorption in them. The spectrum of TDE2 was then corrected by subtracting a scaled version of the telluric spectrum. Finally the spectrum was corrected for dust extinction in the Milky Way \citep{Schlegel98} using the \citet{cardelli89} CCM extinction law with $R_V=3.1$. For a more detailed description of the data reduction we refer to \citet{Ostman11}.  Because the spectrum of TDE2 was not obtained in parallactic angle, it is affected by differential atmospheric refraction. The wavelength dependent flux loss was estimated and corrected for, taking into account the seeing, slit width, airmass, the angle from parallactic angle and the wavelength at which the object was centered on the slit \citep{Owens67}.

\begin{figure}
     \includegraphics[ trim = 0mm 12mm 0mm 55mm, width=.5 \textwidth]{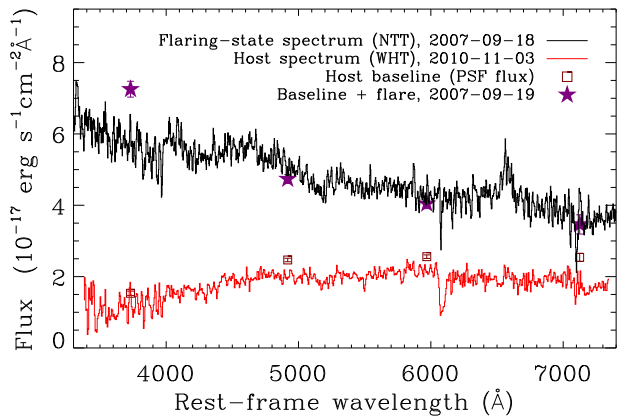}
     \caption{Host and flaring-state spectra for TDE2:  NTT spectrum in the flaring state (black) and WHT spectrum of the host (red). We also show the SDSS PSF \citep{stoughton02} flux of the host in the quiescent state (open squares) and the SDSS flux during the flare (purple stars). We see that the photometric and spectroscopic flux calibration agree reasonably well. The narrow ${\rm H{\alpha}}$ emission is consistent with the level expected from star-formation as implied by the pre-flare FUV luminosity \citep{Kennicutt98}. A zoom on H$\alpha$ is shown in Fig. \ref{fig:TDE2_zoom}. }\label{fig:NTT_spec-fit}
 \end{figure}

Figure \ref{fig:NTT_spec-fit} shows the host and flaring-state spectra together with the SDSS photometry; a zoom on H$\alpha$ is shown in Fig. \ref{fig:TDE2_zoom}. The H$_{\alpha}$ line luminosity in the host spectrum, measured by fitting the observed flux to a Gaussian whose width is given by the PSF, is $L_{{\rm H}\alpha}=1.2 \pm 0.2 \times 10^{40} \, {\rm erg}\,{\rm s}^{-1}$; we simultaneously fit for the [NII] emission lines which yields $\log({\rm[NII]}\lambda 6583/{\rm H}{\alpha}) = -0.4\pm0.2$. Using the redshift obtained from this simultaneous fit as an initial guess, we measure the flux of other well-known emission and absorption lines in the spectrum; only lines measured above the 3-$\sigma$ level, that are displaced by no more than 2\% from the initial redshift, are considered real detections. The results of running this procedure for the host and flare spectrum are shown in Table \ref{tab:spec}. Using all detected lines in the NTT and WHT spectrum we obtain $z=0.2515\pm0.0036$. From the WHT spectrum we obtain $L_{[\rm OIII]} <  3.6 \times 10^{39}\, \mbox{erg}\,\mbox{s}^{-1}$.

Because TDE2 occurred at the center of its host, the NTT spectrum contains a large galaxy component that has to be subtracted to obtain the ``pure'' flare spectrum. After subtraction, no narrow (i.e., unresolved) lines remain (Fig. \ref{fig:TDE2_zoom}, bottom panel). However at the rest-frame wavelength of $\mbox{H}_\alpha$ we identify a broad feature which can be fitted by a single Gaussian of $\sigma=75\pm 5\, \mbox{\AA}$ or ${\rm FWHM} = 8\times10^3 \, \mbox{km}\,\mbox{s}^{-1}$ and has an equivalent width of $ 87\pm 5\,\mbox{\AA}$.

Besides the NTT spectrum, the initial identification of TDE2 as a possible SN also triggered radio observations. We reduced the data from two 8.5 GHz VLA observations, obtained as part of project AS 887 -- the only VLA observations of this field after this event.  The first VLA observation was 7 days after the first SDSS detection of TDE2 and the second was 85 days later.  For both observations, we calibrated the recorded visibilities to flux density using data from short observations of 3C48, and used visibility data of PMN J2323-0317 for phase calibration \citep{Baars77}.  Imaging the two observations separately yielded no detection of TDE2 at either epoch. Combining the visibility data of  TDE2, we obtained an image with an RMS noise of $\sim35 \, \mu$Jy per beam using natural weighting. No source was detected at the location of TDE2, allowing us to place a 3-$\sigma$ upper limit of $\sim0.1$ mJy on its 8.5 GHz flux density.

\begin{deluxetable*}{l c c c c }

\tablewidth{0pt}
\tabletypesize{\footnotesize}
\tablecolumns{           5}
\tablecaption{Narrow Lines\label{tab:spec}}
\tablehead{\colhead{Line name} & \colhead{EW host} & \colhead{EW flare} & \colhead{Flux host} & \colhead{Flux flare} \\
\colhead{} & \colhead{(\AA)} & \colhead{(\AA)} & \colhead{($10^{-16}\, \mbox{erg}\, \mbox{s}^{-1} \mbox{cm}^{-2}$)} & \colhead{($10^{-16}\, \mbox{erg}\, \mbox{s}^{-1} \mbox{cm}^{-2}$)} \\}
\startdata
$\mbox{H}_\alpha$ & $4.2 \pm 0.8$ & $4.1 \pm 0.7$ & $0.8 \pm 0.1$ & $1.8 \pm 0.3$ \\
$\mbox{[NII] 6585}$ & $1.4 \pm 0.7$ & $2.8 \pm 0.7$ & $0.2 \pm 0.1$ & $1.3 \pm 0.3$ \\
Ca H & $-5.7 \pm 1.8$ & $-2.8 \pm 0.7$ & $-0.8 \pm 0.3$ & $-1.5 \pm 0.4$ \\
Ca K & $-7.6 \pm 1.3$ & $-3.3 \pm 0.7$ & $-1.1 \pm 0.2$ & $-1.9 \pm 0.4$ \\
$\mbox{H}_\delta    $ & $-3.4 \pm 1.1$ & $-$ & $-0.6 \pm 0.2$ & $-$ \\

\enddata
\tablecomments{Narrow (i.e., unresolved) lines detected in the host (WHT) and flaring-state (NTT) spectrum. Line fluxes are obtained by fitting a Gaussian profiles whose width is given by the PSF to the observed flux. Only lines detected above 3$\sigma$ are considered.}

\end{deluxetable*}

\subsubsection{Interpretation of Host and Flare Spectra}\label{sssec:interp}
The WHT host spectrum is consistent with the conclusion based on photometry in Section \ref{sec:rel_flux}, that the flare is not due to extreme variability of a hidden AGN. Such an AGN would reveal itself by a higher [NII]$\lambda 6583$ to H$\alpha$ ratio than observed. Using the BPT diagram (Eq. \ref{eq:BPT}), we see that the observed ratio is consistent with the narrow H$\alpha$ originating entirely from star formation. Moreover, using the \citet{Kennicutt98} relation, we conclude that the star formation rate inferred from the narrow ${\rm H}{\alpha}$ luminosity  ($0.5 \, M_\odot \, {\rm yr}^{-1}$) is consistent with the rate expected from the pre-flare FUV luminosity ($0.7 \, M_\odot \, {\rm yr}^{-1}$).  

\begin{figure}
     \includegraphics[ trim = 15mm 5mm 0mm 15mm, clip, width=.5 \textwidth]{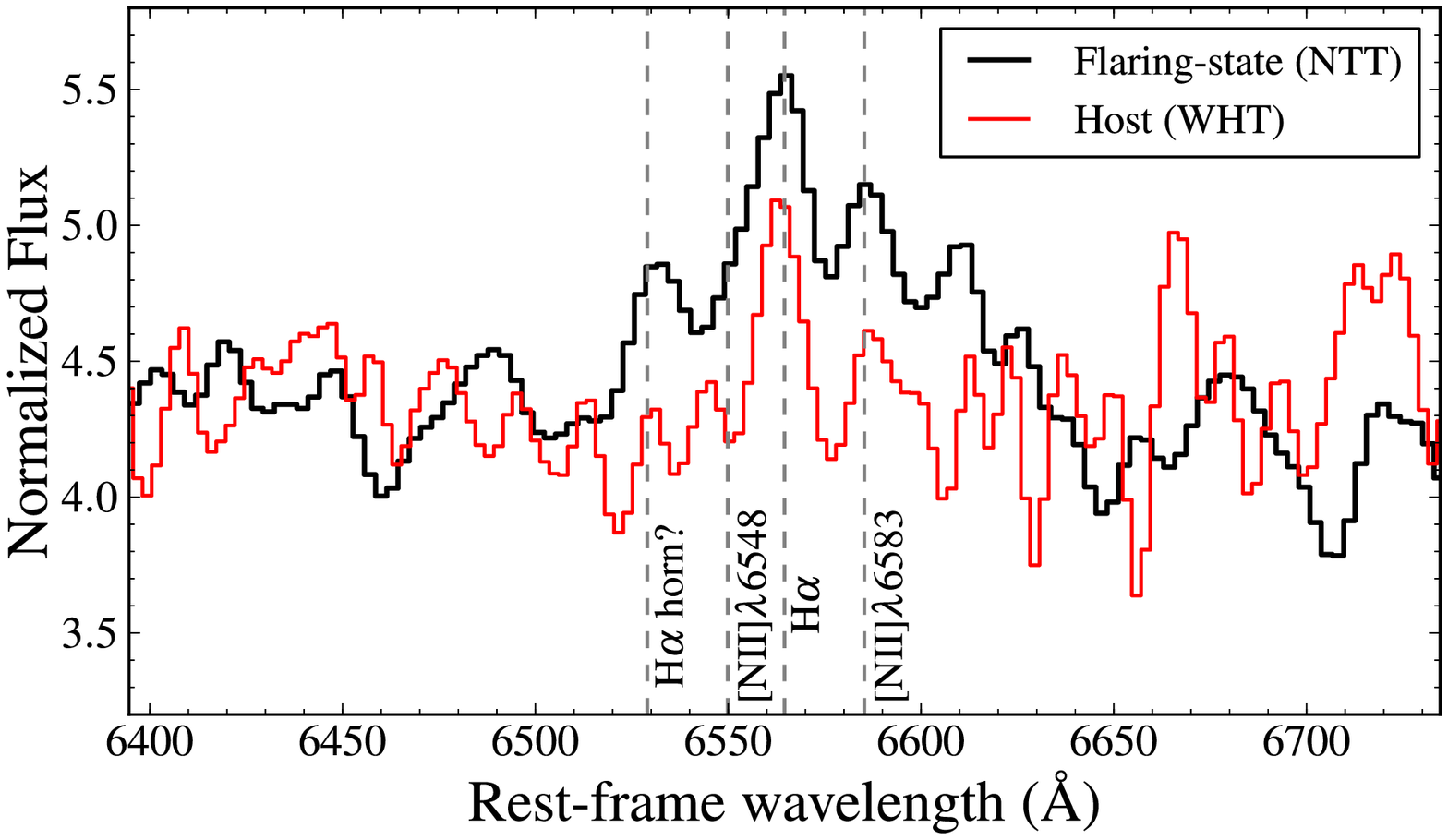} \\ 
     \includegraphics[ trim = 15mm 5mm 0mm 0mm, clip, width=.5 \textwidth]{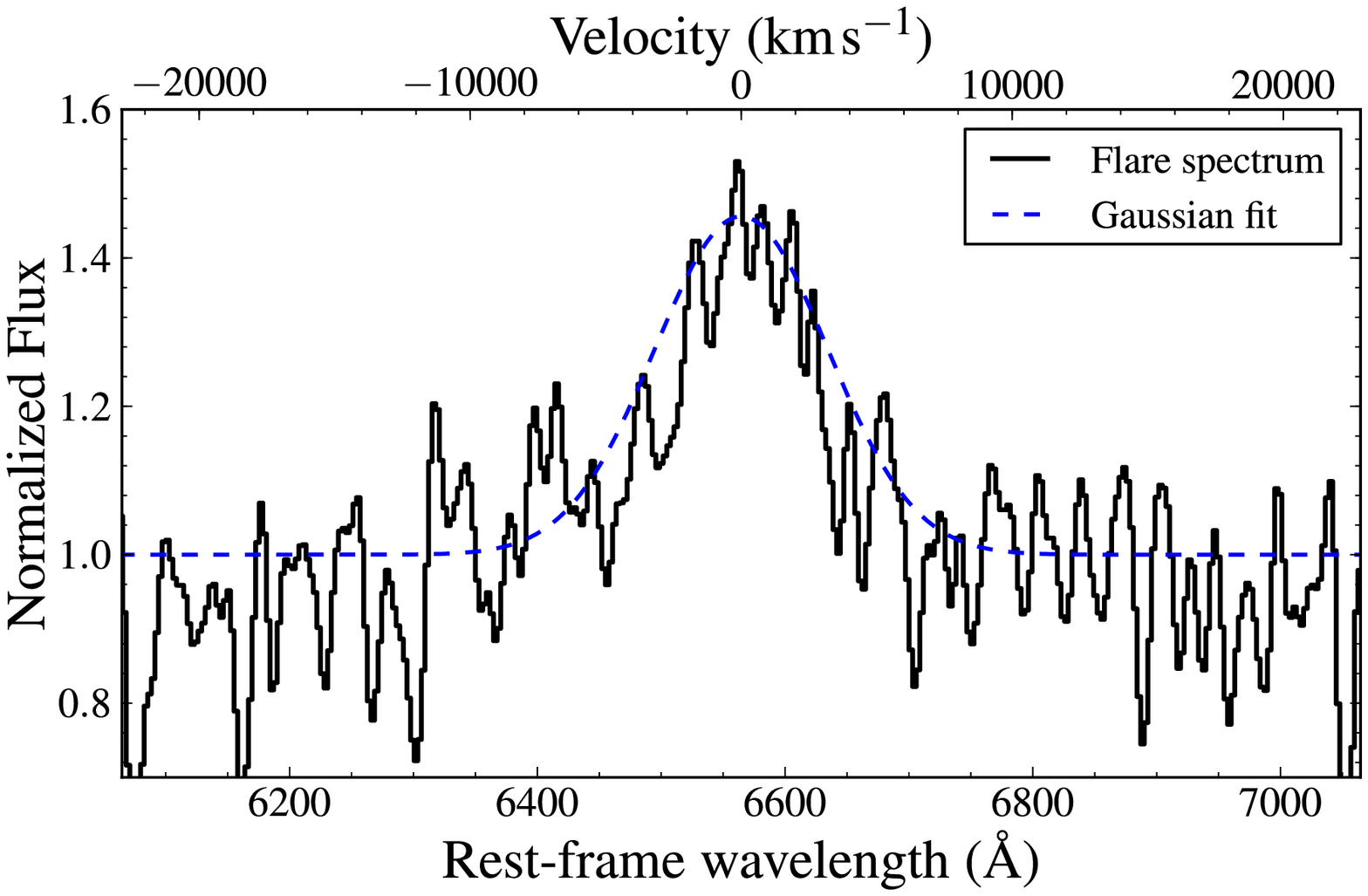} 
     \caption{Zoom on the H$\alpha$ emission of TDE2 for the flaring-state and host spectrum (top) and the host-subtracted spectrum (bottom). The host spectrum is arbitrarily rescaled for comparison to the shape of the flare spectrum. For display purposes, all spectra are  smoothed using a Gaussian filter with $\sigma=1.5\,\mbox{\AA}$. The intermediate width ${\rm H}\alpha$ line in the host-subtracted spectra can be fit with a single Gaussian to find a FWHM of $8\times 10^3\, {\rm km}\,{\rm s}^{-1}$. The feature at $6529.7\, \mbox{\AA}$ is detected at the 2$\sigma$ level (EW $=1.5\pm 0.7\, \mbox{\AA}$); it may be interpreted as weak, blueshifted H$\alpha$ emission (with a velocity of $1.5\times 10^3 \,{\rm km}\,{\rm s}^{-1}$), but could also be an artifact.  The full range of the host-subtracted spectrum is shown in Fig. \ref{fig:tde2_cpbx}.}\label{fig:TDE2_zoom}
 \end{figure}

Employing the widely-used cross correlation method \citep{Tonry79} for identifying SNe, using SNID \citep{Blondin07} on the host-subtracted spectrum, gives the best match to be an early SN type IIn spectrum, but this match is not very convincing: the degree of cross-correlation of the best-matching SN spectrum is similar to the value obtained for a template galaxy or AGN spectrum and is close to the cut-off value of the SNID software. A further comparison of TDE2 to type IIn SN is presented in Section \ref{sec:sn_compare}. 

An exciting signature of a TDE is line emission from tidal disruption debris which is shifted by the very high Keplerian velocities in the vicinity of the supermassive black hole.  However, the strength of this line emission is predicted to be much weaker than the flare continuum \citet{bogdanovic04}, and is likely below the detection threshold of the TDE2 flare spectrum. We note that the narrow line at $\Delta \lambda \approx 30\, \mbox{\AA} \, (1.5\times 10^3 \,{\rm km}\,{\rm s}^{-1})$ blueward of H$\alpha$ (Fig. \ref{fig:TDE2_zoom}, top panel) is a candidate to be such a shifted line, but it is detected only at the 2$\sigma$ level, with an equivalent width of $1.5\pm 0.7\, \mbox{\AA}$.

The interpretation of the intermediate width line in the host-subtracted flare spectrum (Fig. \ref{fig:TDE2_zoom}, bottom panel) is not straightforward. \citet{StrubbeQuataert11} calculate that the line-broadening due to the line-of-sight velocities of the super-Eddington outflows is probably too broad to be detectable. However the stellar debris orbiting the black hole at eccentric orbits could also produce a broad component \citep{Komossa09}. Since intermediate width ${\rm H}_{\alpha}$ lines are a well-known property of many AGN and some type II SNe, their detection in the host-subtracted spectrum of TDE2 is not particularly constraining to the nature of this flare. 

\subsubsection{CRTS observations}\label{sssec:CRTS}
The Catalina Real-Time Transient Survey (CRTS) analyzes data from the Catalina Sky Survey which repeatedly covers $>$30,000 square degrees on the sky, in order to search for optical transients with timescales of minutes to years \citep{DrakeCRTS09}. The Catalina Sky Survey (CSS) Schmidt Telescope and Mount Lemmon (MLS) telescopes are located north of Tucson, Arizona and survey the northern sky using unfiltered 4k $\times$ 4k CCD cameras. 
Images from the MLS telescope cover 1.1 deg$^2$ and reach $V=$21.5, while images from the CSS cover 8 deg$^2$ and reach $V=19.5$. On a clear night, these two telescopes cover $\sim$ 1500 deg$^2$ of sky in sequences of four 30 s exposures. Although CRTS began in 2007 November archival CSS and MLS data dates back to 2004.

CRTS observations of TDE2 are available both in between and beyond the SDSS observational seasons. This data and a Keck spectrum of the host will be presented in a forthcoming paper, with full calibration and optimized difference imaging so they can be quantitatively combined with other observations.  
Here we only show the mean ``unfiltered" CSS difference magnitude (within $\sim 20\%$ of the SDSS $r$-band) of the three CSS detections of the TDE2, obtained 95 to 80 days prior to the first SDSS detection. These data are shown in Figure \ref{fig:lc} to:
\begin{itemize}
\item date the first SDSS observation and NTT spectrum as being at least 95 and 100 days respectively after the peak of the flare, and
\item show that (Fig. \ref{fig:lc}) the peak luminosity of the flare is considerably higher than observed by SDSS, by approximately one magnitude.
\end{itemize}

\bigskip

\section{Comparison of tidal flares to AGN flares and supernovae}\label{sec:compare}

In this section we quantify the likelihood that TDE1 and TDE2 are examples of some already-observed phenomenon.  

\subsection{Comparison of TDEs to AGN flares}\label{sec:agn_compare}

Our pipeline was designed to exclude AGN flares based on the host properties (Sections \ref{sec:AGN_rejection}, \ref{sec:tde1} and \ref{sec:tde2}).  Here we examine the properties of the flare observations themselves to understand how often the behavior displayed by TDE1,2 may occur by chance in the AGN population.  We use two attributes of variable AGNs that we measure in our data -- the range of variability per season and the spectrum of the most extreme flares over the entire Stripe 82 observing period  -- to quantitatively compare TDE1,2 to AGNs.  First, we consider the likelihood for an AGN to have a flare meeting our selection requirements in one season, yet be as quiet as TDE1,2 in the other observed seasons.  Next, we consider the spectrum of flare amplitudes in the AGN population and compare to those of TDE1,2.  Finally we compare the spectral properties of the TDE1,2 flares to AGN flares.

\subsubsection{AGN Variability}\label{sssec:var}
\paragraph{Probability of an isolated flare episode in QSOs}

\begin{figure}
\begin{centering}
 \includegraphics[trim =00mm 00mm 00mm 72mm,  width=.5 \textwidth]{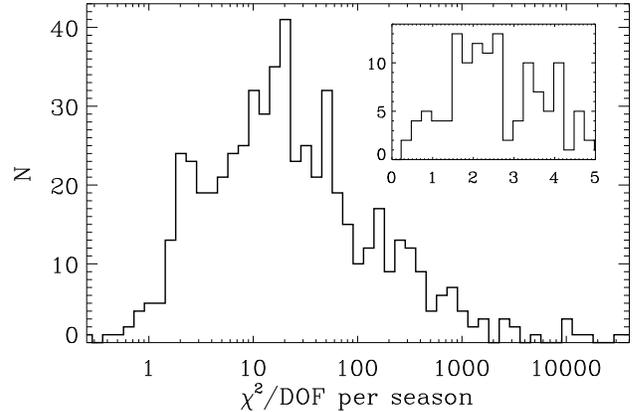} 
\caption{The $\chi^2 / {\rm DOF}$ distribution of all non-flare seasons with three or more observations ($\chi^2_s$), for the spectroscopic QSOs and Seyfert galaxies. The values of $\chi^2_s$ in the non-flare seasons of TDE1,2 are respectively \{1.4, 2.6, 0.6, 1.7, 1.9\} and \{1.0, 1.6, 1.0, 2.0\}. The inset zooms in on the first few bins of the histogram, using a linear bin size.  }\label{fig:chisqs}
\end{centering}
\end{figure}
As discussed Section \ref{sec:rel_flux}, we rejected three candidates which pass all cuts except that their flux in other seasons is not consistent with being constant . We can use the measure of variability developed there to determine the likelihood that an AGN is as quiet as TDE1,2 in all but one observing season.   We have a sample of variable AGNs with the same selection criteria as TDE1,2 -- our sample of flares that are in QSOs and identified Seyfert galaxies -- with which we can quantify the flux excursions in ``off" seasons.  To do this, we introduce $\chi^2_s$: the $\chi^2/{\rm DOF}$ per season, $s$.  The median flux excluding the season that contains the main flare is used as a model for the light curve in other seasons.  We calculate $\chi^2_s$ for all seasons (other than the one with the primary flare) having three or more detections, since that is the minimum number of detections for the TDE candidate sample.   
Fig. \ref{fig:chisqs} shows the distribution of these $\chi^2_s$ values.  
Let $f_j^{(1,2)}$ be the fraction of AGNs with as low or lower value of $\chi^2_s$ as TDE1,2 in season $j$.  Our estimate for the likelihood for an AGN to have as little activity in the off seasons as  displayed by TDE1,2 is then 
\begin{equation}\label{eq:agn_prob} 
P_{\rm AGN}^{(1,2)}\leq  \prod_{j}^{N_s}\;2\, f_j^{(1,2)} \quad.
\end{equation}
Here the product runs over the non-flare seasons with three or more detections; $N_s = 5,4$ for TDE1,2.  The factor 2 is inserted in Eq. (\ref{eq:agn_prob}) because the mean value of $f_j $ for AGNs is 0.5.  Using the relative photometry light curves introduced in Section \ref{sec:AGN_rejection}, we find $P_{\rm AGN} \leq 2\times10^{-6},\, 2 \times 10^{-5}$ for TDE1,2 respectively.

This estimate of the chance probability for a variable AGN to have several quiet years surrounding a major flare, and thus to be able to mimic TDE1,2's variability properties, assumes the flux variability in the years surrounding the major flare is uncorrelated.  This is a reasonable first approximation and can be improved by studying the statistics of the AGN fluctuations in the years near a major flare.  It is important to emphasize that we are {\em not} making the assumption that the natural flux variability of AGNs in seasons near a major flare is the same as in randomly chosen seasons long before or after a major flare:  if there is an enhancement or suppression of flux excursions before or after major AGN flares, it is captured in our measured $\chi^2_s$ distribution, which is measured in AGNs in years surrounding a flare selected by the same criteria as for TDE1,2.    

\paragraph{Probability of comparably large flares in QSOs}
\begin{figure}
\begin{centering}
  \includegraphics[ trim = 0mm 4mm 12mm 24mm, width=.5 \textwidth]{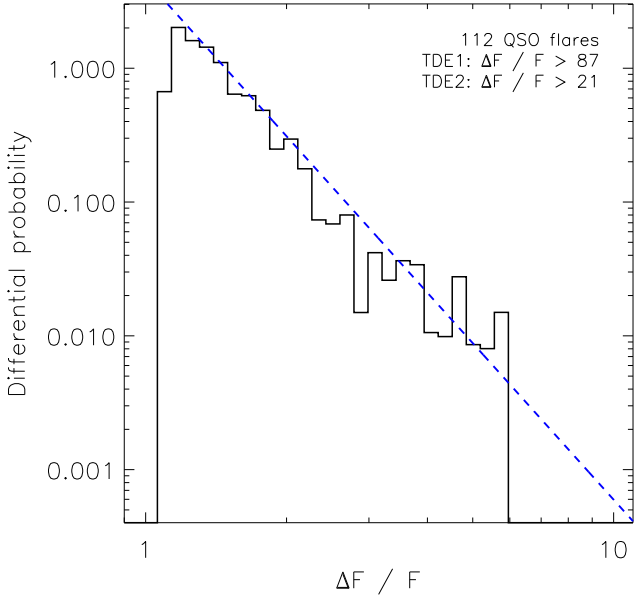}
\caption{The flux increase of QSO flares with respect to the baseline of the flare for the $g$, $r$ and $i$-bands. We fit the histogram with $P(\Delta F/F) \propto (\Delta F / F)^{\alpha}$ for all bins with a flux increase larger than $10\%$ and obtain $\alpha=-3.9\pm0.2$. 
The probability of finding a flare as large or larger than a given $\Delta F/F$ is obtained by integrating this fit and multiplying by  112/1304, the ratio of flaring QSO to all QSO that have been searched for flares.
For reference, we also give the lower limits on the flux increase of TDE1,2 with respect to a hypothetical AGN baseline flux as derived from the (upper limit on) the [OIII] luminosity.}\label{fig:deltaf}
\end{centering}
\end{figure}

We can use another property of the flaring QSO sample to obtain a second, independent probability measure that TDE1 or TDE2 are AGN flares. For this, we quantify the spectrum of flux increase at the peak of QSO flares relative to their baseline flux.  An upper limit on the [OIII] line luminosity of a galaxy can be converted to an upper limit on the baseline luminosity of its active nucleus at  $5000\,\mbox{\AA}$ in the rest-frame  \citep{heckman04} . The observed luminosities of TDE1,2 at this wavelength imply a minimum flux increase with respect to the baseline state of the accretion disk of a factor 87, 21 respectively.  A flux increase of this magnitude is extremely unlikely for an AGN: out of the 1304 extended QSOs in Stripe 82 that we monitored for flares, the largest flux increase measured in the $g$, $r$ or $i$ bands, is a factor of 5.  The spectrum of $\Delta F/ F$ for the QSO flares meeting our selection criteria is shown in Fig. \ref{fig:deltaf}; it is a power-law with slope $\alpha = - 3.9 \pm 0.2$.  Using this power-law fit to calculate the probability of as large a flux excursion as seen in TDE1,2, if they were variable QSOs, gives $P(\Delta F/ F > 87) = 3 \times 10^{-7}$, $P(\Delta F/ F > 21) =2 \times 10^{-5}$, respectively.  

\paragraph{Applicability of the above estimates}
The probability estimates above are valid if there is only one type of AGN variability.  Evidence that a single phenomenon is responsible for the variability observed in most accreting super-massive black holes can be found in the literature. Using  $\sim 9000$ spectroscopically confirmed QSOs of Stripe 82, \citet{Macleod10} showed that a damped random walk model can explain quasar light curves at an impressive fidelity level ($0.01-0.02$ mag), indicating that one single physical process, e.g., turbulent magnetic fields within the accretion disk \citep{Kelly09}, is the dominate source of the variability. This analysis does not apply to BL Lacs, which show larger amplitude fluctuations (up to $\Delta  m\sim1$) than QSOs on all timescales \citep{Bauer09}, but as we shall see, the properties of TDE1,2 do not place them in this class.

The violent fluctuations in BL Lac type AGNs are very likely due to a fluctuating contribution from a relativistic jet. A good example is the flare found by \citet{VandenBerk02}. The first SDSS photometrical observations showed a red source ($g-r=0.3$); the SDSS spectrum obtained about a year later, showed this sources had faded by $\approx2.5$ mag and revealed a galaxy spectrum. Followup observations by \citet{Gal-Yam02} showed broad $\mbox{H}_\alpha$ in the host spectrum, and a second flare with a blue continuum. The radically different SED for the second flare, combined with a detection at 1.4 GHz in FIRST \citep{becker95} and other radio catalogs that revealed radio flux variability, lead \citet{Gal-Yam02} to conclude the flare originated from a radio-loud AGN, probably in the BL Lac class. We note that the \citet{VandenBerk02} flare is somewhat similar to R1 and R2, the flares we rejected based on additional variability of the host (section \ref{sec:rel_flux}), which are also detected in FIRST at $F_{1.4 \rm\, GHz} > 90$ mJy and have similar red colors. TDE1,2 are nothing like the \citet{VandenBerk02} flare: their post-flare spectra show no broad lines, the flare SED is nearly constant, additional seasons of observations in Stripe 82 show no additional variability, and they are not detected in FIRST (or targeted VLA observations for TDE2). 

To quantify the difference between the host of the \citet{VandenBerk02} flare and TDE1,2, we use the \mbox{3-$\sigma$} upper limit on the radio luminosities ($L_\nu<1\times 10^{29}\,{\rm egs}\,{\rm s}^{-1}{\rm Hz}^{-1}$ at 1.4 GHz for TDE1 from FIRST and $L_\nu<2\times 10^{29}\,{\rm egs}\,{\rm s}^{-1}{\rm Hz}^{-1}$ at 8.5 GHz for TDE2) to compute the optical-to-radio spectral indices, $\alpha_{\rm ro} = -\log(F_{\rm r}/F_{\rm o})/\log(\nu_{\rm r}/\nu_{\rm o})<0.08,  0.04$ for TDE1,2 using the $g$-band peak flux for $F_{\rm o}$.  From SDSS observations, optically selected radio-loud BL Lacs have a Gaussian distribution with $\left<\alpha_{\rm ro}\right>=0.42\pm 0.08$ \citep{Plotkin10b}, hence the low radio-to-optical ratio of the hosts of TDE1 or TDE2 is not consistent with a blazar-origin of these flares; they would have to members of a new class of radio-quiet BL Lacs with violent flares.

About 80 radio-quiet objects ($\alpha_{\rm ro}<0.2$) with spectra that resemble BL Lacs at optical wavelengths (i.e., no or very weak emission lines, blue continuum), are known at $z<2.2$  \citep{Plotkin10,Plotkin10b}. Using Stripe 82 data, \citet{Plotkin10} conclude that the level of optical variability of these AGNs is consistent with other radio-quiet quasars: $\chi^2/\mbox{DOF}>10$ for all Stripe 82 observations in the $g$-band (which can be compared to Fig. \ref{fig:non-flare-chi2}) and no excursions from the mean flux greater than $\Delta m =0.5$. Hence all of these peculiar AGNs would be identified by our pipeline and excluded based on their additional variability, and none show flares that are large enough to hide the underlying AGN.  We conclude that on the radio-quiet branch of AGN, there is no evidence for a different, more violent mode of variability as seen in radio-loud BL Lacs, but more data is needed to fully exclude the existence of such a mode.

\subsubsection{Colors of TDE and QSO flares compared}

\begin{figure}
\begin{center}
  \includegraphics[trim = 0mm 4mm 12mm 24mm,  width=.5 \textwidth]{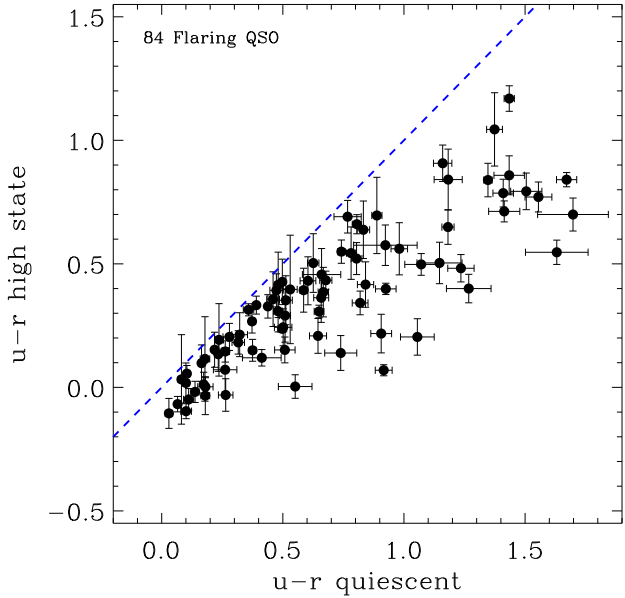}
\caption{Colors of spectroscopic QSO. 
The quiescent state is calculated from the mean PSF flux of the non-flare nights. The high state is obtained by adding the flux of the quiescent state to the difference image and taking the mean colors. No color change during a flare is indicated by the striped blue line. We see that QSOs in their high state are slightly bluer; this effect is most pronounced for flares from relatively red QSO (quiescent  $u-r>0.5$) which can be explained by stellar contamination. For reference, the $u-r$ colors of TDE1,2 measured in the difference image are $-0.62\pm0.06$, $-0.36\pm0.03$.
   }\label{fig:qso-color-change}
\end{center}
\end{figure}

Figure \ref{fig:cc}b shows that the colors of the TDE flares, as measured in the difference image, fall outside the locus that contains 97\% of all spectroscopically confirmed low-redshift ($z<0.8$) QSOs in SDSS \citep{schneider07}.  Here we investigate whether, when they flare, QSOs may change color radically enough that our TDEs' SEDs could be consistent with that of a flaring QSO.  To explore this question, we compare in Fig. \ref{fig:qso-color-change} the $u-r$ colors of QSOs in the flaring and baseline states.  QSOs are bluer in their high state than in the quiescent state, but the change is not nearly enough to push a QSO out of the QSO-locus sufficiently to match the colors of TDE1 or TDE2. A large color change is observed only for flares from QSOs with a relatively red baseline, i.e., where the galaxy contribution is relatively large and is not subtracted, whereas for the TDE flares the galactic contribution has been subtracted.  

\subsubsection{Comparison of TDEs and Seyfert flares}
Comparing the properties of the TDE1,2 flares to the flares of QSOs is powerful because the large number of QSOs allows the dispersion in the QSO flare properties to be measured.  However the TDE hosts are clearly not QSOs so we also compare specifically to the three examples in our flare sample in which the host galaxy is a Seyfert as defined in Section \ref{sec:AGNspec} (galaxies having broad lines or satisfying the BPT criteria, not labeled QSO by SDSS).  In Seyfert galaxies, the AGN contributes only a moderate fraction of the total flux, unlike in QSOs for which the AGN dominates by definition. Hence for Seyferts the difference image flux gives a better measure of the flux of the AGN, and we therefore compare the properties of the difference images of the TDEs with those of the three flares with identified Seyfert hosts which pass our selection criteria. The magnitude and color decay rates of the three Seyfert flares cluster around zero; the bluest Seyfert flare has $u-g=g-r=-0.2$  which is roughly as blue as TDE2.  To obtain more thorough comparison of the broadband properties, we fit each of the five mean SEDs (i.e., the average color) of the difference image to black body spectra.  The Seyfert flares all have poor fits to a black body ($\chi^2/{\rm DOF}=5.9,\,  7.6.\, 22.7$) while TDE1,2 are both well-fit by black-body spectra ($\chi^2/{\rm DOF}=1.9,\, 0.4$), at optical wavelengths.

The light curves of the Seyfert flares are also different from those of TDE1,2.  They are more symmetric (i.e., rising and falling at similar rates) or show more substructure, and they also show variability in the non-flare season. To illustrate this, the light curve of a Seyfert galaxy that hosted at nuclear flare is shown in the bottom right panel of Fig. \ref{fig:rel_lc}. 

\subsection{Comparison of TDE1,2 to Supernovae}\label{sec:sn_compare}
In this section we show that TDE1 and TDE2 are unlike any off-center SN recovered by our pipeline, based on the properties of the flares.  We examine the possibility that TDE1,2 could be exotic SNe, but find no examples of SNe which resemble the TDEs in all significant respects.  We calculate the likelihood that TDE1,2 are the most similar type of known SNe -- SNIIn -- and are by chance as close as observed to the centers of their galaxies to be $\lesssim 0.08~\%$.  Finally, we consider the possibility that TDE1,2 are examples of a new type of extreme SN which occur only near the centers of galaxies, but this requires a thousand-fold or greater enhancement in the rate of such explosions in stars in the nuclear region compared to elsewhere.

\subsubsection{Constraints from late-time UV emission}
The residual UV radiation detected 781, 485 days after the first optical detection of TDE1,2 is incompatible with observed UV properties of SNe. To estimate their maximum possible NUV flux at these times, if TDE1,2 were SNe, we assume that when first detected in the optical, they had the bluest NUV to $g$-band ratio ever observed in any SN: ${\rm NUV}-g=-1$ \citep{bianchi07}.  We further assume that the UV flux falls off with the slowest linear decay rate of UV magnitude that has been measured for a large sample of Ia, Ib/c and type II SN \citep{Brown09}: $dm_{\rm NUV}/dt=0.05\,{\rm mag}\,{\rm day}^{-1}$, which is similar to the UV decay rate of SN 2008es \citep{Gezari09b,Miller09}. This yields upper limits on the NUV fluxes for TDE1,2 under the SN hypothesis which are far below those observed: apparent magnitude of $m_{\rm NUV, \, SN} > 59,42$ compared to  $m_{\rm NUV, \, TDE} =23.0, 21.1$.  We see that a linear decay, as commonly observed for normal SNe, vastly under predicts the UV flux, while a power law decay, which is predicted for stellar tidal disruption events \citep{Rees88}, can explain the large residual UV flux, years after the flare event.  

The sample of \citet{Brown09} does not include type IIn SNe.  To obtain an estimate of the limit of the NUV magnitude for this type of stellar explosion we use the slowest $B$-band decay measured for the \citet{Kiewe10} sample of type IIn SNe ($dm_{B}/dt = 0.016\,{\rm mag}\,{\rm day}^{-1}$), to find $m_{\rm NUV, \, IIn} >32, 26$.  This upper limit on the NUV flux for type IIn SN is orders of magnitude below the observed late-time UV flux for both TDE1 and TDE2. 

\begin{figure}
\begin{center}
 \includegraphics[trim =0mm 5mm 5mm 55mm, width=.5 \textwidth]{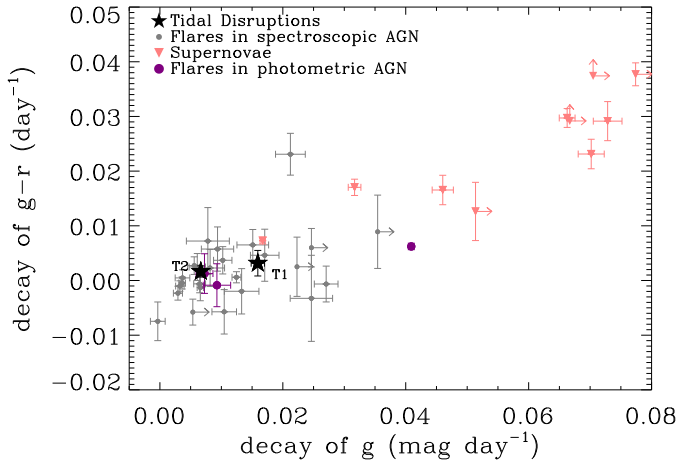}
 \caption{The change in magnitude and color measured after the peak of the flare. In this plot we only show the flares that passed the quality cut for the final TDE study (see Section \ref{sec:pot_tde}). The decay rate (color evolution) is obtained from the slope of the best-fit line to the flare magnitude (color) after the peak as function of time. All identified SNe show significant cooling, as expected, and their light curves decay faster than TDE1 and TDE2, except for one: a type II-P SN whose mean color, $u-g = 1.3$, is far redder than the TDEs. One flare from the photometric AGN sample is more like a SN than an AGN flare; this is consistent with the 0.9 SNe expected in the sample for the TDE analysis. }\label{fig:g_decay-gr_decay}
\end{center}
\end{figure}

\subsubsection{Comparison of TDE1,2 to flares in our SN sample}\label{sssec:compare}
In Section \ref{sec:h_d} we defined a sample of 85 flares which are clearly off-center from their host (Eq. \ref{eq:sn_cut}).
This sample allows up to compare the properties of TDE1,2 to normal SNe (exotic SNe are discussed below).
Because this sample is obtained by a cut on the host-flare distance only, their properties are representative of the properties of SNe that appear in the nuclear sample, which would not be the case if we compared to SNe identified in spectroscopic follow-up campaigns.  We note that requiring at least three $u$-band detections, as is imposed for the potential TDE sample, biases the selection toward bluer-than-average flares, likely increasing the fraction of type IIn SNe. Even so, the TDE1 and TDE2 flares are significantly different from all SNe flares we observed with respect to the mean colors and cooling rate measured using the mean $u-g$, $g-r$ colors and the slope of $u-r$, respectively (Figs. \ref{fig:cc}). Also the decay rate measured using the slope of $g$ is lower for TDE1,2 compared what is observed for the SN in the sample that meet the TDE quality cuts (Fig. \ref{fig:g_decay-gr_decay}), except for one: a II-P SN \citep[2006fg,][]{D'Andrea10}. However, as expected, this SN decays much faster in the $u$-band and is also red, $u-g = 1.3$.  

\subsubsection{Comparison of TDEs to type IIn and Exotic SNe}\label{sssec:IIn}
We saw above that the colors and cooling rates of the flares in our SN sample do not resemble those of TDE1,2, but this does not apply to all type IIn SNe and some exotic SNe.  Here we review such SNe and compare them to TDE1,2.

\paragraph{SNe IIn from the Caltech Core-Collapse Project}
Recently, \citet{Kiewe10} presented four type IIn from the Caltech Core-Collapse Project \citep{Gal-Yam07} which aims at producing an unbiased sample (i.e., essentially every young core-collapse SN that is observable was followed-up). At their peak, all four SNe presented by \citet{Kiewe10} have similar or even bluer $B-V$ colors and a similar decay rate as TDE1,2.  Two of them, 2005bx and 2005cp, show this decay rate over the full extent of their light curves (observed for 50 and 119 days after their peak, respectively), while for the other two the flux drops rapidly after about 2 months. SN 2005bx shows rapid cooling ($d(B-V)/dt=0.04\, \mbox{day}^{-1}$), while SN 2005cp has nearly constant color, $d(B-V)/dt=0.002 \,\mbox{day}^{-1}$, compared to  $d(B-V)/dt=0.003,0.004 \,\mbox{day}^{-1}$ for TDE1,2.  The presence of one among four type IIn SNe in the small but unbiased sample of \citet{Kiewe10}, which has similar colors and cooling with respect to the TDEs, thus shows that flare colors and cooling alone (Fig. \ref{fig:cc}) is insufficient to reject \emph{all} SNe. 

\begin{figure}
\begin{center}
 \includegraphics[trim =15mm 0mm 0mm 15mm, clip, width=.5 \textwidth]{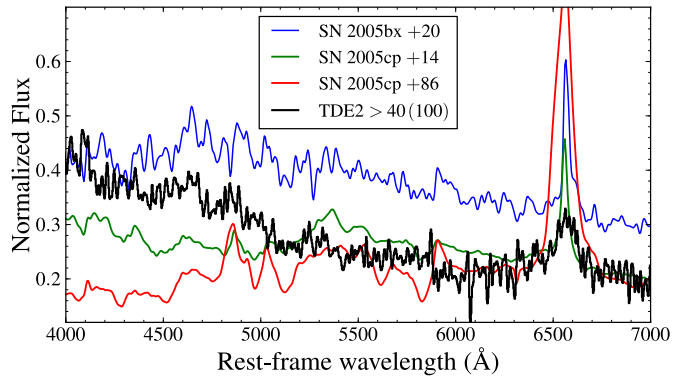}
 \caption{Comparison of the TDE2 flare spectrum (host subtracted) to the two type IIn SNe from \citet{Kiewe10} which are most similar to TDE1,2 in terms of cooling rate and color. The spectra are normalized to be equal at long wavelengths.  The legend lists the time difference with respect to the peak of the light curve.  For TDE2, the location of the peak in the light curve is unknown but we have a lower limit based on the first CSS detection of the flare (see Fig. \ref{fig:lc} and discussion in Section \ref{sssec:CRTS}). The bump at $\mbox{H}_\alpha$ in the TDE2 spectrum can be fit by a Gaussian with FWHM $\sim 8\times 10^{3}\: \mbox{km}\,\mbox{s}^{-1}$. The blue continuum of the pre-peak spectrum of SN 2005cp is similar to that of TDE2, but this SN spectrum shows more narrow and much stronger H$\alpha$ emission. The P-Cygni absorption profiles, a classic feature of SNe spectra, are clearly inconsistent with the TDE2 spectrum. } \label{fig:tde2_cpbx}
\end{center}
\end{figure}

We now compare the luminosity and spectral features observed for TDE2 to the \citet{Kiewe10} type IIn SNe which most resemble them.  With $M_B=-18.9, -18.0$, SN 2005bx, cp are significantly less luminous at the peak than TDE2, whose equivalent $M_B$ value is $-20.3$ when first observed, an unknown time after the peak.   In Fig. \ref{fig:tde2_cpbx} we compare the host-subtracted flare spectrum of TDE2 to the spectra of SN2005bx and 2005cp.  The emission lines and P-Cygni absorption profiles, a classic feature of SNe spectra, are absent in the TDE2 spectrum. 
While there is substantial dispersion within the sample of spectra of SN IIn, at all times in their evolution, the SNe spectra show stronger H$\alpha$ emission than in TDE2. 
Furthermore, the spectrum of TDE2 was taken at least $\sim 40$ days after the peak -- if the peak occurs between the CSS and CRTS observations -- and otherwise the spectrum is at least 100 days post-peak, yet the blue continuum shape seen in spectrum of TDE2 is observed for only tens of days after the peak for the type IIn SN in sample of \citet{Kiewe10}. 

\paragraph{Comparison of TDEs to exotic SNe} \label{sssec:exotic}
We examined above the SNe IIn in the small but unbiased \citet{Kiewe10} sample.  Here we consider specific SNe reported in the literature; due to selection biases these tend to be particularly luminous and unusual.  Only exceptionally blue and slowly decaying, relatively luminous SNe could mimic our TDEs. An example of a slowly decaying and luminous SN is 2006gy  \citep{Quimby06}, which was a rather bright explosion ($M_R = -21.3\pm0.1$) that is considered  to be an interaction-dominated type II SN \citep{Agnoletto09}.  Immediately after its peak, the decay of 2006gy, at $0.2\, {\rm mag}\,{\rm day}^{-1}$, is much faster than in either of our TDEs (see Fig. \ref{fig:g_decay-gr_decay}).  
At 270 days after its peak, the decay rate of 2006gy slows down to $0.004\,{\rm mag}\,{\rm day}^{-1}$, which is similar to known SNe of type II such as 1995G (IIn) \citep{Pastorello02} and 1999E (IIa, hybrid class) \citep{Rigon03}.  
This might lead one to speculate that our TDEs can be explained as emission from the late-time tail of the light curve of type II SNe. However, by the time their decay of flux becomes similar to our TDEs', the colors of SN 1995G, 1999E and 2006gy are red: $B-V>0.8$.  This is very different from TDE1,2 whose colors can be converted  \citep{Jester05} to find $B-V=-0.12, -0.02$. Furthermore, the spectrum of 2006gy at $t>270$ days is characterized by large and broad emission lines, which is inconsistent with the flare spectrum of TDE2 (Fig. \ref{fig:NTT_spec-fit}). 

An example of a blue and UV bright SN is 2008es  \citep{Gezari09b,Miller09}. However the fast cooling of this SN implies a mean color of $B-V=0.6$ over the 60 day period after the peak, which is far more red than our TDEs averaged over a similar period.  By day 91 the UV had faded 5 magnitudes fainter than the optical peak, in contrast to the $\sim 2$ magnitude decrement seen in the TDEs after $\sim$ 800, 400 days.
We can also compare the TDEs to the transient SCP 06F6  \citep{Barbary09}.  This has been suggested  by \citet{Quimby09} to be a high-redshift example of a new class of blue transients that mark the deaths of the most massive stars. Using $z=1.19$ for the redshift of SCP 06F6 \citep{Quimby09},  we extract a rest-frame $u$-band decay rate of $0.08\,{\rm mag}\,{\rm day}^{-1}$.  This is an order of magnitude faster than the decay rate of TDE1 and TDE2 in their rest-frame.  Furthermore, the proposed new class of SNe should originate from faint, metal-poor galaxies  \citep{Quimby09}, which are very different from the hosts of TDE1 and TDE2.  

Finally, we consider SN 2003ma \citep{Rest09}, an extreme type IIn SN. The peak magnitude of the SN, $M_R =-21.5$, is similar to that observed in TDE2, and its decay rate during the first 300 days after the peak ($\sim 0.008\, {\rm mag}\,{\rm day}^{-1}$ in the $VR_{\rm SM}$ band, centered at $6250\mbox{\AA}$) is also similar to TDE1,2. The SN was detected at $d \approx 460 \pm 85$ pc projected distance from its host. Like the TDE2 spectrum, the spectrum of 2003ma shows no emission and no P-Cygni absorption up to the last detection at 1423 days after the peak of the SNe. However the  intermediate width H$\alpha$ emission is much stronger than what is observed for TDE2.  Another difference is that SN 2003ma is redder than TDE1,2: its color averaged over the first 50-100 days after the peak is $B-R\approx 0.35$ compared to $B-R= -0.35,-0.18$ for TDE1,2. The SN also shows strong evolution of the foreground-extinction-corrected colors: $V -I \approx 0$ at peak to $V -I > 1$ one year after the peak.  There were no UV observations of SN2003ma but already by day 213 the B magnitude was 3.5 magnitudes below the peak in the I band, suggesting that by days 400-800 no UV would be seen.  We note that the host galaxy of 2003ma is a starburst galaxy \citep{Rest09}, as expected for the presumably rather massive ($M>10M_\odot$) progenitor of the SN; the star formation rate, derived from the narrow line H$\alpha$ emission of the host of SN 2003ma is one (five) orders of magnitude greater than the (upper limit on) the star formation rate from the TDE2 (TDE1) host spectrum. 

We conclude that TDE1,2 are not members of a \emph{known} type of exotic SNe. However new types of SNe are discovered on a yearly basis so we have to consider the possibility that we discovered a new exotic class of stellar explosion. In the following section we will use the distance between the host and the flare to evaluate this possibility. 

\subsubsection{Rejection of known SNe}\label{sssec:spatial}

\paragraph{Geometric Rejection of stellar-distributed objects in nuclear sample}
Although we rule out that TDE1,2 are ordinary SNe on the basis of the properties of the flares, it is of interest to know the chance probability for two flares whose progenitors follow the stellar distribution to be found as close to the nucleus as observed.  We assume that the rate of ordinary SNe is proportional to the stellar light.  This is justified by the good quality of the fit to the host-flare distance distribution (Fig. \ref{fig:sn_mc-0}) and the existing literature on SN distributions. \citet{Fruchter06} conclude that, while gamma-ray bursts are more concentrated on the brightest regions, the probability of a SN type Ia exploding in a particular pixel is roughly proportional to the surface brightness of the galaxy at that pixel. This result was confirmed by \citet{Kelly08} using also type II SNe; they conclude that both type Ia and type II SNe follow the galaxy light measured in the $g$-band, with a clear exception being the rarer SNe Ic associated with long-duration gamma-ray bursts.   (The observation by  \citet{AndersonJames08}  that SN type II do not trace star formation estimated from $\mbox{H}_\alpha + \mbox{[NII]}$ emission is not relevant for this work, because our model relies on stellar light, not star formation.  Indeed \citet{AndersonJames09} conclude that, except for a central deficit, type II SNe seem to follow the $R$-band light, while SNIb/c appear more centrally concentrated with respect to the stellar light.)

Taking the SN distribution to be given by the fit to the $d$ distribution shown in Fig. \ref{fig:sn_mc-0}, the expected contamination in the TDE sample is 0.9 SNe, in the mean.   
The probability that TDE1 and TDE2 are ordinary SNe found by chance as close as observed to their hosts is the product of two factors.  First, the Poisson probability of finding two or more SNe, in the TDE candidate (nuclear) sample, when 0.9 are expected; this probability is 23\%.  Furthermore and independently, if TDE1,2 were ordinary SNe in the nuclear flare sample, the separations from the centers of their hosts would follow the distribution of stellar light within $d < 0\farcs 2$. The chance that two objects, drawn randomly from this distribution, have $d$ values less than those of the TDEs, is $\approx 7$\%.   Thus the probability that the TDEs are actually ordinary SNe in the sample of flares passing our selection criteria, yet are so perfectly centered on their hosts as they are, is $\approx 1.6$~\%.  This is the purely geometric suppression factor that must be applied to any hypothesized exotic type of flare whose progenitor follows the stellar light distribution.

\paragraph{Likelihood of SNe IIn}
We have seen above in Section \ref{sssec:IIn}, that individual cases can be found of SNe IIn which come close to matching particular properties of TDE1,2, but that when the ensemble of observations -- luminosity, late time UV emission, spectrum, color evolution --  are considered, TDE1,2 bear no resemblance to any SN observed to date.  Nevertheless, the properties of extreme SNe depend strongly upon their environment, so one might think that the TDEs could simply be the latest exotic SNe, whose properties differ from any SN seen earlier.  In such a scenario, TDE1,2 would be on the tail of the distribution of SNe IIn, which are themselves a small fraction of all SNe.  Only 17\% of SNe in a flux limited search such as ours are type II \citep{Li10}, of which about 29\% are type IIn and extreme examples are still more rare.  
The radial distribution of type II SNe is approximately the same as for all SNe, c.f., Fig. 8 in \citep{Leaman10}, so the 1.6\% geometrical penalty for the SNe occurring so close to their galaxy centers applies.  Thus the probability of finding two type IIn SNe hosts and such central locations as TDE1,2 is $<0.08~\%$, and is still lower for more exotic types of SNe for which TDE1,2 might be first examples.

We note that SN2006gy is a type IIn SN which would have passed our pipeline selection criteria had it been at the redshift of TDE1,2:  its proximity to the center of its galaxy would have placed it in the nuclear sample, being resolved only due to its low redshift and high-resolution imaging as at $\approx 350$ pc \citep{Smith07}, and its host would not have been excluded by the QSO locus.  SN2006gy is readily recognized as a SN and distinguished from TDE1,2 based on the color-evolution of the flare, as discussed in Section \ref{sssec:exotic} above, but it is a concrete reminder that geometrical background rejection alone is not sufficient to eliminate SNe in a sufficiently large sample of TDE candidates, and in practice flare properties need to be considered.  
 
\subsubsection{New SN type, only found near the centers of galaxies}\label{sssec:nucSNe}
Since the properties of TDE1,2 are unlike any known SN, one could entertain the idea that there may be some special type of previously-undetected SNe which occur exclusively as close or closer to the center of galaxies than TDE1,2.    
In that case the above statistical arguments would not apply.  To assess the viability of such an option, in this section we make the ansatz that TDE1,2 are examples of a new class of centrally concentrated SNe and see if the properties of the required class are reasonable.  The distance between TDE1,2 and the center of its host is $d=0\farcs 058,\, 0\farcs 068$ with a mean accuracy of $\approx 0\farcs06$; this corresponds to 0.14, 0.26 kpc in projection (but consistent with zero), respectively. 

Nuclear star clusters (NCs) are a rather mysterious phenomenon, observed in all Hubble-types \citep{bokerNCS10} and studied in considerable detail in the Milky Way \citep{genzel_GCreview_2010}.   The nuclear star clusters reported by \citet{Walcher05} in nine late-type spirals range from $ 8 \times 10^5 - 6 \times 10^7 \, M_\odot$ with a median $\approx 10^{6.5} \, M_\odot$; the mean age of the last star formation burst is 34 Myr with the youngest stellar population having a mean mass of $10^{5.5} \, M_\odot$\citep{Walcher06}.  Having a radius of $\approx$ 5 pc, NC's provide a concrete example of a possible scenario with concentrated sources.   

If every galaxy contains a nuclear star cluster, we can infer the minimum required rate of TDE-like explosions as follows.  The average stellar mass of the galaxies monitored in our search is  $\approx 10^{10}\, M_\odot$, so $\approx 3 \times 10^{-4}$ of all stars could be in nuclear star clusters.   The total number of SNe that would have been detected in the TDE search if we had not imposed a cut on the host-flare distance or additional quality cuts is $\approx 150$.\footnote{The total number of SNe that would be detected if the $d<1"$ requirement were not imposed can be estimated by integrating the fit to the $d$-distribution (Fig. \ref{fig:sn_mc-0}) to $d\to\infty$ and taking into account number of SN lost by requiring two detections after peak of the flare (in the $g$ and $r$ band only, to avoid a color bias) and restricting to host outside the QSO locus.} Thus if a new class of ``nuclear" SNe is to explain the existence of two SNe we require the average rate per star of TDE-like SNe in a nuclear star cluster to be a factor $\kappa$ larger than the rate of normal SNe per star, where
\begin{equation}\label{eq:kappa}
\kappa ~ 3 \times 10^{-4}  \approx 2 \,/  \,150 ~~,
\end{equation}
giving $\kappa \approx 45 $. 

This factor, $\kappa = 45$, is the enhancement factor by which stars in the nuclear star cluster must explode as a SN of the new type, compared to having a normal SN explosion, if the TDE candidates are to be explained as first cases of a new class of nuclear SNe.  Much more challenging theoretically, is to explain the {\em absence} of TDE-like SNe at larger radii, where none are observed.  With $\approx 3000$ times as many stars in an average galaxy as in its nuclear star cluster, we would expect to have seen $\approx 6000$ TDE-like events at larger radii if the rate per star were the same, requiring a suppression in the rate-per-star by a factor of at least $2.3/6000$ compared to the rate in the nuclear star cluster (2.3 being the 90\% CL upper limit on the number if 0 are seen).    

Although NC's contain on average younger, more massive stars than in the ensemble of monitored galaxies, a mechanism which produces a factor $\sim 60,000$ contrast between the rates of this new type of SNe in stars in a nuclear star cluster, compared to their rate elsewhere in the galaxy, may be difficult to devise. We note that the unusual SN 2006gy, which occurred at about 350~pc  from the center of its host \citep{Smith07}, could be an example of a ``galatic nucleus SN'', but a larger sample of similar SNe is needed to asses this possibility. 

\subsubsection{Summary of comparison of TDE1,2 to SNe}\label{sssec:SNsum}
We have compared TDE1 and TDE2 to ``ordinary" SNe, to an unbiased sample of type IIn SNe, and to particular exotic SNe.  Ordinary SNe have entirely different properties than the TDEs in their color, color evolution and decay rates.  Although some SNe IIn and exotic SNe resemble the TDEs in some respects, e.g., rate of change in magnitude and color, there are no examples of SNe whose properties are not significantly different than the TDEs in at least some important respects.  
Flares from known SNe types including IIn's follow the stellar distribution to a good approximation, with the result that the probability of two SNe IIn flares occurring so close to the nucleus as TDE1 and TDE2, is $\lesssim 0.08~\%$.  

Thus if TDE1,2 are SNe, they must be the first examples of a new exotic type of flare which occurs only very near the center of galaxies, to evade the $2 \times 10^{-2}$ geometrical penalty and explain the failure to detect many more events at larger radii.  Since known SNe are essentially local phenomenon -- depending only on the progenitor star and its immediate environment -- and there are so many more stars in the bulk of the galaxy than in any physical system near the galactic center, a model along these lines requires that either the progenitors of TDE-like SNe are concentrated by a large factor near the nucleus of the galaxy or some process in the nucleus stimulates TDE-like explosions at a much higher rate than in ordinary SNe.  We considered nuclear star clusters as a possible example, and found a minimum required enhancement factor of $\approx 45$.  Equally or more challenging, the rate of these new SNe must be at least a factor $\approx 3000$ {\em lower} for stars outside the nuclear star cluster than for stars in it.  

Given the evidence above, we conclude that TDE1 and TDE2 -- the two events which survived our pipeline cuts -- are unlikely to be supernovae.

\bigskip

 
\section{Discussion}\label{sec:discussion}
In this section we first compare the observed properties of TDE1,2 with properties of TDE candidates in the literature, and then compare with theoretical predictions.

\subsection{Comparison to TDE candidates in the literature}
\citet{Gezari09} presented two candidate TDEs discovered with \textsl{GALEX} that had simultaneous weekly optical difference imaging.  The optical colors and light curve shape of TDE1,2 are similar to the $g$, $r$, and $i$ and NUV light curves of those two \textsl{GALEX} TDE candidates. The black body temperatures of those \textsl{GALEX} candidates, measured by fitting to both UV and optical data, are $\sim 5\times 10^4\, {\rm K}$.  This is to be compared to $\sim 2\times 10^{4} \, {\rm K}$ for TDE1,2, obtained by fitting to optical data only.   Thus the observed optical properties of our two TDEs are similar to each other and to the \textsl{GALEX} TDE candidates  Given the small number of examples, the differences can be attributed either to variations within the TDE population or to the differences in the selection biases of the observations.  

Unfortunately, simultaneous optical imaging is not available for the X-ray selected TDE candidates, hence a comparison is possible only for derived properties. As discussed in \citet{Gezari09}, the black body temperatures derived from the X-ray spectra are higher than those obtained for the UV and optical-selected TDE candidates, which can be explained by the geometry of the emitting region (i.e., the high energy photons are produced closer to the black hole).

\subsection{Comparison of flare properties to theory}\label{sec:theory}
Predicting the properties of optical TDE flares is nontrivial, and the range of variation in observed flare properties for a given black hole mass is expected to be large due to the variety of possible pericentric distances, masses and radii of the disrupted star, black hole spin, and viewing angles.  Therefore, a much larger sample of TDEs will be needed -- selected as done here based on host and geometrical criteria rather than flare properties, with spectra and much more finely sampled light curves and simultaneous observations outside the optical band -- before the full complexity of the tidal disruption phenomenon can be even superficially explored.   

Here we give a preliminary comparison of the most basic properties of the flares -- summarized in Table \ref{tab:tde_prop} -- to the predictions of two theoretical models,  \citet[][LU97]{loebUlmer97} and \citet[][SQ09]{strubbe_quataert09}.   LU97 is a simple model based on thermal emission at the Eddington luminosity, while SQ09 follows the evolution of the tidal debris and makes detailed numerical predictions for emission during and after the super-Eddington period, in three illustrative examples.  

The most massive black hole in the SQ09 examples has $M_{\rm BH} = 10^7 \, M_\odot$ -- in the middle of the estimated  black hole mass range for the TDE1 host galaxy, but at the very low end of the range for TDE2.   For this example, SQ09 predict a peak value of $\nu L_\nu \approx {\rm few} \times 10^{41} {\rm erg/s}$ in the $g$-band.  This is 1 (2) orders of magnitude below the {\em observed, i.e., post-peak} maxima of TDE1 and TDE2, respectively.  Furthermore, SQ09 predicts a temperature which is considerably larger than found with a black body fit to the optical SED.  

The discrepancy in the SQ09 modeling is particularly severe for TDE2 because i) according to SQ09 Fig. 4, the peak luminosity decreases for larger $M_{\rm BH}$ and the central mass estimate for the TDE2 black hole is higher than in the SQ09 example, ii) we know from the CSS pre-SDSS observation that the flare is at least $\approx 40$ and possibly greater than 90 days old when seen by SDSS, depending on whether the peak occurs after or before the CSS observation, and iii) the observed $g$-band luminosity is a factor $\approx 100$ higher than predicted at the peak for this example.  (Interestingly, the peak optical luminosities and blackbody temperatures of the TDE candidates discovered by \textsl{GALEX} \citep[][Table 2]{Gezari09} are also inconsistent with SQ09 predictions.) 

Table \ref{tab:tde_prop} reveals that the color and time evolution of TDE1 and TDE2 are quite similar to each other.  The pipeline selection process did not place constraints on the color and, although too-short flares would not be accepted, the lifetimes of TDE1,2 are far longer than required to pass the selection criteria.   Thus it is reasonable to think that these properties are typical for optically discovered TDE flares.  Luminosities on the other hand are naturally biased toward the high end of the distribution.  Since the volume of detectability of a flare increases as $L^{3/2}$, we cannot exclude the existence of a population of dimmer flares.  

This raises the question of how to account for the observed flare properties. The earlier and simpler model of LU97, based on thermal emission at the Eddington luminosity, correctly predicts the observed temperature (within uncertainties) and its slow evolution, but predicts a much {\em higher} luminosity than observed.  LU97 also predicts that the luminosity decays much more slowly than observed.  The LU97 model can be reconciled with observation if their hypothesized optically thick envelope subtends only a fraction of the $4 \pi$ solid angle, and this fraction decreases with time.  Or perhaps SQ09 are on the right track but they adopted parameter choices which need modification, or better modeling of the radiatively driven wind is needed \citep{Lodato_Rossi10}.   Or some important aspect of the process may be missed completely; for instance the presence of a weak, pre-existing accretion disk might significantly enhance the power of the flare, as proposed by \citet{fg08} as an explanation for the correlation reported between ultra-high energy cosmic rays and AGNs \citep{Auger07,Auger08} whose luminosities are too low to accelerate protons to the observed energies \citep{fg08,Zaw09}.  

\emph{Note added:}  After the properties of TDE1,2 were presented in the preliminary archive version of this work \citep{vanVelzen10}, \citet{StrubbeQuataert11} explored a larger parameter space in their model and found that TDE1 can be readily explained adopting adjusted parameter assumptions.  They report that the luminosity of TDE2 can be explained in their framework as well, but agree with our observation that its color and slow decay are more suggestive of the state considered by LU97.

The luminosity and temperature of a tidal disruption flare, and their time evolution, depend on the (unknown) time between disruption and first observation, and on unknown or poorly-known parameters of the black hole ($M_{\rm BH}$ and spin) and the initial star and the orbit and viewing angle.  This means that the range of flare types is enormous and one might despair of being able to test theoretical models.  Remarkably, however, the {\em ratio} of the cooling rate to the bolometric luminosity decay, $(d \,{\rm ln\,T} / dt)/ (d \,{\rm ln\,} L\, / dt )$, is independent of all these unknown parameters and also independent of the time since disruption, in the SQ09 model during the super-Eddington phase.  Combining equations 2, 11, and 13 of SQ09 we derive that
\begin{equation} \label{eq:SQ}
(d \,{\rm ln\,T} / dt)/ (d \,{\rm ln\,} L\, / dt ) = -5/4
\end{equation}
for any TDE, at any time during the super-Eddington phase.   This is certainly not the evolution observed for TDE1,2 in the $g$-band, for which we measure  $(d \,{\rm ln\,T} / dt)/ (d \,{\rm ln\,} L_g\, / dt ) = 0 \pm 0.2$ and $0.2 - 0.7$ respectively.  Whether this is further evidence of a problem with the SQ09 model requires detailed modeling since the evolution of the $g$-band luminosity may not follow that of the bolometric luminosity, and a single black-body may not correctly describe the SED.  We present the relation Eq. \ref{eq:SQ} here because of its power to test the picture of the wind-driven super-Eddington phase independently of the initial conditions of the disruption event.  Simultaneous measurement of a larger portion of the SED, to allow Eq. (\ref{eq:SQ}) to be tested, would therefore be highly beneficial in future observational campaigns to explore the TDE phenomenon.

\bigskip

\section{Future Surveys}\label{sec:future}

\begin{deluxetable*}{l c c l  r}
  \tablewidth{400pt}
  \tablecolumns{5}
  \tablecaption{ Detection rates of other optional surveys.\label{tab:scale}}
  \tablehead{\colhead{Survey Name} & \colhead{Cadence} & \colhead{$F_{\rm lim}$} &  \colhead{$f_{\rm sky}$} & \colhead{$\dot{N}_{\rm obs}$} \\
    {} & {} & {(mag)} & {} & {(${\rm yr}^{-1}$)}}  
    \startdata
    CSS [1]                                                 & 14 days& 19.5  & 0.6 & 5 \\ 
    MLS [1]                                                 & 14 days& 21.5  & 0.09 & 12 \\ 
    QUEST [2]                                 & hours to years & 20.5 & 0.36 &12\\
    Palomar Transient Factory [3]               & 5 days& 21.0  & 0.2 & 13 \\ 
    Pan-STARRS Medium Deep Survey [4]  & 4 days & 24.8 & 0.0012& 15 \\ 
    Pan-STARRS $3\pi$\ Survey [4]        & 6 months& 23.5& 0.75&  1557 \\ 
    Large Synoptic Survey Telescope [5]   & 3 days   & 24.5 & 0.5& 4131
    \enddata
    \tablecomments{The survey plus reference used to obtain or estimate the cadence, flux limit ($F_{\rm lim}$) and fraction of the sky covered ($f_{\rm sky}$) are listed. We scale the detection rate using, Eq. \ref{eq:NdotObs} and $\dot{N}_{\rm obs}=1.9\,{\rm yr}^{-1}$ for the analysis presented here. We have used 300 ${\rm deg^2}$ as the angular area for Stripe 82. Since the cadence of the observations of Stripe 82 decreases towards the edges, \citet{sesar07} have used 290 ${\rm deg^2}$ for this area. However the total area of Stripe 82 that is imaged is 312 ${\rm deg^2}$; we thus adopted 300 ${\rm deg^2}$ as a reasonable value to obtain $f_{\rm sky}$.}
    \tablerefs{[1] \citet{DrakeCRTS09}. [2] \citet{Hadjiyska11}.  [3] \citet{Law09}. [4] \citet{Gezari08}. [5] \citet{Ivezic08}}.
\end{deluxetable*}

We estimate below the \emph{detection} rate of TDEs which can be expected in current and future optical transient surveys, for the pipeline used here and similar observational conditions and cadence as SDSS Stripe 82.  This estimate differs from earlier estimates such as \citep{Gezari09,strubbe_quataert09} because those ignore the cost in event-rate implied by cuts needed to insure clean and unambiguous detections, and our estimate here incorporates the observed properties of TDEs rather than relying on models.  

The total effective time spanned by the SDSS Stripe 82 data, $\tau_{\rm obs}$, is just the total time between observations within a season;  the $>9$ month gap between seasons is not included in $\tau_{\rm obs}$. To account for the difference in sampling across Stripe 82 we calculate $\tau_{\rm obs}$ in bins of width 3.6 degree along right ascension. The mean $\tau_{\rm obs}$ is $1.03\,{\rm yr}$ with a mean cadence of 7.5 days.  Thus two detected TDEs corresponds to a TDE detection rate of $\dot{N}_{\rm obs}=1.9 \, {\rm yr}^{-1}$.  We can scale this detection rate to current and future optical surveys of similar cadence and selection criteria using the flux limit, $F_{\rm lim}$, and faction of the sky observed, $f_{\rm sky}$: 
\begin{equation}\label{eq:NdotObs}
  \dot{N}_{\rm obs} \propto f_{\rm sky} \, F_{\rm lim}^{-3/2} \quad.  
\end{equation}
This yields $\dot{N}_{\rm obs} =  13,\:14,\:4180\,\, {\rm yr}^{-1}$ for the PTF,  Pan-STARRS Medium Deep Survey and LSST, respectively. In Table \ref{tab:scale} we list the adopted values of $F_{\rm lim}$, $f_{\rm sky}$.  These detection rates are lower bounds on the actual number of TDEs which can be observed (if the observational quality is equal to that of SDSS and these TDEs have typical luminosities), because the cadence can be optimized and the pipeline made more efficient, to maximize detections for any targeted light-curve type in a dedicated survey.  

Future optical surveys will be predominantly photometric and will generally not have a large fraction of hosts for which spectra have been obtained, as we have for SDSS Stripe 82.  In fact, this need not prevent obtaining a TDE candidate sample for followup with O(1) false positives, if the angular resolution allows adequate rejection of non-nuclear flares.  The first line of defense against SNe contamination is good resolution.  The purity of the nuclear sample is determined by the accuracy with which the flare-host separation can be measured, because SN background increases very rapidly as the resolution is compromised.  

When the goal is rapid, intensive spectral and multi-wavelength follow-up rather than discovery in archival data, the appropriate selection strategy changes from the one used for here for TDE1,2.  The first priority is to be confident that a flare passing the selection criteria has a very high chance of being a TDE and low chance of being uninteresting.  Elements of such a strategy are:\\
$\bullet$ QSOs can be suppressed in the target sample by excluding galaxies within the QSO locus in Fig. \ref{fig:gal_qso_locus};  galaxies showing continuing irregular variability during monitoring can be excluded as presumptive AGNs without spectroscopic follow-up, if spectroscopic resources are limited. \\
$\bullet$ TDE1,2 both fall in a``TDE-locus" based on photometric properties of the flares alone (Fig. \ref{fig:cc}), allowing powerful rejection of AGN flares and SNe without spectroscopic followup.  Fig. \ref{fig:g_decay-gr_decay} shows that flares from the hosts which are identified as QSOs based on their photometric properties only, have properties similar to the flares from spectroscopically confirmed QSOs.  Moreover requiring a flare to fall in a ``TDE-locus" of very blue flares -- $u-g \leq -0.1$ and $g-r \leq -0.2$ (Fig. \ref{fig:cc}) -- reduces the contamination of SNe in a TDE search by a factor of $\gtrsim 50$.  Thus photometry alone, without prior spectroscopy, is sufficient to reduce the variable-AGN contamination of the TDE candidate sample, and rejects almost all SNe. \\
$\bullet$  The only SNe which are observed in the TDE-locus in flare color are of type IIn, but these occur only very rarely in E/S0 galaxies \citep{Li10}. Thus without introducing a bias in the selection of TDEs, targeting can be restricted to early-type galaxies.  This reduces the contamination of SNe IIn, the most troublesome SN contaminant in a TDE search, based on color alone.  The appropriate trade-off between stringency of rejection of SNIIn and loss of real TDEs will be determined by the specifics of the survey and follow-up resources.

\bigskip

\section{Summary}\label{sec:summary}
We have presented here the first two compelling, optically-discovered, stellar tidal disruption candidates.  This work demonstrates the feasibility of discovering TDEs in optical synoptic surveys without imposing selection criteria depending on the properties of the flare.

Our pipeline rejection of non-TDEs is based on geometrical criteria to eliminate supernovae and on host properties to eliminate variable AGN flares, rather than being based on properties of the flares themselves, in order to minimize selection bias.  The pipeline rejection gives an {\em a priori} probability of the flares being SNe or variable AGNs of less than 3\%.  
 Thanks to the very large sample of galaxies in SDSS Stripe 82, with a large number of them having spectra, we have excellent data on the properties of host galaxies and their variability.  This allows us to remove 90\% of the QSO hosts by a color-color cut on the host galaxy.  The most serious remaining background comes from variable emission from hidden active nuclei.  These cases are excluded by the variability of their hosts in the non-peak seasons, eliminating 3 of the 5 TDE candidates which survived the SNe geometrical and QSO color-locus cuts.  Based on the variability observed in QSOs and identified Seyferts, we estimate the probability that a QSO or Seyfert which satisfies our flare selection criteria in some season, shows as little variability in the other seasons as TDE1,2, to be $< 2 \times 10^{-5}$.  

Further SDSS observations, follow-up spectra and \textsl{GALEX} observations provide powerful {\em a posteriori} evidence that TDE1,2 are neither SNe or variable AGNs.  The host spectra are consistent with the hosts not having active nuclei. Although no requirement was placed on the flare properties in the selection process, the TDE flares are very distinctive in comparison to SNe and flares in variable QSOs and AGNs.  The lower limit on their increase in flux compared to a possible AGN contribution to the baseline flux is far greater in the TDEs than any QSO flare in the Stripe 82 data (Fig. \ref{fig:deltaf}).  The mean color and cooling rate, as well as the decay rate, of the TDE flares are significantly different from any SN in our sample, as shown in Figs. \ref{fig:cc} and \ref{fig:qso-color-change}. The TDE flares are significantly bluer than QSOs and QSO flares, although like QSO flares their color evolves very slowly compared to SNe.  In particular, \textsl{GALEX}  recorded a level of UV emission from TDE2 $\approx$ 800 days after the flare that is orders of magnitude greater than in any known SN.
\textsl{GALEX} observations of TDE2 combined with color information, show that TDE2's flare is unlike every known SN flare in at least one respect.  The closest resemblance is to type IIn SNe and a few exotic SNe.  

Serendipitously, a spectrum of TDE2 was taken during the flare, a few days after the first SDSS detection of the flaring state.  We have recently taken a spectrum of the host galaxy, giving us by subtraction a spectrum of the flare itself. We compared of this spectrum to spectra from an unbiased sample of type IIn SNe, finding that these SNe spectra show stronger H$\alpha$ emission. 

The properties of the flares argue against the possibility that TDE1,2 are an unusual but known type of SN.  The most TDE1,2-like SNe are type IIn's, but the probability of finding two of those, and in such central locations as TDE1,2, is $\lesssim 0.08~\%$, and is still lower for more exotic types of SNe.  A final option, that these are the first examples of a new class of SNe occurring only at the centers of galaxies is shown to require a thousand-fold or greater enhancement in the rate of such events in stars near the nucleus compared to the rate in stars located elsewhere.

With only two examples of probable TDEs to study, it is still very much early-days for testing models in the literature.  Nevertheless, the events have already enabled refinement of the recent detailed modeling of the process by \citet[][SQ09]{strubbe_quataert09}:  the observed luminosities of TDE1,2 are at least 1-2 orders of magnitude higher than predicted there, and the temperatures determined by a black body fit to the optical SED are considerably lower than in the simulations, for the assumed parameter choices.  However with adjusted parameter choices the features of TDE1 can be explained and the luminosity of TDE2 as well \citep{StrubbeQuataert11}.  We show that, independently of initial conditions and the time since disruption, the SQ09 model in its bright, super-Eddington phase predicts that the temperature of a TDE flare increases while its bolometric luminosity decreases, with a specific relationship between the rates which is independent of initial conditions or time since flare.  This cannot be tested directly for these TDEs because most of the luminosity is predicted to be emitted above optical frequencies, but when such measurements become available they will provide a decisive test of the SQ09 model.  The earlier and simpler model of \citet{loebUlmer97}, based on thermal emission at the Eddington luminosity, can be reconciled with observation if their hypothesized optically thick envelope subtends only a fraction of the $4 \pi$ solid angle, and this fraction decreases with time.  

Our work demonstrates that candidate tidal disruption events can be efficiently identified using photometric imaging alone.  We conclude based on our observed TDE rate and pipeline efficiency, that current and next-generation optical synoptic surveys should contain hundreds or thousands of TDEs.  We have shown that a TDE candidate sample with {\emph O(1)} purity can be identified using host selection, geometric resolution and color alone.  With such a sample, the cost of excluding imposters with a follow-up observation the next night or later that night on another instrument is sufficiently low, that a campaign to create a large sample of tidal disruption events with high-frequency, multi-wavelength observations is feasible.  This will allow the tidal disruption phenomenon to be explored in full detail, opening an exciting new chapter in black hole astrophysics.
\bigskip

\acknowledgments
GRF and SvV acknowledge valuable conversations with M. Blanton, A. Filippenko, J. Frieman, J. Greene, A. Gruzinov, D. Hogg, S. Komossa, M. Modjaz, J. Moustakas, N. Smith, L. Strubbe, N. Zakamska and G. Zhu;  in addition we thank P. Massey for assistance with obtaining the TDE1 spectrum,  A. Goobar and R. Nichol for cooperation in use of the TDE2 spectrum, A. Gal-Yam and O. Yaron for cooperation in use of CCCP data, P. Groot and K. Verbeek for help with the WHT spectrum, and G. Djorgovski and E. Beshore for cooperation in use of CRTS and CSS data. We are grateful to the anonymous referee for the comments that improved the manuscript and made the presentation more balanced The research of SvV was supported in part by the Huygens Scholarship Programme. The research of GRF was supported in part by NSF-PHY-0701451. 

Funding for the SDSS and SDSS-II has been provided by the Alfred P. Sloan Foundation, the Participating Institutions, the National Science Foundation, the U.S. Department of Energy, the National Aeronautics and Space Administration, the Japanese Monbukagakusho, the Max Planck Society, and the Higher Education Funding Council for England. The SDSS Web Site is http://www.sdss.org/.  The SDSS is managed by the Astrophysical Research Consortium for the Participating Institutions. The Participating Institutions are the American Museum of Natural History, Astrophysical Institute Potsdam, University of Basel, University of Cambridge, Case Western Reserve University, University of Chicago, Drexel University, Fermilab, the Institute for Advanced Study, the Japan Participation Group, Johns Hopkins University, the Joint Institute for Nuclear Astrophysics, the Kavli Institute for Particle Astrophysics and Cosmology, the Korean Scientist Group, the Chinese Academy of Sciences (LAMOST), Los Alamos National Laboratory, the Max-Planck-Institute for Astronomy (MPIA), the Max-Planck-Institute for Astrophysics (MPA), New Mexico State University, Ohio State University, University of Pittsburgh, University of Portsmouth, Princeton University, the United States Naval Observatory, and the University of Washington. 

This paper includes data gathered with the 6.5 meter Magellan Telescopes located at Las Campanas Observatory, Chile, with the ESO New Technology Telescope at the La Silla observatory, and with the William Herschel Telescope, operated on the island of La Palma by the Isaac Newton Group in the Spanish Observatorio del Roque de los Muchachos of the Inst\'{\i}tuto de Astrof\'{\i}sica de Canarias. The CSS survey is funded by the National Aeronautics and Space Administration under Grant No. NNG05GF22G issued through the Science Mission Directorate Near-Earth Objects Observations Program. J.L.P. is supported by NSF grant AST-0707982. The CRTS survey is supported by the U.S. National Science Foundation under grants AST-0909182.  We acknowledge NASA's support for construction, operation, and science analysis for the GALEX mission, developed in cooperation with Centre National d'Etudes Spatiales of France and the Korean Ministry of Science and Technology. The National Radio Astronomy Observatory is a facility of the National Science Foundation operated under cooperative agreement by Associated Universities, Inc.  
%

\bibliography{flares}

\begin{thebibliography}{118}
\expandafter\ifx\csname natexlab\endcsname\relax\def\natexlab#1{#1}\fi

\bibitem[{{Abazajian} {et~al.}(2009){Abazajian}, {Adelman-McCarthy},
  {Ag{\"u}eros}, {Allam}, {Allende Prieto}, {An}, {Anderson}, {Anderson},
  {Annis}, {Bahcall}, {Bailer-Jones}, {Barentine}, {Bassett}, {Becker},
  {Beers}, {Bell}, {Belokurov}, {Berlind}, {Berman}, {Bernardi}, {Bickerton},
  {Bizyaev}, {Blakeslee}, {Blanton}, {Bochanski}, {Boroski}, {Brewington},
  {Brinchmann}, {Brinkmann}, {Brunner}, {Budav{\'a}ri}, {Carey}, {Carliles},
  {Carr}, {Castander}, {Cinabro}, {Connolly}, {Csabai}, {Cunha}, {Czarapata},
  {Davenport}, {de Haas}, {Dilday}, {Doi}, {Eisenstein}, {Evans}, {Evans},
  {Fan}, {Friedman}, {Frieman}, {Fukugita}, {G{\"a}nsicke}, {Gates},
  {Gillespie}, {Gilmore}, {Gonzalez}, {Gonzalez}, {Grebel}, {Gunn},
  {Gy{\"o}ry}, {Hall}, {Harding}, {Harris}, {Harvanek}, {Hawley}, {Hayes},
  {Heckman}, {Hendry}, {Hennessy}, {Hindsley}, {Hoblitt}, {Hogan}, {Hogg},
  {Holtzman}, {Hyde}, {Ichikawa}, {Ichikawa}, {Im}, {Ivezi{\'c}}, {Jester},
  {Jiang}, {Johnson}, {Jorgensen}, {Juri{\'c}}, {Kent}, {Kessler}, {Kleinman},
  {Knapp}, {Konishi}, {Kron}, {Krzesinski}, {Kuropatkin}, {Lampeitl},
  {Lebedeva}, {Lee}, {Lee}, {Leger}, {L{\'e}pine}, {Li}, {Lima}, {Lin}, {Long},
  {Loomis}, {Loveday}, {Lupton}, {Magnier}, {Malanushenko}, {Malanushenko},
  {Mandelbaum}, {Margon}, {Marriner}, {Mart{\'{\i}}nez-Delgado}, {Matsubara},
  {McGehee}, {McKay}, {Meiksin}, {Morrison}, {Mullally}, {Munn}, {Murphy},
  {Nash}, {Nebot}, {Neilsen}, {Newberg}, {Newman}, {Nichol}, {Nicinski},
  {Nieto-Santisteban}, {Nitta}, {Okamura}, {Oravetz}, {Ostriker}, {Owen},
  {Padmanabhan}, {Pan}, {Park}, {Pauls}, {Peoples}, {Percival}, {Pier}, {Pope},
  {Pourbaix}, {Price}, {Purger}, {Quinn}, {Raddick}, {Fiorentin}, {Richards},
  {Richmond}, {Riess}, {Rix}, {Rockosi}, {Sako}, {Schlegel}, {Schneider},
  {Scholz}, {Schreiber}, {Schwope}, {Seljak}, {Sesar}, {Sheldon}, {Shimasaku},
  {Sibley}, {Simmons}, {Sivarani}, {Smith}, {Smith}, {Smol{\v c}i{\'c}},
  {Snedden}, {Stebbins}, {Steinmetz}, {Stoughton}, {Strauss}, {Subba Rao},
  {Suto}, {Szalay}, {Szapudi}, {Szkody}, {Tanaka}, {Tegmark}, {Teodoro},
  {Thakar}, {Tremonti}, {Tucker}, {Uomoto}, {Vanden Berk}, {Vandenberg},
  {Vidrih}, {Vogeley}, {Voges}, {Vogt}, {Wadadekar}, {Watters}, {Weinberg},
  {West}, {White}, {Wilhite}, {Wonders}, {Yanny}, {Yocum}, {York}, {Zehavi},
  {Zibetti}, \& {Zucker}}]{Abazajian09}
{Abazajian}, K.~N., {et~al.} 2009, \apjs, 182, 543

\bibitem[{{Agnoletto} {et~al.}(2009){Agnoletto}, {Benetti}, {Cappellaro},
  {Zampieri}, {Turatto}, {Mazzali}, {Pastorello}, {Della Valle}, {Bufano},
  {Harutyunyan}, {Navasardyan}, {Elias-Rosa}, {Taubenberger}, {Spiro}, \&
  {Valenti}}]{Agnoletto09}
{Agnoletto}, I., {et~al.} 2009, \apj, 691, 1348

\bibitem[{{Aller} \& {Richstone}(2002)}]{Aller_Richstone02}
{Aller}, M.~C., \& {Richstone}, D. 2002, \aj, 124, 3035

\bibitem[{{Anderson} \& {James}(2008)}]{AndersonJames08}
{Anderson}, J.~P., \& {James}, P.~A. 2008, \mnras, 390, 1527

\bibitem[{{Anderson} \& {James}(2009)}]{AndersonJames09}
---. 2009, \mnras, 399, 559

\bibitem[{{Baars} {et~al.}(1977){Baars}, {Genzel}, {Pauliny-Toth}, \&
  {Witzel}}]{Baars77}
{Baars}, J.~W.~M., {Genzel}, R., {Pauliny-Toth}, I.~I.~K., \& {Witzel}, A.
  1977, \aap, 61, 99

\bibitem[{{Bade} {et~al.}(1996){Bade}, {Komossa}, \& {Dahlem}}]{Bade96}
{Bade}, N., {Komossa}, S., \& {Dahlem}, M. 1996, \aap, 309, L35

\bibitem[{{Baldwin} {et~al.}(1981){Baldwin}, {Phillips}, \&
  {Terlevich}}]{baldwin81}
{Baldwin}, J.~A., {Phillips}, M.~M., \& {Terlevich}, R. 1981, \pasp, 93, 5

\bibitem[{{Barbary} {et~al.}(2009){Barbary}, {Dawson}, {Tokita}, {Aldering},
  {Amanullah}, {Connolly}, {Doi}, {Faccioli}, {Fadeyev}, {Fruchter},
  {Goldhaber}, {Goobar}, {Gude}, {Huang}, {Ihara}, {Konishi}, {Kowalski},
  {Lidman}, {Meyers}, {Morokuma}, {Nugent}, {Perlmutter}, {Rubin}, {Schlegel},
  {Spadafora}, {Suzuki}, {Swift}, {Takanashi}, {Thomas}, \&
  {Yasuda}}]{Barbary09}
{Barbary}, K., {et~al.} 2009, \apj, 690, 1358

\bibitem[{{Bauer} {et~al.}(2009){Bauer}, {Baltay}, {Coppi}, {Ellman}, {Jerke},
  {Rabinowitz}, \& {Scalzo}}]{Bauer09}
{Bauer}, A., {Baltay}, C., {Coppi}, P., {Ellman}, N., {Jerke}, J.,
  {Rabinowitz}, D., \& {Scalzo}, R. 2009, \apj, 696, 1241

\bibitem[{{Becker} {et~al.}(1995){Becker}, {White}, \& {Helfand}}]{becker95}
{Becker}, R.~H., {White}, R.~L., \& {Helfand}, D.~J. 1995, \apj, 450, 559

\bibitem[{{Bianchi} {et~al.}(2007){Bianchi}, {Rodriguez-Merino}, {Viton},
  {Laget}, {Efremova}, {Herald}, {Conti}, {Shiao}, {Gil de Paz}, {Salim},
  {Thakar}, {Friedman}, {Rey}, {Thilker}, {Barlow}, {Budav{\'a}ri}, {Donas},
  {Forster}, {Heckman}, {Lee}, {Madore}, {Martin}, {Milliard}, {Morrissey},
  {Neff}, {Rich}, {Schiminovich}, {Seibert}, {Small}, {Szalay}, {Wyder},
  {Welsh}, \& {Yi}}]{bianchi07}
{Bianchi}, L., {et~al.} 2007, \apjs, 173, 659

\bibitem[{{Blanton} \& {Roweis}(2007)}]{blanton07}
{Blanton}, M.~R., \& {Roweis}, S. 2007, \aj, 133, 734

\bibitem[{{Blanton} {et~al.}(2001){Blanton}, {Dalcanton}, {Eisenstein},
  {Loveday}, {Strauss}, {SubbaRao}, {Weinberg}, {Anderson}, {Annis}, {Bahcall},
  {Bernardi}, {Brinkmann}, {Brunner}, {Burles}, {Carey}, {Castander},
  {Connolly}, {Csabai}, {Doi}, {Finkbeiner}, {Friedman}, {Frieman}, {Fukugita},
  {Gunn}, {Hennessy}, {Hindsley}, {Hogg}, {Ichikawa}, {Ivezi{\'c}}, {Kent},
  {Knapp}, {Lamb}, {Leger}, {Long}, {Lupton}, {McKay}, {Meiksin}, {Merelli},
  {Munn}, {Narayanan}, {Newcomb}, {Nichol}, {Okamura}, {Owen}, {Pier}, {Pope},
  {Postman}, {Quinn}, {Rockosi}, {Schlegel}, {Schneider}, {Shimasaku},
  {Siegmund}, {Smee}, {Snir}, {Stoughton}, {Stubbs}, {Szalay}, {Szokoly},
  {Thakar}, {Tremonti}, {Tucker}, {Uomoto}, {Vanden Berk}, {Vogeley},
  {Waddell}, {Yanny}, {Yasuda}, \& {York}}]{blanton01}
{Blanton}, M.~R., {et~al.} 2001, \aj, 121, 2358

\bibitem[{{Blondin} \& {Tonry}(2007)}]{Blondin07}
{Blondin}, S., \& {Tonry}, J.~L. 2007, \apj, 666, 1024

\bibitem[{{Bogdanovi{\'c}} {et~al.}(2004){Bogdanovi{\'c}}, {Eracleous},
  {Mahadevan}, {Sigurdsson}, \& {Laguna}}]{bogdanovic04}
{Bogdanovi{\'c}}, T., {Eracleous}, M., {Mahadevan}, S., {Sigurdsson}, S., \&
  {Laguna}, P. 2004, \apj, 610, 707

\bibitem[{B{\"o}ker(2010)}]{bokerNCS10}
B{\"o}ker, T. 2010, in The Impact of HST on European Astronomy, ed. W.~Burton,
  L.~L. Christensen, \& F.~D. Macchetto, Astrophysics and Space Science
  Proceedings (Springer Netherlands), 99--104

\bibitem[{{Bramich} {et~al.}(2008){Bramich}, {Vidrih}, {Wyrzykowski}, {Munn},
  {Lin}, {Evans}, {Smith}, {Belokurov}, {Gilmore}, {Zucker}, {Hewett},
  {Watkins}, {Faria}, {Fellhauer}, {Miknaitis}, {Bizyaev}, {Ivezi{\'c}},
  {Schneider}, {Snedden}, {Malanushenko}, {Malanushenko}, \& {Pan}}]{bramich08}
{Bramich}, D.~M., {et~al.} 2008, \mnras, 386, 887

\bibitem[{{Brown} {et~al.}(2009){Brown}, {Holland}, {Immler}, {Milne},
  {Roming}, {Gehrels}, {Nousek}, {Panagia}, {Still}, \& {Vanden
  Berk}}]{Brown09}
{Brown}, P.~J., {et~al.} 2009, \aj, 137, 4517

\bibitem[{{Bruzual} \& {Charlot}(2003)}]{Bruzual_Charlot03}
{Bruzual}, G., \& {Charlot}, S. 2003, \mnras, 344, 1000

\bibitem[{{Cappelluti} {et~al.}(2009){Cappelluti}, {Ajello}, {Rebusco},
  {Komossa}, {Bongiorno}, {Clemens}, {Salvato}, {Esquej}, {Aldcroft},
  {Greiner}, \& {Quintana}}]{Cappelluti09}
{Cappelluti}, N., {et~al.} 2009, \aap, 495, L9

\bibitem[{{Cardelli} {et~al.}(1989){Cardelli}, {Clayton}, \&
  {Mathis}}]{cardelli89}
{Cardelli}, J.~A., {Clayton}, G.~C., \& {Mathis}, J.~S. 1989, \apj, 345, 245

\bibitem[{{Chambers}(2007)}]{Chambers07}
{Chambers}, K.~C. 2007, in Bulletin of the American Astronomical Society,
  Vol.~38, Bulletin of the American Astronomical Society, 995

\bibitem[{{D'Andrea} {et~al.}(2010){D'Andrea}, {Sako}, {Dilday}, {Frieman},
  {Holtzman}, {Kessler}, {Konishi}, {Schneider}, {Sollerman}, {Wheeler},
  {Yasuda}, {Cinabro}, {Jha}, {Nichol}, {Lampeitl}, {Smith}, {Atlee},
  {Bassett}, {Castander}, {Goobar}, {Miquel}, {Nordin}, {{\"O}stman}, {Prieto},
  {Quimby}, {Riess}, \& {Stritzinger}}]{D'Andrea10}
{D'Andrea}, C.~B., {et~al.} 2010, \apj, 708, 661

\bibitem[{{Dekker} {et~al.}(1986){Dekker}, {Delabre}, \& {D'Odorico}}]{SPIE86}
{Dekker}, H., {Delabre}, B., \& {D'Odorico}, S. 1986, in Society of
  Photo-Optical Instrumentation Engineers (SPIE) Conference Series, Vol. 627,
  Society of Photo-Optical Instrumentation Engineers (SPIE) Conference Series,
  ed. D.~L. {Crawford}, 339--348

\bibitem[{{Donley} {et~al.}(2002){Donley}, {Brandt}, {Eracleous}, \&
  {Boller}}]{donley02}
{Donley}, J.~L., {Brandt}, W.~N., {Eracleous}, M., \& {Boller}, T. 2002, \aj,
  124, 1308

\bibitem[{{Drake} {et~al.}(2009){Drake}, {Djorgovski}, {Mahabal}, {Beshore},
  {Larson}, {Graham}, {Williams}, {Christensen}, {Catelan}, {Boattini},
  {Gibbs}, {Hill}, \& {Kowalski}}]{DrakeCRTS09}
{Drake}, A.~J., {et~al.} 2009, \apj, 696, 870

\bibitem[{{Esquej} {et~al.}(2008){Esquej}, {Saxton}, {Komossa}, {Read},
  {Freyberg}, {Hasinger}, {Garc{\'{\i}}a-Hern{\'a}ndez}, {Lu}, {Zaur{\'{\i}}n},
  {S{\'a}nchez-Portal}, \& {Zhou}}]{Esquej08}
{Esquej}, P., {et~al.} 2008, \aap, 489, 543

\bibitem[{{Farrar} \& {Gruzinov}(2009)}]{fg08}
{Farrar}, G.~R., \& {Gruzinov}, A. 2009, \apj, 693, 329

\bibitem[{{Ferrarese} \& {Merritt}(2000)}]{Ferrarese_Merritt00}
{Ferrarese}, L., \& {Merritt}, D. 2000, \apjl, 539, L9

\bibitem[{{Frank} \& {Rees}(1976)}]{Frank_Rees76}
{Frank}, J., \& {Rees}, M.~J. 1976, \mnras, 176, 633

\bibitem[{{Frieman} {et~al.}(2008){Frieman}, {Bassett}, {Becker}, {Choi},
  {Cinabro}, {DeJongh}, {Depoy}, {Dilday}, {Doi}, {Garnavich}, {Hogan},
  {Holtzman}, {Im}, {Jha}, {Kessler}, {Konishi}, {Lampeitl}, {Marriner},
  {Marshall}, {McGinnis}, {Miknaitis}, {Nichol}, {Prieto}, {Riess}, {Richmond},
  {Romani}, {Sako}, {Schneider}, {Smith}, {Takanashi}, {Tokita}, {van der
  Heyden}, {Yasuda}, {Zheng}, {Adelman-McCarthy}, {Annis}, {Assef},
  {Barentine}, {Bender}, {Blandford}, {Boroski}, {Bremer}, {Brewington},
  {Collins}, {Crotts}, {Dembicky}, {Eastman}, {Edge}, {Edmondson}, {Elson},
  {Eyler}, {Filippenko}, {Foley}, {Frank}, {Goobar}, {Gueth}, {Gunn},
  {Harvanek}, {Hopp}, {Ihara}, {Ivezi{\'c}}, {Kahn}, {Kaplan}, {Kent},
  {Ketzeback}, {Kleinman}, {Kollatschny}, {Kron}, {Krzesi{\'n}ski}, {Lamenti},
  {Leloudas}, {Lin}, {Long}, {Lucey}, {Lupton}, {Malanushenko}, {Malanushenko},
  {McMillan}, {Mendez}, {Morgan}, {Morokuma}, {Nitta}, {Ostman}, {Pan},
  {Rockosi}, {Romer}, {Ruiz-Lapuente}, {Saurage}, {Schlesinger}, {Snedden},
  {Sollerman}, {Stoughton}, {Stritzinger}, {Subba Rao}, {Tucker}, {Vaisanen},
  {Watson}, {Watters}, {Wheeler}, {Yanny}, \& {York}}]{frieman08}
{Frieman}, J.~A., {et~al.} 2008, \aj, 135, 338

\bibitem[{{Fruchter} {et~al.}(2006){Fruchter}, {Levan}, {Strolger},
  {Vreeswijk}, {Thorsett}, {Bersier}, {Burud}, {Castro Cer{\'o}n},
  {Castro-Tirado}, {Conselice}, {Dahlen}, {Ferguson}, {Fynbo}, {Garnavich},
  {Gibbons}, {Gorosabel}, {Gull}, {Hjorth}, {Holland}, {Kouveliotou}, {Levay},
  {Livio}, {Metzger}, {Nugent}, {Petro}, {Pian}, {Rhoads}, {Riess}, {Sahu},
  {Smette}, {Tanvir}, {Wijers}, \& {Woosley}}]{Fruchter06}
{Fruchter}, A.~S., {et~al.} 2006, \nat, 441, 463

\bibitem[{{Fukugita} {et~al.}(1996){Fukugita}, {Ichikawa}, {Gunn}, {Doi},
  {Shimasaku}, \& {Schneider}}]{fukugita96}
{Fukugita}, M., {Ichikawa}, T., {Gunn}, J.~E., {Doi}, M., {Shimasaku}, K., \&
  {Schneider}, D.~P. 1996, \aj, 111, 1748

\bibitem[{{Gadotti}(2009)}]{Gadotti09}
{Gadotti}, D.~A. 2009, \mnras, 393, 1531

\bibitem[{{Gal-Yam} {et~al.}(2007){Gal-Yam}, {Cenko}, {Fox}, {Leonard}, {Moon},
  {Sand}, \& {Soderberg}}]{Gal-Yam07}
{Gal-Yam}, A., {Cenko}, S.~B., {Fox}, D.~B., {Leonard}, D.~C., {Moon}, D.,
  {Sand}, D.~J., \& {Soderberg}, A.~M. 2007, in American Institute of Physics
  Conference Series, Vol. 924, The Multicolored Landscape of Compact Objects
  and Their Explosive Origins, ed. {T.~di Salvo, G.~L.~Israel, L.~Piersant,
  L.~Burderi, G.~Matt, A.~Tornambe, \& M.~T.~Menna}, 297--303

\bibitem[{{Gal-Yam} {et~al.}(2002){Gal-Yam}, {Ofek}, {Filippenko}, {Chornock},
  \& {Li}}]{Gal-Yam02}
{Gal-Yam}, A., {Ofek}, E.~O., {Filippenko}, A.~V., {Chornock}, R., \& {Li}, W.
  2002, \pasp, 114, 587

\bibitem[{{Gebhardt} {et~al.}(2000){Gebhardt}, {Bender}, {Bower}, {Dressler},
  {Faber}, {Filippenko}, {Green}, {Grillmair}, {Ho}, {Kormendy}, {Lauer},
  {Magorrian}, {Pinkney}, {Richstone}, \& {Tremaine}}]{Gebhardt00}
{Gebhardt}, K., {et~al.} 2000, \apjl, 539, L13

\bibitem[{{Genzel} {et~al.}(2010){Genzel}, {Eisenhauer}, \&
  {Gillessen}}]{genzel_GCreview_2010}
{Genzel}, R., {Eisenhauer}, F., \& {Gillessen}, S. 2010, Reviews of Modern
  Physics, 82, 3121

\bibitem[{{Gezari} {et~al.}(2006){Gezari}, {Martin}, {Milliard}, {Basa},
  {Halpern}, {Forster}, {Friedman}, {Morrissey}, {Neff}, {Schiminovich},
  {Seibert}, {Small}, \& {Wyder}}]{Gezari06}
{Gezari}, S., {et~al.} 2006, \apjl, 653, L25

\bibitem[{{Gezari} {et~al.}(2008){Gezari}, {Basa}, {Martin}, {Bazin},
  {Forster}, {Milliard}, {Halpern}, {Friedman}, {Morrissey}, {Neff},
  {Schiminovich}, {Seibert}, {Small}, \& {Wyder}}]{Gezari08}
---. 2008, \apj, 676, 944

\bibitem[{{Gezari} {et~al.}(2009{\natexlab{a}}){Gezari}, {Halpern}, {Grupe},
  {Yuan}, {Quimby}, {McKay}, {Chamarro}, {Sisson}, {Akerlof}, {Wheeler},
  {Brown}, {Cenko}, {Rau}, {Djordjevic}, \& {Terndrup}}]{Gezari09b}
---. 2009{\natexlab{a}}, \apj, 690, 1313

\bibitem[{{Gezari} {et~al.}(2009{\natexlab{b}}){Gezari}, {Heckman}, {Cenko},
  {Eracleous}, {Forster}, {Gon{\c c}alves}, {Martin}, {Morrissey}, {Neff},
  {Seibert}, {Schiminovich}, \& {Wyder}}]{Gezari09}
---. 2009{\natexlab{b}}, \apj, 698, 1367

\bibitem[{{Graham} {et~al.}(2001){Graham}, {Erwin}, {Caon}, \&
  {Trujillo}}]{Graham01}
{Graham}, A.~W., {Erwin}, P., {Caon}, N., \& {Trujillo}, I. 2001, \apjl, 563,
  L11

\bibitem[{{Hadjiyska} {et~al.}(2011){Hadjiyska}, {Rabinowitz}, {Baltay},
  {Zinn}, {Coppi}, {Ellman}, \& {Miller}}]{Hadjiyska11}
{Hadjiyska}, E.~I., {Rabinowitz}, D., {Baltay}, C., {Zinn}, R., {Coppi}, P.,
  {Ellman}, N., \& {Miller}, L.~R. 2011, in Bulletin of the American
  Astronomical Society, Vol.~43, American Astronomical Society Meeting
  Abstracts \#217, 433.18



\bibitem[{{Haines} {et~al.}(2008){Haines}, {Gargiulo}, \&
  {Merluzzi}}]{Haines08}
{Haines}, C.~P., {Gargiulo}, A., \& {Merluzzi}, P. 2008, \mnras, 385, 1201

\bibitem[{{H{\"a}ring} \& {Rix}(2004)}]{haringRix04}
{H{\"a}ring}, N., \& {Rix}, H. 2004, \apjl, 604, L89

\bibitem[{{Heckman}(1980)}]{heckman80}
{Heckman}, T.~M. 1980, \aap, 87, 152

\bibitem[{{Heckman} {et~al.}(2004){Heckman}, {Kauffmann}, {Brinchmann},
  {Charlot}, {Tremonti}, \& {White}}]{heckman04}
{Heckman}, T.~M., {Kauffmann}, G., {Brinchmann}, J., {Charlot}, S., {Tremonti},
  C., \& {White}, S.~D.~M. 2004, \apj, 613, 109

\bibitem[{{Hills}(1975)}]{Hills75}
{Hills}, J.~G. 1975, \nat, 254, 295

\bibitem[{{Ho}(1999)}]{Ho99}
{Ho}, L.~C. 1999, \apj, 516, 672

\bibitem[{{Ho}(2004)}]{Ho04}
{Ho}, L.~C., ed. 2004, {Carnegie Observatories Astrophysics Series, Vol. 1:
  Coevolution of Black Holes and Galaxies} (Cambridge: Cambridge Univ. Press)

\bibitem[{{Holtzman} {et~al.}(2008){Holtzman}, {Marriner}, {Kessler}, {Sako},
  {Dilday}, {Frieman}, {Schneider}, {Bassett}, {Becker}, {Cinabro}, {DeJongh},
  {Depoy}, {Doi}, {Garnavich}, {Hogan}, {Jha}, {Konishi}, {Lampeitl},
  {Marshall}, {McGinnis}, {Miknaitis}, {Nichol}, {Prieto}, {Riess}, {Richmond},
  {Romani}, {Smith}, {Takanashi}, {Tokita}, {van der Heyden}, {Yasuda}, \&
  {Zheng}}]{Holtzman08}
{Holtzman}, J.~A., {et~al.} 2008, \aj, 136, 2306

\bibitem[{{Ivezi{\'c}} {et~al.}(2008){Ivezi{\'c}}, {Tyson}, {Allsman},
  {Andrew}, \& {Angel}}]{Ivezic08}
{Ivezi{\'c}}, Z., {Tyson}, J.~A., {Allsman}, R., {Andrew}, J., \& {Angel}, R.
  2008, ArXiv:0805.2366

\bibitem[{{Ivezi{\'c}} {et~al.}(2001){Ivezi{\'c}}, {Tabachnik}, {Rafikov},
  {Lupton}, {Quinn}, {Hammergren}, {Eyer}, {Chu}, {Armstrong}, {Fan},
  {Finlator}, {Geballe}, {Gunn}, {Hennessy}, {Knapp}, {Leggett}, {Munn},
  {Pier}, {Rockosi}, {Schneider}, {Strauss}, {Yanny}, {Brinkmann}, {Csabai},
  {Hindsley}, {Kent}, {Lamb}, {Margon}, {McKay}, {Smith}, {Waddel}, {York}, \&
  {the SDSS Collaboration}}]{ivezic01}
{Ivezi{\'c}}, {\v Z}., {et~al.} 2001, \aj, 122, 2749

\bibitem[{{Jeong} {et~al.}(2009){Jeong}, {Yi}, {Bureau}, {Davies},
  {Falc{\'o}n-Barroso}, {van de Ven}, {Peletier}, {Bacon}, {Cappellari}, {de
  Zeeuw}, {Emsellem}, {Krajnovi{\'c}}, {Kuntschner}, {McDermid}, {Sarzi}, \&
  {van den Bosch}}]{jeong09}
{Jeong}, H., {et~al.} 2009, \mnras, 398, 2028

\bibitem[{{Jester} {et~al.}(2005){Jester}, {Schneider}, {Richards}, {Green},
  {Schmidt}, {Hall}, {Strauss}, {Vanden Berk}, {Stoughton}, {Gunn},
  {Brinkmann}, {Kent}, {Smith}, {Tucker}, \& {Yanny}}]{Jester05}
{Jester}, S., {et~al.} 2005, \aj, 130, 873

\bibitem[{{Kauffmann} {et~al.}(2003){Kauffmann}, {Heckman}, {Tremonti},
  {Brinchmann}, {Charlot}, {White}, {Ridgway}, {Brinkmann}, {Fukugita}, {Hall},
  {Ivezi{\'c}}, {Richards}, \& {Schneider}}]{kauffmann03}
{Kauffmann}, G., {et~al.} 2003, \mnras, 346, 1055

\bibitem[{{Kelly} {et~al.}(2009){Kelly}, {Bechtold}, \&
  {Siemiginowska}}]{Kelly09}
{Kelly}, B.~C., {Bechtold}, J., \& {Siemiginowska}, A. 2009, \apj, 698, 895

\bibitem[{{Kelly} {et~al.}(2008){Kelly}, {Kirshner}, \& {Pahre}}]{Kelly08}
{Kelly}, P.~L., {Kirshner}, R.~P., \& {Pahre}, M. 2008, \apj, 687, 1201

\bibitem[{{Kennicutt}(1998)}]{Kennicutt98}
{Kennicutt}, Jr., R.~C. 1998, \araa, 36, 189

\bibitem[{{Kiewe} {et~al.}(2010){Kiewe}, {Gal-Yam}, {Arcavi}, {Leonard},
  {Emilio Enriquez}, {Cenko}, {Fox}, {Moon}, {Sand}, \& {Soderberg}}]{Kiewe10}
{Kiewe}, M., {et~al.} 2010, ArXiv:1010.2689

\bibitem[{{Komossa}(2002)}]{Komossa02}
{Komossa}, S. 2002, in Lighthouses of the Universe: The Most Luminous Celestial
  Objects and Their Use for Cosmology, ed. {M.~Gilfanov, R.~Sunyeav, \&
  E.~Churazov}, 436

\bibitem[{{Komossa} \& {Bade}(1999)}]{KomossaBade99}
{Komossa}, S., \& {Bade}, N. 1999, \aap, 343, 775

\bibitem[{{Komossa} {et~al.}(2009){Komossa}, {Zhou}, {Rau}, {Dopita},
  {Gal-Yam}, {Greiner}, {Zuther}, {Salvato}, {Xu}, {Lu}, {Saxton}, \&
  {Ajello}}]{Komossa09}
{Komossa}, S., {et~al.} 2009, \apj, 701, 105

\bibitem[{{Law} {et~al.}(2009){Law}, {Kulkarni}, {Dekany}, {Ofek}, {Quimby},
  {Nugent}, {Surace}, {Grillmair}, {Bloom}, {Kasliwal}, {Bildsten}, {Brown},
  {Cenko}, {Ciardi}, {Croner}, {Djorgovski}, {van Eyken}, {Filippenko}, {Fox},
  {Gal-Yam}, {Hale}, {Hamam}, {Helou}, {Henning}, {Howell}, {Jacobsen},
  {Laher}, {Mattingly}, {McKenna}, {Pickles}, {Poznanski}, {Rahmer}, {Rau},
  {Rosing}, {Shara}, {Smith}, {Starr}, {Sullivan}, {Velur}, {Walters}, \&
  {Zolkower}}]{Law09}
{Law}, N.~M., {et~al.} 2009, \pasp, 121, 1395

\bibitem[{{Leaman} {et~al.}(2010){Leaman}, {Li}, {Chornock}, \&
  {Filippenko}}]{Leaman10}
{Leaman}, J., {Li}, W., {Chornock}, R., \& {Filippenko}, A.~V. 2010, \mnras, 412, 1419

\bibitem[{{Li} {et~al.}(2010){Li}, {Leaman}, {Chornock}, {Filippenko},
  {Poznanski}, {Ganeshalingam}, {Wang}, {Modjaz}, {Jha}, {Foley}, \&
  {Smith}}]{Li10}
{Li}, W., {et~al.} 2010,\mnras, 412, 1441

\bibitem[{{Lidskii} \& {Ozernoi}(1979)}]{Lidskii_Ozernoi79}
{Lidskii}, V.~V., \& {Ozernoi}, L.~M. 1979, Soviet Astronomy Letters, 5, 16

\bibitem[{{Lodato} {et~al.}(2009){Lodato}, {King}, \& {Pringle}}]{Lodato09}
{Lodato}, G., {King}, A.~R., \& {Pringle}, J.~E. 2009, \mnras, 392, 332

\bibitem[{{Lodato} \& {Rossi}(2011)}]{Lodato_Rossi10}
{Lodato}, G., \& {Rossi}, E. 2011, \mnras, 410, 359

\bibitem[{{Loeb} \& {Ulmer}(1997)}]{loebUlmer97}
{Loeb}, A., \& {Ulmer}, A. 1997, \apj, 489, 573

\bibitem[{{Lupton} {et~al.}(2001){Lupton}, {Gunn}, {Ivezi{\'c}}, {Knapp}, \&
  {Kent}}]{lupton_gunn01}
{Lupton}, R., {Gunn}, J.~E., {Ivezi{\'c}}, Z., {Knapp}, G.~R., \& {Kent}, S.
  2001, Astronomical Data Analysis Software and Systems X, 238, 269

\bibitem[{{MacLeod} {et~al.}(2010){MacLeod}, {Ivezi{\'c}}, {Kochanek},
  {Koz{\l}owski}, {Kelly}, {Bullock}, {Kimball}, {Sesar}, {Westman}, {Brooks},
  {Gibson}, {Becker}, \& {de Vries}}]{Macleod10}
{MacLeod}, C.~L., {et~al.} 2010, \apj, 721, 1014

\bibitem[{{Maksym} {et~al.}(2010){Maksym}, {Ulmer}, \& {Eracleous}}]{Maksym10}
{Maksym}, P., {Ulmer}, M.~P., \& {Eracleous}, M. 2010, \apj, 772, 1035

\bibitem[{{Maoz} {et~al.}(2005){Maoz}, {Nagar}, {Falcke}, \& {Wilson}}]{Maoz05}
{Maoz}, D., {Nagar}, N.~M., {Falcke}, H., \& {Wilson}, A.~S. 2005, \apj, 625,
  699

\bibitem[{{Marconi} \& {Hunt}(2003)}]{Marconi_Hunt03}
{Marconi}, A., \& {Hunt}, L.~K. 2003, \apjl, 589, L21

\bibitem[{{Miller} {et~al.}(2009){Miller}, {Chornock}, {Perley},
  {Ganeshalingam}, {Li}, {Butler}, {Bloom}, {Smith}, {Modjaz}, {Poznanski},
  {Filippenko}, {Griffith}, {Shiode}, \& {Silverman}}]{Miller09}
{Miller}, A.~A., {et~al.} 2009, \apj, 690, 1303

\bibitem[{{Morrissey} {et~al.}(2007){Morrissey}, {Conrow}, {Barlow}, {Small},
  {Seibert}, {Wyder}, {Budav{\'a}ri}, {Arnouts}, {Friedman}, {Forster},
  {Martin}, {Neff}, {Schiminovich}, {Bianchi}, {Donas}, {Heckman}, {Lee},
  {Madore}, {Milliard}, {Rich}, {Szalay}, {Welsh}, \& {Yi}}]{Morrissey07}
{Morrissey}, P., {et~al.} 2007, \apjs, 173, 682

\bibitem[{{Murayama} \& {Taniguchi}(1998)}]{Murayama98}
{Murayama}, T., \& {Taniguchi}, Y. 1998, \apjl, 497, L9

\bibitem[{{Oke}(1974)}]{oke74}
{Oke}, J.~B. 1974, \apjs, 27, 21

\bibitem[{{{\"O}stman} {et~al.}(2011){{\"O}stman}, {Nordin}, {Goobar},
  {Amanullah}, {Smith}, {Sollerman}, {Stanishev}, {Stritzinger}, {Bassett},
  {Davis}, {Edmondson}, {Frieman}, {Garnavich}, {Lampeitl}, {Leloudas},
  {Marriner}, {Nichol}, {Romer}, {Sako}, {Schneider}, \& {Zheng}}]{Ostman11}
{{\"O}stman}, L., {et~al.} 2011, \aap, 526, A28

\bibitem[{{Owens}(1967)}]{Owens67}
{Owens}, J.~C. 1967, Appl. Opt., 6, 51

\bibitem[{{Pastorello} {et~al.}(2002){Pastorello}, {Turatto}, {Benetti},
  {Cappellaro}, {Danziger}, {Mazzali}, {Patat}, {Filippenko}, {Schlegel}, \&
  {Matheson}}]{Pastorello02}
{Pastorello}, A., {et~al.} 2002, \mnras, 333, 27

\bibitem[{{Phinney}(1989)}]{Phinney89}
{Phinney}, E.~S. 1989, in IAU Symposium, Vol. 136, The Center of the Galaxy,
  ed. {M.~Morris}, 543--553

\bibitem[{{Pier} {et~al.}(2003){Pier}, {Munn}, {Hindsley}, {Hennessy}, {Kent},
  {Lupton}, \& {Ivezi{\'c}}}]{pier03}
{Pier}, J.~R., {Munn}, J.~A., {Hindsley}, R.~B., {Hennessy}, G.~S., {Kent},
  S.~M., {Lupton}, R.~H., \& {Ivezi{\'c}}, {\v Z}. 2003, \aj, 125, 1559

\bibitem[{{Plotkin} {et~al.}(2010{\natexlab{a}}){Plotkin}, {Anderson},
  {Brandt}, {Diamond-Stanic}, {Fan}, {MacLeod}, {Schneider}, \&
  {Shemmer}}]{Plotkin10}
{Plotkin}, R.~M., {Anderson}, S.~F., {Brandt}, W.~N., {Diamond-Stanic}, A.~M.,
  {Fan}, X., {MacLeod}, C.~L., {Schneider}, D.~P., \& {Shemmer}, O.
  2010{\natexlab{a}}, \apj, 721, 562

\bibitem[{{Plotkin} {et~al.}(2010{\natexlab{b}}){Plotkin}, {Anderson},
  {Brandt}, {Diamond-Stanic}, {Fan}, {Hall}, {Kimball}, {Richmond},
  {Schneider}, {Shemmer}, {Voges}, {York}, {Bahcall}, {Snedden}, {Bizyaev},
  {Brewington}, {Malanushenko}, {Malanushenko}, {Oravetz}, {Pan}, \&
  {Simmons}}]{Plotkin10b}
{Plotkin}, R.~M., {et~al.} 2010{\natexlab{b}}, \aj, 139, 390

\bibitem[{{Pojmanski}(2007)}]{Pojmanski07}
{Pojmanski}, G. 2007, IAUCirc., 8875, 1

\bibitem[{{Quimby}(2006)}]{Quimby06}
{Quimby}, R. 2006, Central Bureau Electronic Telegrams, 644, 1

\bibitem[{{Quimby} {et~al.}(2011){Quimby}, {Kulkarni}, {Kasliwal}, {Gal-Yam},
  {Arcavi}, {Sullivan}, {Nugent}, {Thomas}, {Howell}, {Bildsten}, {Bloom},
  {Theissen}, {Law}, {Dekany}, {Rahmer}, {Hale}, {Smith}, {Ofek}, {Zolkower},
  {Velur}, {Walters}, {Henning}, {Bui}, {McKenna}, {Poznanski}, {Cenko}, \&
  {Levitan}}]{Quimby09}
{Quimby}, R.~M., {et~al.} 2011, Nature, 487, 487

\bibitem[{{Ramirez-Ruiz} \& {Rosswog}(2009)}]{Ramirez-Ruiz09}
{Ramirez-Ruiz}, E., \& {Rosswog}, S. 2009, \apjl, 697, L77

\bibitem[{{Rees}(1988)}]{Rees88}
{Rees}, M.~J. 1988, \nat, 333, 523

\bibitem[{{Rest} {et~al.}(2009){Rest}, {Foley}, {Gezari}, {Narayan}, {Draine},
  {Olsen}, {Huber}, {Matheson}, {Garg}, {Welch}, {Becker}, {Challis},
  {Clocchiatti}, {Cook}, {Damke}, {Meixner}, {Miknaitis}, {Minniti}, {Morelli},
  {Nikolaev}, {Pignata}, {Prieto}, {Smith}, {Stubbs}, {Suntzeff}, {Walker},
  {Wood-Vasey}, {Zenteno}, {Wyrzykowski}, {Udalski}, {Szymanski}, {Kubiak},
  {Pietrzynski}, {Soszynski}, {Szewczyk}, {Ulaczyk}, \& {Poleski}}]{Rest09}
{Rest}, A., {et~al.} 2009, \apj, 729, 88

\bibitem[{{Richards} {et~al.}(2004){Richards}, {Nichol}, {Gray}, {Brunner},
  {Lupton}, {Vanden Berk}, {Chong}, {Weinstein}, {Schneider}, {Anderson},
  {Munn}, {Harris}, {Strauss}, {Fan}, {Gunn}, {Ivezi{\'c}}, {York},
  {Brinkmann}, \& {Moore}}]{Richards04}
{Richards}, G.~T., {et~al.} 2004, \apjs, 155, 257

\bibitem[{{Rigon} {et~al.}(2003){Rigon}, {Turatto}, {Benetti}, {Pastorello},
  {Cappellaro}, {Aretxaga}, {Vega}, {Chavushyan}, {Patat}, {Danziger}, \&
  {Salvo}}]{Rigon03}
{Rigon}, L., {et~al.} 2003, \mnras, 340, 191

\bibitem[{{Sako} {et~al.}(2008){Sako}, {Bassett}, {Becker}, {Cinabro},
  {DeJongh}, {Depoy}, {Dilday}, {Doi}, {Frieman}, {Garnavich}, {Hogan},
  {Holtzman}, {Jha}, {Kessler}, {Konishi}, {Lampeitl}, {Marriner}, {Miknaitis},
  {Nichol}, {Prieto}, {Riess}, {Richmond}, {Romani}, {Schneider}, {Smith},
  {Subba Rao}, {Takanashi}, {Tokita}, {van der Heyden}, {Yasuda}, {Zheng},
  {Barentine}, {Brewington}, {Choi}, {Dembicky}, {Harnavek}, {Ihara}, {Im},
  {Ketzeback}, {Kleinman}, {Krzesi{\'n}ski}, {Long}, {Malanushenko},
  {Malanushenko}, {McMillan}, {Morokuma}, {Nitta}, {Pan}, {Saurage}, \&
  {Snedden}}]{sako08}
{Sako}, M., {et~al.} 2008, \aj, 135, 348

\bibitem[{Saxton}{et~al.}(2011)]{Saxton11}
Saxton~R., Read~A., Esquej~P., Miniutti~G., and Alvarez~E., 2011, ArXiv:1106.3507

\bibitem[{{Schlegel} {et~al.}(1998){Schlegel}, {Finkbeiner}, \&
  {Davis}}]{Schlegel98}
{Schlegel}, D.~J., {Finkbeiner}, D.~P., \& {Davis}, M. 1998, \apj, 500, 525

\bibitem[{{Schneider} {et~al.}(2007){Schneider}, {Hall}, {Richards}, {Strauss},
  {Vanden Berk}, {Anderson}, {Brandt}, {Fan}, {Jester}, {Gray}, {Gunn},
  {SubbaRao}, {Thakar}, {Stoughton}, {Szalay}, {Yanny}, {York}, {Bahcall},
  {Barentine}, {Blanton}, {Brewington}, {Brinkmann}, {Brunner}, {Castander},
  {Csabai}, {Frieman}, {Fukugita}, {Harvanek}, {Hogg}, {Ivezi{\'c}}, {Kent},
  {Kleinman}, {Knapp}, {Kron}, {Krzesi{\'n}ski}, {Long}, {Lupton}, {Nitta},
  {Pier}, {Saxe}, {Shen}, {Snedden}, {Weinberg}, \& {Wu}}]{schneider07}
{Schneider}, D.~P., {et~al.} 2007, \aj, 134, 102

\bibitem[{{Sesar} {et~al.}(2007){Sesar}, {Ivezi{\'c}}, {Lupton}, {Juri{\'c}},
  {Gunn}, {Knapp}, {DeLee}, {Smith}, {Miknaitis}, {Lin}, {Tucker}, {Doi},
  {Tanaka}, {Fukugita}, {Holtzman}, {Kent}, {Yanny}, {Schlegel}, {Finkbeiner},
  {Padmanabhan}, {Rockosi}, {Bond}, {Lee}, {Stoughton}, {Jester}, {Harris},
  {Harding}, {Brinkmann}, {Schneider}, {York}, {Richmond}, \& {Vanden
  Berk}}]{sesar07}
{Sesar}, B., {et~al.} 2007, \aj, 134, 2236

\bibitem[{{Smith} {et~al.}(2002){Smith}, {Tucker}, {Kent}, {Richmond},
  {Fukugita}, {Ichikawa}, {Ichikawa}, {Jorgensen}, {Uomoto}, {Gunn}, {Hamabe},
  {Watanabe}, {Tolea}, {Henden}, {Annis}, {Pier}, {McKay}, {Brinkmann}, {Chen},
  {Holtzman}, {Shimasaku}, \& {York}}]{smith02}
{Smith}, J.~A., {et~al.} 2002, \aj, 123, 2121

\bibitem[{{Smith} {et~al.}(2007){Smith}, {Li}, {Foley}, {Wheeler}, {Pooley},
  {Chornock}, {Filippenko}, {Silverman}, {Quimby}, {Bloom}, \&
  {Hansen}}]{Smith07}
{Smith}, N., {et~al.} 2007, \apj, 666, 1116

\bibitem[{{Stoughton} {et~al.}(2002){Stoughton}, {Lupton}, {Bernardi},
  {Blanton}, {Burles}, {Castander}, {Connolly}, {Eisenstein}, {Frieman},
  {Hennessy}, {Hindsley}, {Ivezi{\'c}}, {Kent}, {Kunszt}, {Lee}, {Meiksin},
  {Munn}, {Newberg}, {Nichol}, {Nicinski}, {Pier}, {Richards}, {Richmond},
  {Schlegel}, {Smith}, {Strauss}, {SubbaRao}, {Szalay}, {Thakar}, {Tucker},
  {Vanden Berk}, {Yanny}, {Adelman}, {Anderson}, {Anderson}, {Annis},
  {Bahcall}, {Bakken}, {Bartelmann}, {Bastian}, {Bauer}, {Berman},
  {B{\"o}hringer}, {Boroski}, {Bracker}, {Briegel}, {Briggs}, {Brinkmann},
  {Brunner}, {Carey}, {Carr}, {Chen}, {Christian}, {Colestock}, {Crocker},
  {Csabai}, {Czarapata}, {Dalcanton}, {Davidsen}, {Davis}, {Dehnen},
  {Dodelson}, {Doi}, {Dombeck}, {Donahue}, {Ellman}, {Elms}, {Evans}, {Eyer},
  {Fan}, {Federwitz}, {Friedman}, {Fukugita}, {Gal}, {Gillespie}, {Glazebrook},
  {Gray}, {Grebel}, {Greenawalt}, {Greene}, {Gunn}, {de Haas}, {Haiman},
  {Haldeman}, {Hall}, {Hamabe}, {Hansen}, {Harris}, {Harris}, {Harvanek},
  {Hawley}, {Hayes}, {Heckman}, {Helmi}, {Henden}, {Hogan}, {Hogg}, {Holmgren},
  {Holtzman}, {Huang}, {Hull}, {Ichikawa}, {Ichikawa}, {Johnston}, {Kauffmann},
  {Kim}, {Kimball}, {Kinney}, {Klaene}, {Kleinman}, {Klypin}, {Knapp},
  {Korienek}, {Krolik}, {Kron}, {Krzesi{\'n}ski}, {Lamb}, {Leger},
  {Limmongkol}, {Lindenmeyer}, {Long}, {Loomis}, {Loveday}, {MacKinnon},
  {Mannery}, {Mantsch}, {Margon}, {McGehee}, {McKay}, {McLean}, {Menou},
  {Merelli}, {Mo}, {Monet}, {Nakamura}, {Narayanan}, {Nash}, {Neilsen},
  {Newman}, {Nitta}, {Odenkirchen}, {Okada}, {Okamura}, {Ostriker}, {Owen},
  {Pauls}, {Peoples}, {Peterson}, {Petravick}, {Pope}, {Pordes}, {Postman},
  {Prosapio}, {Quinn}, {Rechenmacher}, {Rivetta}, {Rix}, {Rockosi}, {Rosner},
  {Ruthmansdorfer}, {Sandford}, {Schneider}, {Scranton}, {Sekiguchi}, {Sergey},
  {Sheth}, {Shimasaku}, {Smee}, {Snedden}, {Stebbins}, {Stubbs}, {Szapudi},
  {Szkody}, {Szokoly}, {Tabachnik}, {Tsvetanov}, {Uomoto}, {Vogeley}, {Voges},
  {Waddell}, {Walterbos}, {Wang}, {Watanabe}, {Weinberg}, {White}, {White},
  {Wilhite}, {Wolfe}, {Yasuda}, {York}, {Zehavi}, \& {Zheng}}]{stoughton02}
{Stoughton}, C., {et~al.} 2002, \aj, 123, 485

\bibitem[{{Strubbe} \& {Quataert}(2009)}]{strubbe_quataert09}
{Strubbe}, L.~E., \& {Quataert}, E. 2009, \mnras, 400, 2070


\bibitem[{{Strubbe} \& {Quataert}(2011)}]{StrubbeQuataert11}
---. 2011, \mnras, 415, 168

\bibitem[{{The Pierre Auger Collaboration}(2007)}]{Auger07}
{The Pierre Auger Collaboration}. 2007, Science, 318, 938

\bibitem[{{The Pierre Auger Collaboration}(2008)}]{Auger08}
---. 2008, Astroparticle Physics, 29, 188

\bibitem[{{Tonry} \& {Davis}(1979)}]{Tonry79}
{Tonry}, J., \& {Davis}, M. 1979, \aj, 84, 1511

\bibitem[{{Tundo} {et~al.}(2007){Tundo}, {Bernardi}, {Hyde}, {Sheth}, \&
  {Pizzella}}]{Tundo07}
{Tundo}, E., {Bernardi}, M., {Hyde}, J.~B., {Sheth}, R.~K., \& {Pizzella}, A.
  2007, \apj, 663, 53

\bibitem[{{Ulmer}(1999)}]{Ulmer99}
{Ulmer}, A. 1999, \apj, 514, 180

\bibitem[{{van Velzen} {et~al.}(2010){van Velzen}, {Farrar}, {Gezari},
  {Morrell}, {Zaritsky}, {Ostman}, {Smith}, \& {Gelfand}}]{vanVelzen10}
{van Velzen}, S., {Farrar}, G.~R., {Gezari}, S., {Morrell}, N., {Zaritsky}, D.,
  {Ostman}, L., {Smith}, M., \& {Gelfand}, J. 2010, ArXiv:1009.1627

\bibitem[{{Vanden Berk} {et~al.}(2001){Vanden Berk}, {Richards}, {Bauer},
  {Strauss}, {Schneider}, {Heckman}, {York}, {Hall}, {Fan}, {Knapp},
  {Anderson}, {Annis}, {Bahcall}, {Bernardi}, {Briggs}, {Brinkmann}, {Brunner},
  {Burles}, {Carey}, {Castander}, {Connolly}, {Crocker}, {Csabai}, {Doi},
  {Finkbeiner}, {Friedman}, {Frieman}, {Fukugita}, {Gunn}, {Hennessy},
  {Ivezi{\'c}}, {Kent}, {Kunszt}, {Lamb}, {Leger}, {Long}, {Loveday}, {Lupton},
  {Meiksin}, {Merelli}, {Munn}, {Newberg}, {Newcomb}, {Nichol}, {Owen}, {Pier},
  {Pope}, {Rockosi}, {Schlegel}, {Siegmund}, {Smee}, {Snir}, {Stoughton},
  {Stubbs}, {SubbaRao}, {Szalay}, {Szokoly}, {Tremonti}, {Uomoto}, {Waddell},
  {Yanny}, \& {Zheng}}]{VandenBerk01}
{Vanden Berk}, D.~E., {et~al.} 2001, \aj, 122, 549

\bibitem[{{Vanden Berk} {et~al.}(2002){Vanden Berk}, {Lee}, {Wilhite},
  {Beacom}, {Lamb}, {Annis}, {Abazajian}, {McKay}, {Kron}, {Kent}, {Hurley},
  {Kehoe}, {Wren}, {Henden}, {York}, {Schneider}, {Adelman}, {Brinkmann},
  {Brunner}, {Csabai}, {Harvanek}, {Hennessy}, {Ivezi{\'c}}, {Kleinman},
  {Kleinman}, {Krzesinski}, {Long}, {Neilsen}, {Newman}, {Snedden},
  {Stoughton}, {Tucker}, \& {Yanny}}]{VandenBerk02}
---. 2002, \apj, 576, 673

\bibitem[{{Walcher} {et~al.}(2006){Walcher}, {B{\"o}ker}, {Charlot}, {Ho},
  {Rix}, {Rossa}, {Shields}, \& {van der Marel}}]{Walcher06}
{Walcher}, C.~J., {B{\"o}ker}, T., {Charlot}, S., {Ho}, L.~C., {Rix}, H.,
  {Rossa}, J., {Shields}, J.~C., \& {van der Marel}, R.~P. 2006, \apj, 649, 692

\bibitem[{{Walcher} {et~al.}(2005){Walcher}, {van der Marel}, {McLaughlin},
  {Rix}, {B{\"o}ker}, {H{\"a}ring}, {Ho}, {Sarzi}, \& {Shields}}]{Walcher05}
{Walcher}, C.~J., {et~al.} 2005, \apj, 618, 237

\bibitem[{{York} {et~al.}(2000){York}, {Adelman}, {Anderson}, {Anderson},
  {Annis}, {Bahcall}, {Bakken}, {Barkhouser}, {Bastian}, {Berman}, {Boroski},
  {Bracker}, {Briegel}, {Briggs}, {Brinkmann}, {Brunner}, {Burles}, {Carey},
  {Carr}, {Castander}, {Chen}, {Colestock}, {Connolly}, {Crocker}, {Csabai},
  {Czarapata}, {Davis}, {Doi}, {Dombeck}, {Eisenstein}, {Ellman}, {Elms},
  {Evans}, {Fan}, {Federwitz}, {Fiscelli}, {Friedman}, {Frieman}, {Fukugita},
  {Gillespie}, {Gunn}, {Gurbani}, {de Haas}, {Haldeman}, {Harris}, {Hayes},
  {Heckman}, {Hennessy}, {Hindsley}, {Holm}, {Holmgren}, {Huang}, {Hull},
  {Husby}, {Ichikawa}, {Ichikawa}, {Ivezi{\'c}}, {Kent}, {Kim}, {Kinney},
  {Klaene}, {Kleinman}, {Kleinman}, {Knapp}, {Korienek}, {Kron}, {Kunszt},
  {Lamb}, {Lee}, {Leger}, {Limmongkol}, {Lindenmeyer}, {Long}, {Loomis},
  {Loveday}, {Lucinio}, {Lupton}, {MacKinnon}, {Mannery}, {Mantsch}, {Margon},
  {McGehee}, {McKay}, {Meiksin}, {Merelli}, {Monet}, {Munn}, {Narayanan},
  {Nash}, {Neilsen}, {Neswold}, {Newberg}, {Nichol}, {Nicinski}, {Nonino},
  {Okada}, {Okamura}, {Ostriker}, {Owen}, {Pauls}, {Peoples}, {Peterson},
  {Petravick}, {Pier}, {Pope}, {Pordes}, {Prosapio}, {Rechenmacher}, {Quinn},
  {Richards}, {Richmond}, {Rivetta}, {Rockosi}, {Ruthmansdorfer}, {Sandford},
  {Schlegel}, {Schneider}, {Sekiguchi}, {Sergey}, {Shimasaku}, {Siegmund},
  {Smee}, {Smith}, {Snedden}, {Stone}, {Stoughton}, {Strauss}, {Stubbs},
  {SubbaRao}, {Szalay}, {Szapudi}, {Szokoly}, {Thakar}, {Tremonti}, {Tucker},
  {Uomoto}, {Vanden Berk}, {Vogeley}, {Waddell}, {Wang}, {Watanabe},
  {Weinberg}, {Yanny}, \& {Yasuda}}]{york02}
{York}, D.~G., {et~al.} 2000, \aj, 120, 1579

\bibitem[{{Yuan} \& {Akerlof}(2008)}]{yuan_akerlof08}
{Yuan}, F., \& {Akerlof}, C.~W. 2008, \apj, 677, 808

\bibitem[{{Zaw} {et~al.}(2009){Zaw}, {Farrar}, \& {Greene}}]{Zaw09}
{Zaw}, I., {Farrar}, G.~R., \& {Greene}, J.~E. 2009, \apj, 696, 1218

\end{thebibliography}

\end{document}